\documentclass[twocolumn,showpacs,prb,aps,amsmath,amssymb,superscriptaddress,floatfix]{revtex4-1}

\usepackage{graphicx}% Include figure files
\usepackage{epstopdf}

\usepackage{bm}% bold math
\begin{document}

\title{
Spin nematics in frustrated spin-dimer systems with bilayer structure
}
\author{Toshiya Hikihara}
\affiliation{Faculty of Science and Technology, Gunma University,
Kiryu, Gunma 376-8515, Japan}
\author{Takahiro Misawa}
\affiliation{Institute for Solid State Physics, University of Tokyo,
5-1-5 Kashiwanoha, Kashiwa, Chiba 277-8581, Japan}
\author{Tsutomu Momoi}
\affiliation{Condensed Matter Theory Laboratory, RIKEN,
Wako, Saitama 351-0198, Japan}
\affiliation{RIKEN Center for Emergent Matter Science
(CEMS), Wako, Saitama, 351-0198, Japan}
\date{\today}

\begin{abstract}
We study frustrated spin-1/2 dimer systems  in two dimensions with a bilayer structure, where
spins are ferromagnetically coupled in dimers.
Our model includes frustrated two-spin exchange interactions as well as four-spin interaction.
We pay particular attention to the spin nematic phase, which does not exhibit any magnetic (spin-dipole) order but has a spin-quadrupolar long-range order.
Employing a perturbation calculation, a mean-field approximation,
and a numerical many-variable variational Monte Carlo method, we determine ground-state phase diagrams on various two-dimensional lattices.
It is found that the model exhibits the spin nematic phase with ferro-quadrupolar order in a wide parameter region,
in addition to conventional magnetically-ordered phases.
In particular, it is shown that even when the four-spin interactions are absent, frustrated two-spin exchange interactions
can realize the spin nematic phase as a result of strong interdimer correlations.
It is also found that the phase transitions between the spin nematic phase and antiferromagnetic phases can be continuous.
Furthermore, we present some exact arguments that various phases including the spin nematic phase and the vector chiral (p-type nematic) phase emerge  from an SU(4) symmetric point in the model by the addition of  appropriate perturbative interactions.
The spin nematic phase generated from the SU(4) point is connected with the spin nematic phase found numerically
in the system with only two-spin interactions.
\end{abstract}

\maketitle

\section{Introduction}\label{sec:Intro}
%%%% Spin nematic state
The search for spin nematic states has been under active investigation in the last decade.
The spin nematic state is characterized by the absence of any magnetic long-range order (except a trivial magnetization in
an applied magnetic field) and the spontaneous symmetry breaking of spin rotation  accompanied with a spin-quadrupolar long-range order.\cite{Andreev1984}
Because of these peculiar properties,  the spin nematic state  is  a novel, intriguing non-magnetic state with a hidden order.

%%%% Theoretical models
In theoretical studies, there are already many proposals of the spin models that exhibit the spin nematic order at low temperatures.
A typical example in such models is the spin-1 bilinear-biquadratic model,\cite{Blume1969,Chen1971,Papanicolaou1984}
which includes bilinear (two-spin) and biquadratic (four-spin) exchange interactions.
The appearance of the spin nematic phase on the cubic lattice was shown for sufficiently large biquadratic interactions
by a rigorous proof~\cite{Tanaka2001} and quantum Monte Carlo simulations.\cite{Harada2002}
It has been also shown that  the spin-1 bilinear-biquadratic model exhibits the spin-nematic ground-state phases with ferro- or
antiferro-quadrupolar order on various two-dimensional lattices.\cite{TothLMP2012,NiesenC2017,Zhaoetal2012,LauchliMP2006,Liuetal2015}
Another example of the models that show the spin nematic phase is a family of spin-1/2 frustrated ferromagnets which include ferromagnetic (FM) exchange interactions and competing antiferromagnetic (AFM) ones.\cite{Shannon2006,Shindou2011,Chubukov1991,Hikihara2008,Sudan2009,Zhitomirsky2010,Ueda2013,Janson2016}
In this case, the instability leading to the spin nematic phase appears in the saturated state  in an external field,
where two magnons form a bound state.
When the external field decreases below the saturation field, those bound magnon pairs condense,
which leads to spin nematicity.\cite{Momoi2005,Shannon2006}
There are also a few other examples which show  the spin-nematic ground state in anisotropic spin models\cite{Schulz1986,ChenHS2003,DamleS2006}, the Heisenberg model on Shastry-Sutherland
lattice,\cite{MomoiT2000,WangB2018} and models with multiple-spin ring exchanges.\cite{Momoi2005,HikiharaY2008,momoi2012,YokoyamaH2018}

%%%% Lack of cadidate materials (at zero field)
In real materials, however, candidate materials for the spin nematic state are rather limited.
%\sout{and furthermore
%experimental detection of the spin nematic phase itself is not easy}.
The biquadratic interaction in spin-1 systems is not very strong in general,
but the realization of  the spin nematic state in the spin-1 systems  requires relatively strong biquadratic interaction comparable to or larger than the bilinear exchange interaction.
We note that there are some proposals to enhance the ratio of the biquadratic interaction to the bilinear one.\cite{MilaZ2000,TanakaYH2018}
In the case of spin-1/2 frustrated ferromagnets, the spin nematic phase can appear only
in a narrow parameter range at zero field and it appears in a wider parameter space in a high magnetic field,\cite{Sindzingre2009}
which is not easy to access in experiments.
Nevertheless, active studies on spin-1/2 frustrated ferromagnets are on going, for example,
in
the quasi-one-dimensional spin-1/2 frustrated ferromagnets LiCuVO$_4$\cite{Nawa2017,Orlova2017} and Rb$_2$Cu$_2$Mo$_3$O$_{12}$,\cite{Matsuietal2017}
and a two-dimensional kagome compound, volborthite Cu$_3$V$_2$O$_7$(OH)$_2 \cdot$2H$_2$O.\cite{Yoshida2017,Kohama2019}
For stimulating further studies, it is desirable to search still more theoretical models for describing the spin nematic phase
at zero field
in the systems with only two-spin interactions, which are easier to access experimentally.

%%%%Lack of secure starting point for theoretical understanding of S=1/2 spin systems (at zero field)
%Another issue we address here is how to understand and describe the emergence of the spin nematic phase.
Theoretical studies on spin nematics in spin-1/2 systems often face technical difficulties.
The appearance of spin nematic phases is usually understood from the bi-magnon instability in the fully polarized
      state,\cite{Shannon2006,Kecke2007,momoi2012} where some of the induced spin nematic phases are expected to remain down to low magnetization regime.
Nevertheless, even in the spin-1/2 square-lattice $J_1$-$J_2$ model with ferromagnetic $J_1$,
which is  thought to be a typical model for spin
  nematic ordering,\cite{Shannon2006} there still exists a certain controversy if the spin nematic order remains at zero field.\cite{Richter2010,Shindou2011}
  For a better understanding of spin nematic ordering at zero field, it is hence helpful to find some specific spin-1/2 models
  which definitely show spin nematic phases at zero magnetic field.
In this context, the mechanism of spin nematics in spin-1 systems is instructive.
 In the spin-1 bilinear-biquadratic model,  it is known that  an SU(3) symmetry
 %, appearing in specific parameter points where the biquadratic interaction has the same strength as  the bilinear interaction,
  plays an important  role in the emergence of spin nematic phases  at zero field.\cite{Papanicolaou1988,Penc2011}
Referring to this, we also  take into account the effect of four-spin interactions in our spin-1/2 systems,
to elucidate mechanism of spin nematic ordering.
% in our model with only two-spin interactions.
We will see that an SU(4) symmetry appearing in our model at a certain parameter point serves as an origin of the spin-nematic phase.

%%%% Our study
The aim of this work is twofold:
One is to propose a spin-1/2 model that requires neither the four-spin interaction nor magnetic field for realizing the spin nematic state. The other is to clarify the mechanism of this spin nematic ordering by considering four-spin interactions as well.
%role of SU(4) symmetry in the emergence of the spin nematic state.
For these purposes, we introduce in this paper a spin-1/2 frustrated ferromagnetic model in
two dimensions.
The model
 consists of ferromagnetically coupled dimers of $S=1/2$ spins forming a bilayer structure and contains frustrated bilinear Heisenberg exchange interactions and a type of four-spin interaction [see Eq.\ (\ref{eq:Ham}) and Fig.\ \ref{fig:model} in Sec.\ \ref{sec:model}].
The parameter space of the model includes the effective spin-1 bilinear-biquadratic model and an SU(4) symmetric model,
which enables us to show the mechanism of spin nematics.
Using analytical and numerical techniques, we obtain the following results.

%In this paper, we introduce a spin-1/2 frustrated ferromagnetic model
%which can have a spin-nematic ground state at zero field.
%The model consists of coupled dimers of $S=1/2$ spins, which form
%a two-dimensional bilayer lattice of the spins, and includes frustrated
%bilinear Heisenberg exchange interactions.
%To understand the mechanism of this spin nematic ordering we further
%study the effect of  four-spin
%interaction [see Eq.\ (\ref{eq:Ham}) and Fig.\ \ref{fig:model}
%in Sec.\ \ref{sec:model}].
%Using analytical and numerical techniques, we obtain the following results.

First, we analyze the strong ferromagnetic dimer limit.
Performing a perturbative calculation, we map the model to the spin-1 bilinear-biquadratic model.
With the help of former studies on spin-1 systems, we can determine the ground-state phase diagrams
on various lattices, which include the spin nematic phases in wide parameter regions.
In particular, we show that, even when the four-spin interaction is absent, a second-order perturbation %of bilinear exchanges
yields an {\it effective} biquadratic interaction, which leads to the appearance of the spin nematic phase with ferro-quadrupolar order in a parameter regime where all the first-order perturbations are canceled to each other.

Second, we study how the spin nematic phase emerging at the strong dimer limit is affected by decreasing the ferromagnetic intradimer coupling from infinite to
finite values.
We use
a mean-field approximation with product-state ansatz and the many-variable variational Monte Carlo (mVMC)
method.\cite{Misawa2018,mVMC}
In the mean-field approximation, we determine the ground-state phase diagram
as a function of two-spin and
%for the ferromagnetic intradimer-exchange and  negative
four-spin interactions.
In addition to conventional magnetically-ordered phases,  the spin nematic phase with the ferro-quadrupolar order
appears in a wide parameter region.
In this mean-field approximation, the spin nematic phase vanishes when the four-spin interaction is deleted.
The result also reveals that this spin nematic phase connects with the SU(4) symmetric point
in our model.
To further take account of the effects of interdimer correlations,
we study the model with only the two-spin interactions using mVMC method.
The obtained phase diagrams  for the square and triangular lattices show that, for large but finite intradimer interactions, the spin nematic phase emerges in a finite parameter range.
This result is consistent with our aforementioned discussion from the second-order perturbation
and thereby confirms  that our model exhibits the spin nematic phase even when it does not include any four-spin
interaction or external field.

Lastly, we focus on the model around the SU(4) symmetric point.
From an exact symmetry argument, we find that
our model at the SU(4) point has degenerate ground states
including the spin nematic
state and the vector chiral (also known as p-type nematic\cite{Andreev1984}) state.
It also shows that our mean-field approximation indeed presents exact ground states in the SU(4) model.
Thus, the SU(4) model is a source of various exotic phases and one can realize
each of them by adding perturbative interactions.
As an example of such a perturbation, we show that
an appropriate set of Ising-type interactions can yield
the spin nematic phase.

%%%% Organization of the paper
The rest of the paper is organized as follows:
In Sec.\ \ref{sec:model}, we introduce the model Hamiltonian, and also discuss
symmetric properties of the model and the order parameters studied.
In Sec.\ \ref{sec:perturbation}, we present the results of the perturbation calculations from the strong ferromagnetic dimer limit.
We further present the results of the mean-field approximation with a product state and
the mVMC
method in Secs.\ \ref{subsec:Vari_product} and \ref{subsec:mVMC}, respectively.
In Sec.\ \ref{sec:SU4} we discuss the exact arguments on emerging phases in the vicinity of the SU(4)-symmetric model
and also on the effect of perturbations of Ising interactions.
Section\ \ref{sec:conc} is devoted to summary and concluding remarks.
Details of the mean-field calculation, SU(4) transformation on dimers, and nontrivial degeneracies in  the mean-field solutions are discussed in Appendices\ \ref{sec:append_Num_MFA}, \ref{sec:append_su4}, and \ref{sec:append_degeneracy_MFA}, respectively.

\section{Model}\label{sec:model}

Here we introduce the spin model studied in this paper. We also present all local observables
which concern us and symmetries inherent in our
model.

\subsection{Hamiltonian}
\begin{figure}
\begin{center}
\includegraphics[width=70mm]{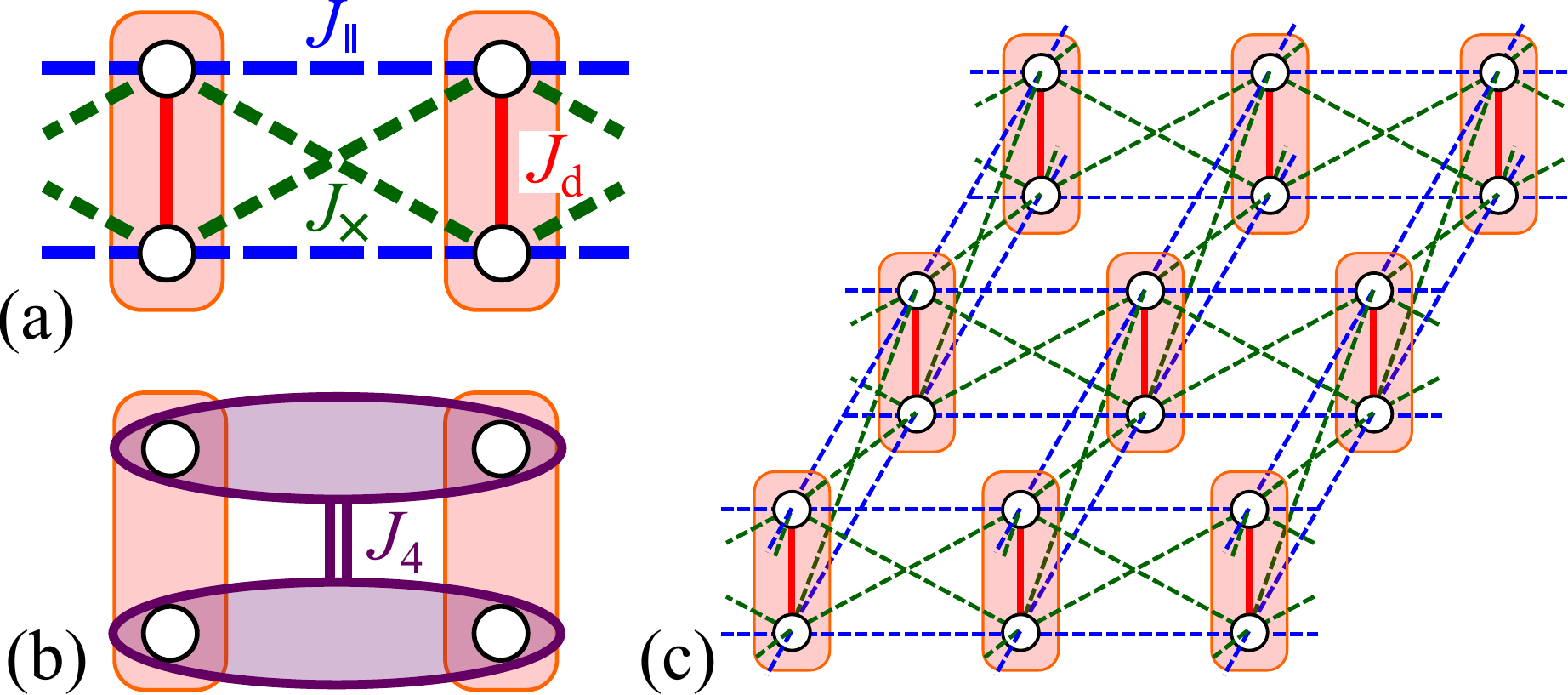}
\caption{
Schematic pictures of the model (\ref{eq:Ham}).
In all panels, circles and rectangles represent the spin-1/2 spins and their dimers, respectively.
(a) Two-spin exchange interactions.
Solid, dashed, and dotted lines represent the exchange interactions of $J_{\rm d}$, $J_\parallel$, and $J_\times$, respectively.
(b) Four-spin interaction $J_4$.
(c) The model in the square lattice.
Only the two-spin interactions are drawn for clarity.
}
\label{fig:model}
\end{center}
\end{figure}

We study the spin-1/2 frustrated quantum magnet consisting of spin-dimer units, which form a two-dimensional bilayer
structure of spins.
The model Hamiltonian has the form
\begin{subequations}
\label{eq:Ham}
\begin{eqnarray}
\mathcal{H}
&=& \mathcal{H}_{\rm d} + \mathcal{H}_\parallel
+ \mathcal{H}_\times + \mathcal{H}_4,
\label{eq:Ham_total} \\
\mathcal{H}_{\rm d}
&=& J_{\rm d} \sum_j {\bm S}_{1,j} \cdot {\bm S}_{2,j},
\label{eq:Ham_dim} \\
\mathcal{H}_\parallel
&=& J_\parallel \sum_{\langle j,j'\rangle} \left(
{\bm S}_{1,j} \cdot {\bm S}_{1,j'} + {\bm S}_{2,j} \cdot {\bm S}_{2,j'}
\right),
\label{eq:Ham_parallel} \\
\mathcal{H}_\times
&=& J_\times \sum_{\langle j,j'\rangle} \left(
{\bm S}_{1,j} \cdot {\bm S}_{2,j'} + {\bm S}_{2,j} \cdot {\bm S}_{1,j'}
\right),
\label{eq:Ham_diagonal} \\
\mathcal{H}_4
&=& J_4 \sum_{\langle j,j'\rangle}
\left( {\bm S}_{1,j} \cdot {\bm S}_{1,j'} \right)
\left( {\bm S}_{2,j} \cdot {\bm S}_{2,j'} \right),
\label{eq:Ham_four-spin}
\end{eqnarray}
\end{subequations}
where ${\bm S}_{l,j}$ is the spin-1/2 operator of the $l$th spin ($l=1,2$) in the $j$th dimer.
The dimer sites, labeled with $j$ or $j'$, form two-dimensional lattices and  the sum $\sum_{\langle j,j'\rangle}$ is taken for the nearest-neighboring sites in the  lattices.
We consider several lattices including the square, honeycomb, triangular, and kagome lattices.
Schematic pictures of the model are shown in Fig.\ \ref{fig:model}.
Throughout this paper we consider the case that the intradimer exchange interaction is ferromagnetic or zero, $J_{\rm d} \le 0$,
where two $S=1/2$ spins
in each dimer dominantly form a spin triplet.

\subsection{Local observables}\label{subsec:OP}
We explore local order parameters measured with the following local observables:
For a parallel spin order on a dimer bond, we use the total spin operators
on each dimer,
\begin{align}\label{eq:total_spin}
T_j^\alpha &= S_{1,j}^\alpha+S_{2,j}^\alpha
\end{align}
with $\alpha=x,y,z$ and, for an antiparallel spin order, N\'eel-spin operators
\begin{align}\label{eq:AF_spin}
N_j^\alpha &= S_{1,j}^\alpha-S_{2,j}^\alpha.
\end{align}
For a spin quadrupolar order on a dimer bond, we use
the five-component spin quadrupolar operators
\begin{subequations}
\label{eq:Quad_orig}
\begin{eqnarray}
Q^{(1)}_j &=& 2\left( S^x_{1,j}S^x_{2,j}-S^y_{1,j}S^y_{2,j}\right),
\label{eq:Quad_1_orig} \\
Q^{(2)}_j &=& \frac{2}{\sqrt{3}} \left( 2 S^z_{1,j}S^z_{2,j} - S^x_{1,j}S^x_{2,j} - S^y_{1,j}S^y_{2,j} \right),
\label{eq:Quad_2_orig} \\
Q^{(3)}_j &=& 2\left( S^x_{1,j}S^y_{2,j}+S^y_{1,j}S^x_{2,j}\right),
\label{eq:Quad_3_orig} \\
Q^{(4)}_j &=& 2\left( S^y_{1,j}S^z_{2,j}+S^z_{1,j}S^y_{2,j}\right),
\label{eq:Quad_4_orig} \\
Q^{(5)}_j &=& 2\left( S^z_{1,j}S^x_{2,j}+S^x_{1,j}S^z_{2,j}\right),
\label{eq:Quad_5_orig}
\end{eqnarray}
\end{subequations}
which act on two $S=1/2$ spins on each dimer.

These quadrupolar operators are a natural extension of the on-site quadrupolar
operators in spin-1 systems;
one can obtain the above operators by
inserting  the total spin operators $T^\alpha_j$ into the spin-1 operators
in the quadrupolar operators of spin-1 systems. (For the definition of the spin-1 quadrupolar operators, see for example Ref.~\onlinecite{Penc2011}.)
This derivation of Eq.~(\ref{eq:Quad_orig})
readily concludes that the commutation relations between the total-spin operators $T_j^\alpha$  and the quadrupolar operators
$Q_j^{(n)}$ are the same as those in spin-1 systems. Hence
the combined set of $T_j^\alpha$ ($\alpha=x,y,z$) and $Q_j^{(n)}$ ($n=1,\cdots,5$)
forms SU(3) group; $T_j^\alpha$ and $Q_j^{(n)}$ are eight-dimensional generators of
su(3) algebra.

Incidentally,
one can obtain the quadrupolar operators (\ref{eq:Quad_orig}) with a minus sign by
substituting the N\'eel-spin operators $N^\alpha_j$ into the $S=1$ spin operators
in the quadrupolar operators of spin-1 systems. From this fact, it also follows that
 the combined set of $N_j^\alpha$ ($\alpha=x,y,z$) and
$-Q_j^{(n)}$ ($n=1,\cdots,5$) also forms another SU(3) group.

\subsection{Symmetries}\label{subsec:symmetries}
In addition to the apparent SU(2) symmetry, the model (\ref{eq:Ham}) has the following higher symmetries in specific parameter spaces.

{\it
SU(3) symmetry: } In the parameter space defined by
\begin{eqnarray}
J_\parallel = J_4/4,\ \ \ \   J_\times = 0
\label{eq:SU3_line}
\end{eqnarray}
with any $J_{\rm d}$,
the total Hamiltonian (\ref{eq:Ham}) has a global SU(3) symmetry.
The eight generators of  SU(3) rotation are given
by $\sum_j T_j^\alpha$ ($\alpha=x,y,z$) and $\sum_j Q_j^{(n)}$ ($n=1,\cdots,5$).
In particular,
in the $J_{\rm d} \to -\infty$ limit,
the model (\ref{eq:Ham}) in the broader parameter space
\begin{eqnarray}\label{eq:SU(3)param}
J_\parallel - \frac{J_4}{4} + J_\times = 0
\end{eqnarray}
also has the same SU(3) symmetry. We note that this space (\ref{eq:SU(3)param})
includes the aforementioned space (\ref{eq:SU3_line}).
In this limit,
as the singlet state is gapped out in each dimer,
the model is reduced to a  spin-1 model. The model in the space (\ref{eq:SU(3)param})
is mapped to the SU(3)
symmetric spin-1 bilinear-biquadratic model (see Sec.\ \ref{sec:perturbation}).

{\it SU(4) symmetry:}
In the case of $J_{\rm d}=0$ among the SU(3) symmetric space (\ref{eq:SU3_line}), i.e.,
\begin{eqnarray}
J_\parallel = J_4/4, \ \ \ \  J_\times = J_{\rm d} = 0,
\label{eq:SU4_line}
\end{eqnarray}
the total Hamiltonian has a global SU(4) symmetry;
$\mathcal{H}$ commutes with the operators $\sum_j S^\alpha_{1,j}$, $\sum_j S^\alpha_{2,j}$ ($\alpha=x,y,z$), and $\sum_j S^\alpha_{1,j} S^\beta_{2,j}$ ($\alpha,\beta=x,y,z$), which  are known as the fifteen generators of SU(4) group.
The SU(3) group mentioned above is a subgroup of this SU(4) group.
This SU(4) symmetry also contains another SU(3) symmetry generated by $\sum_j N_j^\alpha$  and
$-\sum_j Q_j^{(n)}$.
We present our analysis on the SU(4) symmetric model with ferromagnetic coupling ($J_\parallel = J_4/4<0$)
in Sec.\ \ref{sec:SU4} and further describe SU(4) transformation in Appendix~\ref{sec:append_su4}.

We will see in the following sections that
the above high symmetric models in a ferromagnetic coupling regime
are on an exact phase boundary or a multiple point
in the ground-state phase diagram.
The SU(3) symmetric model in the space (\ref{eq:SU3_line})
with ferromagnetic couplings
$J_{\rm d}<0$ and $J_\parallel<0$ is on an exact phase boundary
between the ferromagnetic phase and the spin nematic phase
with ferroquadrupolar order.
Similarly, in the $J_d \rightarrow -\infty$ limit, the SU(3) model with
the parameters (\ref{eq:SU(3)param}) under the ferromagnetic condition
$J_\parallel+J_\times +\frac{J_4}{4} < 0$ is on the same phase boundary,
which is shown in the next section.
The SU(4) symmetric model given by Eq.\ (\ref{eq:SU4_line})
in the ferromagnetic case $J_\parallel < 0$
is at a multiple point where many phases coexist,
which is further discussed in Sec. \ref{sec:SU4}.

\section{Strong ferromagnetic-dimer limit}
\label{sec:perturbation}

In this section, we study the model (\ref{eq:Ham}) in the limit of strong ferromagnetic intradimer coupling,
$J_{\rm d} \to -\infty$.
Treating  the intradimer exchange term $\mathcal{H}_{\rm d}$ as a unperturbed Hamiltonian and the rest of
terms
as a perturbation, we derive an effective Hamiltonian,
which enables us to see the mechanism of spin nematic ordering.

In the ground state of the unperturbed Hamiltonian $\mathcal{H}_{\rm d}$, two $S=1/2$ spins in each dimer form a spin triplet and the ground states are $3^N$-fold degenerate, where $N$ is the number of dimers in the system.
The first-order perturbation induces state transitions between degenerate ground states, whose matrix elements are
written with the effective Hamiltonian
\begin{eqnarray}
\mathcal{H}^{(1)}
&=& \left( \frac{J_\parallel+J_\times}{2}+\frac{J_4}{8}\right)
\sum_{\langle j,j'\rangle} \tilde{\bm S}_j \cdot \tilde{\bm S}_{j'}
\nonumber \\
&& + \frac{J_4}{4} \sum_{\langle j,j'\rangle} \left( \tilde{\bm S}_j \cdot \tilde{\bm S}_{j'}\right)^2 + {\rm const.},
\label{eq:1st_order_Ham}
\end{eqnarray}
where $\tilde{\bm S}_j$ denote the spin-1 operators acting on the $S=1$ triplet sector
on the $j$th dimer.
This first-order perturbation Hamiltonian is nothing but the spin-1 bilinear-biquadratic model.

\begin{table*}
\caption{
Parameter ranges or regions of various phases in the ground state of the $S=1$ bilinear-biquadratic models
($\mathcal{H}_{\rm bb}$ and $\mathcal{H}^{(1)}$) on (a) square and honeycomb lattices, and  on
(b) triangular and kagome lattices.
FM and FQ respectively denote the ferromagnetic and ferro-quadrupolar phases, and AFM and AFQ respectively the
antiferromagnetic and antiferro-quadrupolar phases. PVBS  and TVBC respectively denote  plaquette
valence-bond solid and trimerized valence-bond crystal.
Both AFM3 and AFQ3 represent phases with three-sublattice structure, and
AFM120$^\circ$ does the AFM phase with $120^\circ$ structure.
The results for $\mathcal{H}_{\rm bb}$ are taken from Refs.\ \onlinecite{TothLMP2012,NiesenC2017,Zhaoetal2012,LauchliMP2006,Liuetal2015}.
}
\label{tab:BLBQ}
\begin{ruledtabular}
%\begin{center}
\begin{tabular}{lcccc}
 (a) & & & & \\
\hline
  & \multicolumn{2}{c}{Square} & \multicolumn{2}{c}{Honeycomb}   \\
 Phase & $\mathcal{H}_{\rm bb}$ (Refs.\ \onlinecite{TothLMP2012,NiesenC2017}) & $\mathcal{H}^{(1)}$ &
 $\mathcal{H}_{\rm bb}$ (Ref.\ \onlinecite{Zhaoetal2012}) & $\mathcal{H}^{(1)}$  \\
\hline
 FM       & $\frac{\pi}{2} < \theta < \frac{5}{4}\pi$ & $J_\parallel + J_\times < -\frac{|J_4|}{4}$ &
 $\frac{\pi}{2} < \theta < \frac{5}{4}\pi$ & $J_\parallel + J_\times < -\frac{|J_4|}{4}$ \\
 FQ      & $-\frac{3}{4}\pi < \theta < -\frac{\pi}{2}$ & $\frac{J_4}{4} < J_\parallel + J_\times < -\frac{J_4}{4}$ &
 $-\frac{3}{4}\pi < \theta < -\frac{\pi}{2}$ & $\frac{J_4}{4} < J_\parallel + J_\times < -\frac{J_4}{4}$ \\
 N\'{e}el & $-\frac{\pi}{2} < \theta < 0.189\pi$ & $J_\parallel + J_\times > {\rm max}\left(0.49 J_4,-\frac{J_4}{4}\right)$ &
 $-\frac{\pi}{2} < \theta < 0.19\pi$ & $J_\parallel + J_\times > {\rm max}\left(0.49 J_4,-\frac{J_4}{4}\right)$ \\
 Haldane  & $0.189\pi < \theta < 0.217\pi$ & $0.49J_4 > J_\parallel + J_\times > 0.37 J_4$ &
 - & - \\
 AFM3     & $0.217\pi < \theta < \frac{\pi}{4}$ & $0.37J_4 > J_\parallel + J_\times > \frac{J_4}{4}$ &
 - & - \\
 AFQ3    & $\frac{\pi}{4} < \theta < \frac{\pi}{2}$ & $-\frac{J_4}{4} < J_\parallel + J_\times < \frac{J_4}{4}$ &
 - & - \\
 PVBS     & - & - &
 $0.19\pi < \theta < \frac{\pi}{2}$ & $-\frac{J_4}{4} < J_\parallel + J_\times < 0.49J_4$ \\
\hline
 & & & & \\
 (b) & & & & \\
\hline
  & \multicolumn{2}{c}{Triangular} & \multicolumn{2}{c}{Kagome}   \\
 Phase & $\mathcal{H}_{\rm bb}$ (Ref.\ \onlinecite{LauchliMP2006}) & $\mathcal{H}^{(1)}$ &
 $\mathcal{H}_{\rm bb}$ (Ref.\ \onlinecite{Liuetal2015}) & $\mathcal{H}^{(1)}$  \\
\hline
 FM       & $\frac{\pi}{2} < \theta < \frac{5}{4}\pi$ & $J_\parallel + J_\times < -\frac{|J_4|}{4}$ &
 $\frac{\pi}{2} < \theta < \frac{5}{4}\pi$ & $J_\parallel + J_\times < -\frac{|J_4|}{4}$ \\
 FQ      & $-\frac{3}{4}\pi < \theta < -0.11\pi$ & $\frac{J_4}{4} < J_\parallel + J_\times < -1.6J_4$ &
 $-\frac{3}{4}\pi < \theta < -0.04\pi$ & $\frac{J_4}{4} < J_\parallel + J_\times < -4.2J_4$ \\
 AFM120$^\circ$ & $-0.11\pi < \theta < \frac{\pi}{4}$ & $J_\parallel + J_\times > {\rm max}\left(\frac{J_4}{4},-1.6J_4\right)$ &
  - & - \\
 AFQ3 & $\frac{\pi}{4} < \theta < \frac{\pi}{2}$ & $-\frac{J_4}{4} < J_\parallel + J_\times < \frac{J_4}{4}$ &
  - & - \\
 TVBC  & - & - &
 $-0.04\pi < \theta < 0.37\pi$ & $J_\parallel + J_\times > {\rm max}\left(-0.03J_4,-4.2J_4\right)$ \\
 AFQ & - & - &
$0.37\pi < \theta < \frac{\pi}{2}$ & $-\frac{J_4}{4} < J_\parallel + J_\times < -0.03J_4$ \\
%\hline
\end{tabular}
%\end{center}
\end{ruledtabular}
\end{table*}

The spin-1 bilinear-biquadratic model has been extensively studied on various lattices.
In Table \ref{tab:BLBQ}, we summarize  the obtained ground-state phases.
Here, defining the Hamiltonian
\begin{align}
\mathcal{H}_{\rm bb} = J_{\rm bb} \sum_{\langle j,j'\rangle} \left[ \cos\theta~ \tilde{\bm S}_j \cdot \tilde{\bm S}_{j'} + \sin\theta \left( \tilde{\bm S}_j \cdot \tilde{\bm S}_{j'}\right)^2 \right]
\label{eq:Ham_BLBQ}
\end{align}
with the parameter $\theta$ and $J_{\rm bb} > 0$, we describe the phase diagrams as functions of $\theta$.
For the square lattice the phase diagram  contains at least five phases, i.e., the ferromagnetic (FM), ferro-quadrupolar
(FQ), N\'{e}el, three-sublattice antiferromagnetic (AFM3), and three-sublattice antiferro-quadrupolar  (AFQ3) phases.\cite{TothLMP2012}
The emergence of a quasi-one-dimensional Haldane phase in a narrow region between the N\'{e}el and AFM3 phases was also reported.\cite{NiesenC2017}
The phase diagram for the honeycomb lattice includes the FM, FQ, N\'{e}el, and plaquette valence-bond-solid phases.\cite{Zhaoetal2012}
For the triangular lattice, the  phase diagram contains the FM, FQ, 120$^\circ$-structure antiferromagnetic (120$^\circ$-AFM), and AFQ3 phases.\cite{LauchliMP2006}
For the kagome lattice, the phase diagram was found to include the FM, FQ, antiferro-quadrupolar, and trimerized valence-bond-crystal phases.\cite{Liuetal2015}
The parameter ranges of $\theta$ for these phases are shown  in Table\ \ref{tab:BLBQ} for each lattice.
It is noteworthy that the FQ phases in the geometrically-frustrated (triangular and kagome) lattices appear in wider regions than those in the bipartite (square and honeycomb) lattices because of the suppression of antiferromagnetic ordering in the former lattices.

\begin{figure}
\begin{center}
\includegraphics[width=42mm]{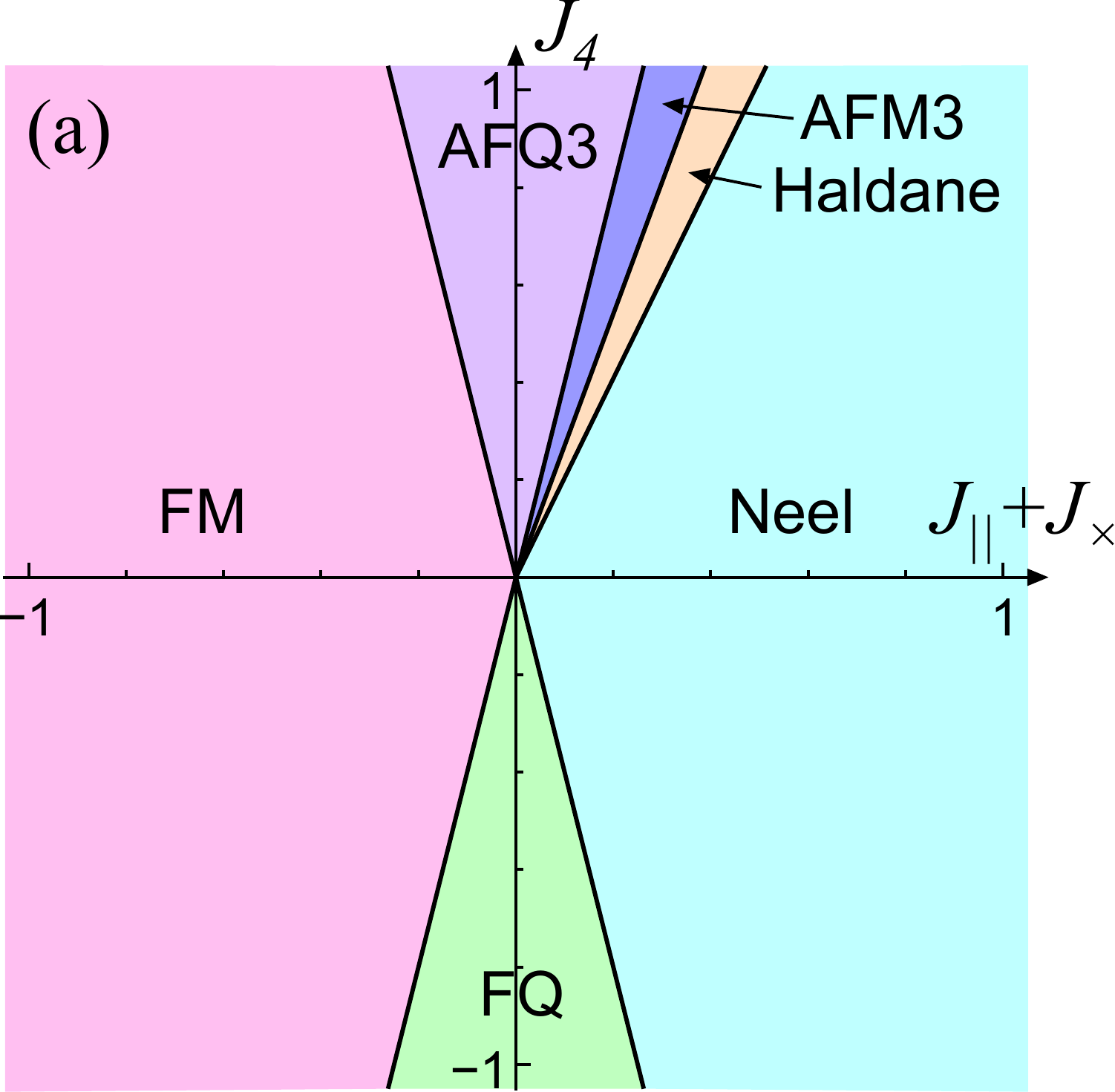}
\includegraphics[width=42mm]{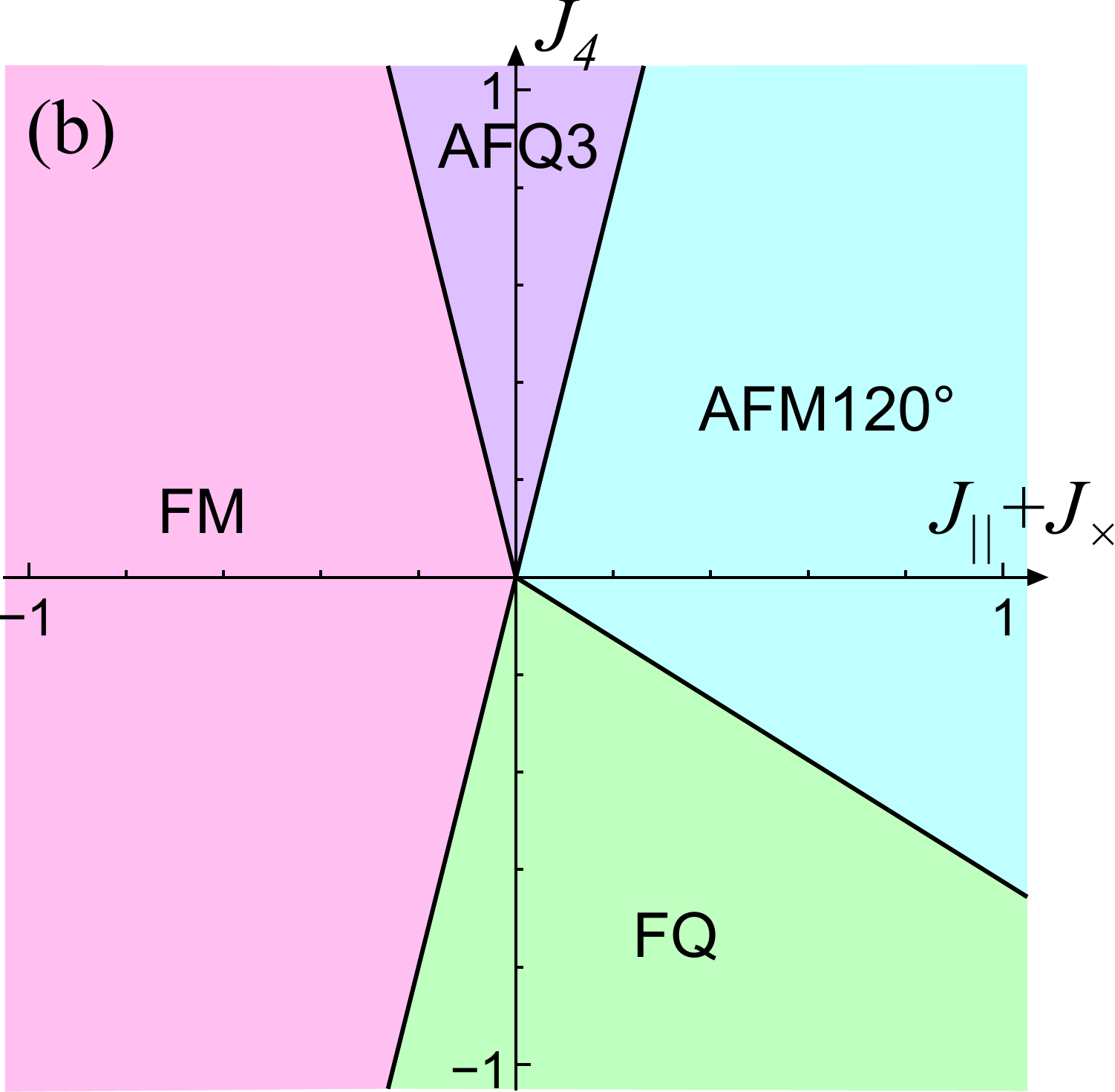}
\caption{
Ground-state phase diagram of the first-order perturbation Hamiltonian $\mathcal{H}^{(1)}$
on the (a) square and (b) triangular lattices. Abbreviations of phases are the same as in Table~\ref{tab:BLBQ}.
}
\label{fig:pd_1st_order}
\end{center}
\end{figure}

From these results, we can derive  the phase diagram for the first-order perturbation Hamiltonian ${\cal H}^{(1)}$
using the relation
$J_{\rm bb}\cos\theta=(J_\parallel+J_\times)/2+J_4/8$
and $J_{\rm bb}\sin\theta=J_4/4$.
The resultant regions of each phase are also presented in Table~\ref{tab:BLBQ} for the lattices considered.
Again, we find that the regions of the FQ phases in the triangular and kagome lattice systems are wider than those in the square and
honeycomb lattice systems.
The phase diagrams in the first-order perturbation for the square and triangular lattices  are shown
in the $J_\parallel+J_\times$ versus $J_4$ plane in Fig.\ \ref{fig:pd_1st_order}. The quadrupolar phases appear in between
the ferromagnetic phase and antiferromagnetic phases.
The parameter regions of the FQ and AFQ3 phases appearing for negative and positive $J_4$, respectively, shrink with decreasing $|J_4|$ and vanish at $J_4=0$ (within the first-order perturbation).

In the case of $J_4=0$, the first-order perturbation Hamiltonian contains only the two-spin exchange interactions and they
vanish at $J_\parallel+J_\times=0$.
In this situation, the second-order perturbation, which induces further effective interactions, becomes relevant.
From a standard procedure of perturbation theory, the second-order perturbation Hamiltonian in the case of $J_\parallel+J_\times=J_4=0$
turns out to have a rather simple form\cite{HikiharaO2010}
\begin{eqnarray}
\mathcal{H}^{(2)} = J^{(2)} \sum_{\langle j,j'\rangle} \left[ \left(\tilde{S}_j \cdot \tilde{S}_{j'}\right)^2 -1 \right]
\label{eq:Ham_2nd_order}
\end{eqnarray}
with
\begin{eqnarray}
J^{(2)} = -\frac{(J_\parallel-J_\times)^2}{8|J_{\rm d}|}
= -\frac{J_\parallel^2}{2|J_{\rm d}|}
= -\frac{J_\times^2}{2|J_{\rm d}|}.
\label{eq:const_2nd_order}
\end{eqnarray}
Note that the coupling constant $J^{(2)}$ is always negative.

The second-order perturbation Hamiltonian (\ref{eq:Ham_2nd_order}) leads us to an important conclusion for the case of
  $J_4=0$.
It is natural to expect that, for strong ferromagnetic $J_{\rm d}$,
the biquadratic interaction in Eq.\ (\ref{eq:Ham_2nd_order}) is still dominant over other interactions
in a finite parameter region around $J_\parallel+J_\times=0$.
As shown in Table~\ref{tab:BLBQ}, the pure-biquadratic model (\ref{eq:Ham_2nd_order}), which is equivalent to the spin-1 bilinear-biquadratic model (\ref{eq:Ham_BLBQ}) with $\theta=-\pi/2$, has the ferro-quadrupolar ground state on the triangular and kagome lattices.
We hence  conclude that, in our model (\ref{eq:Ham}) in the vicinity of $J_\parallel+J_\times=0$ with strong ferromagnetic $J_{\rm d}$,
the dominant effective biquadratic interaction  leads to the ferro-quadrupolar ground state on the geometrically frustrated lattices
even in the case of $J_4=0$, in which the original model includes only the bilinear exchange terms.
We will confirm this conclusion numerically in Sec.\ \ref{subsec:mVMC}.

For the case of the square and honeycomb lattices, the spin-1 biquadratic
Hamiltonian (\ref{eq:Ham_2nd_order}) is just on the phase boundary between the ferro-quadrupolar and N\'{e}el
ordered phases as shown in Table~\ref{tab:BLBQ},
where the ferro-quadrupolar phase spreads to a finite region with ferromagnetic bilinear
interactions.
If we slightly shift the couplings
$J_\parallel$ and $J_\times$ from the phase boundary $J_\parallel + J_\times =0$ into the region
$J_\parallel + J_\times < 0$, they yield a
ferromagnetic bilinear interaction between the effective $S=1$ spins $\tilde{\bm S}_j$ and $\tilde{\bm S}_{j'}$ due to the first-order perturbation.
We hence expect that even when $J_4=0$, our original model (\ref{eq:Ham}) on these bipartite lattices realizes the
ferro-quadrupolar phase in a finite parameter region
in $J_\parallel+J_\times < 0$.
We will confirm  in Sec.\ \ref{subsec:mVMC} that this is also the case.

\section{Mean-Field Approximation with Product-State Ansatz}\label{subsec:Vari_product}
In this section,
we employ a mean-field approximation with product-state ansatz
to determine the ground-state phase diagram  of the model (\ref{eq:Ham}).
We consider the case of $J_4 \le 0$ and $J_{\rm d} \le 0$.
Some details of the method and results are presented also in Appendix\ \ref{sec:append_Num_MFA}.

\subsection{Method}\label{subsec:Vari_product_method}
We employ the approximation in which the ground-state wave function is expressed by
a direct product of dimer states,
\begin{eqnarray}
|\Phi_{\rm DP}\rangle = \prod_j |\varphi\rangle_j,
\label{eq:variational_wf_DP}
\end{eqnarray}
where the dimer states $|\varphi\rangle_j$ can take an arbitrary state spanned with the dimer bases.
We further assume that the wave function has two- and three-sublattice structures, respectively,
for the bipartite (square and honeycomb) lattice systems and the triangular lattice system;
namely, the dimers in the same sublattice are in the same state,
\begin{eqnarray}
|\varphi\rangle_j = |\varphi\rangle_\Lambda = \sum_{\sigma_1, \sigma_2} a_{\Lambda,\sigma_1\sigma_2} |\sigma_1 \sigma_2\rangle
\label{eq:dimerstate_DP}
\end{eqnarray}
for $j\in \Lambda$, where $\Lambda = {\rm A,B}$ $({\rm A,B,C})$ denotes  the two sublattices (three sublattices)
and $ |\sigma_1 \sigma_2\rangle $ denotes the dimer state with the eigenvalues
$S^z_{1,j}=\sigma_1$ and $S^z_{2,j}=\sigma_2$.
Optimizing the coefficients $a_{\Lambda,\sigma_1\sigma_2}$ in Eq.\ (\ref{eq:dimerstate_DP}) variationally, we obtain
the lowest-energy mean-field solution.

To obtain the ground state, we minimize the expectation value of the bond Hamiltonian of the model (\ref{eq:Ham}),
\begin{eqnarray}
\mathcal{H}_{jj'} &=&
\frac{J_{\rm d}}{z} \left( {\bm S}_{1,j} \cdot {\bm S}_{2,j}
+ {\bm S}_{1,j'} \cdot {\bm S}_{2,j'} \right)
\nonumber \\
&&+ J_\parallel \left( {\bm S}_{1,j} \cdot {\bm S}_{1,j'}
+ {\bm S}_{2,j} \cdot {\bm S}_{2,j'} \right)
\nonumber \\
&&+ J_\times \left( {\bm S}_{1,j} \cdot {\bm S}_{2,j'}
+ {\bm S}_{2,j} \cdot {\bm S}_{1,j'} \right)
\nonumber \\
&&+ J_4 \left( {\bm S}_{1,j} \cdot {\bm S}_{1,j'} \right)
\left( {\bm S}_{2,j} \cdot {\bm S}_{2,j'} \right),
\label{eq:Ham_bond}
\end{eqnarray}
where $z$ is the coordination number.
We note that the coordination number $z$ is taken into account only through the coupling constant $J_{\rm d}/z$ in Eq.\ (\ref{eq:Ham_bond}).
Using the resultant ground state, we calculate the expectation values of
local observables defined in Sec.\ \ref{subsec:OP},
\begin{align}\label{eq:OP}
 {\bm  T}^{\rm MF}_\Lambda  &=  \langle {\bm T}_j \rangle,\nonumber\\
 {\bm N}^{\rm MF}_\Lambda  &=  \langle {\bm N}_j \rangle,\nonumber\\
 {\bm Q}^{\rm MF}_\Lambda &=    \langle {\bm Q}_j \rangle
\end{align}
for any $j\in \Lambda$ for each sublattice $\Lambda$.

The similar mean-field approximation with site-decoupled wave functions has been applied to the spin-1 bilinear-biquadratic model (\ref{eq:Ham_BLBQ}) on the square, honeycomb, and triangular lattices.
This approximation provides pretty accurate results for $\pi/2 < \theta < 2\pi$:
For the square\cite{TothLMP2012} and honeycomb\cite{Zhaoetal2012} lattices, the mean-field approximation yields the phase diagram for $\pi/2 < \theta < 2\pi$ which is completely the same as those obtained by other numerical approaches such as exact diagonalization and tensor renormalization group technique.
For the triangular lattice,\cite{LauchliMP2006} the phase diagram obtained by the mean-field approximation is essentially the same as that by the exact diagonalization; the only discrepancy appears in
the phase boundary between the ferro-quadrupolar and 120$^\circ$-AFM phases, where the ferro-quadrupolar phase region becomes narrower in the mean-field approximation.
On the other hand, for $0 < \theta < \pi/2$, the mean-field approximation is rather unreliable at least for the bipartite lattices since the direct-product wave function is not able to describe the Haldane phase on the square lattice and the plaquette valence-bond-crystal phase on the honeycomb lattice, in which the entanglement between different dimers is essential.
In our calculation, we hence restrict ourselves to explore the parameter region of $J_4 \le 0$, which corresponds, in the limit $J_{\rm d} \to -\infty$, to the region of $\pi \le \theta \le 2\pi$, where the mean-field approximation is expected to be reliable.
In the following, setting
\begin{align}
J_\parallel = -1,
\end{align}
we determine the ground-state phase diagram in $J_\times$ versus $J_4$ planes with $J_4 \le 0$ for several fixed values of $J_{\rm d}/z$.

\subsection{Two-sublattice case}\label{subsec:Vari_product_two_sublattice}

\begin{figure}
\begin{center}
\includegraphics[width=70mm]{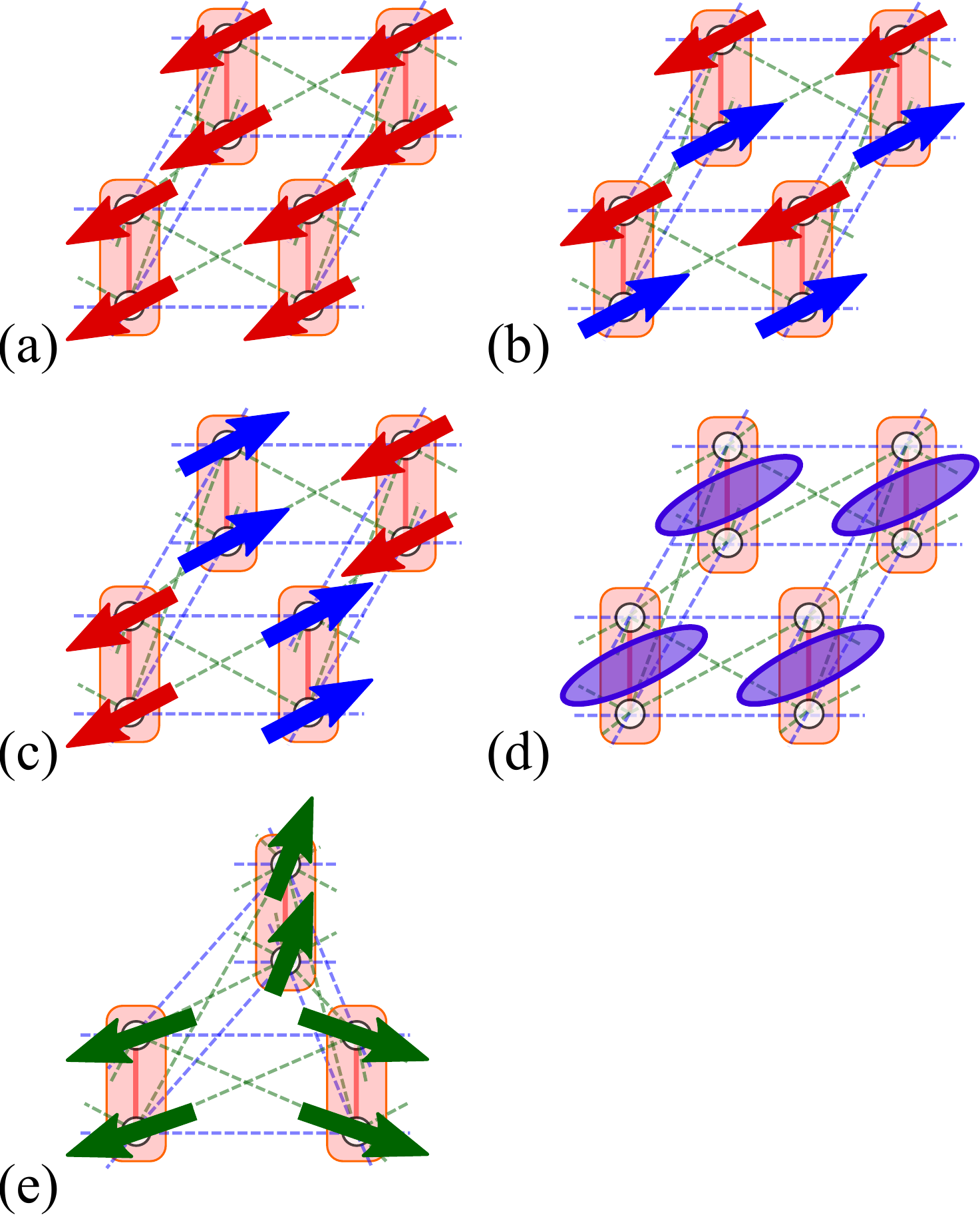}
\caption{
Schematic illustration of spin structures appearing in the model (\ref{eq:Ham})
in
(a) FM phase,
(b) A-type AFM phase,
(c) C-type AFM phase,
(d) spin nematic phase with ferro-quadrupolar order, and
(e) C-type 120$^\circ$-AFM phase.
FM, A-type AFM, and spin nematic phases appear in both two- and three-sublattice cases,
while the C-type AFM (C-type 120$^\circ$-AFM) phase appears only in the two-sublattice (three-sublattice) case.
}
\label{fig:state}
\end{center}
\end{figure}

First, we discuss the case of the two-sublattice structure.
From the expectation values of local observables we studied,
we found  four distinct phases.
These phases are characterized with the expectation values
${\bm  T}^{\rm MF}_\Lambda$, ${\bm N}^{\rm MF}_\Lambda$, and  ${\bm Q}^{\rm MF}_\Lambda$
on two sublattices $\Lambda={\rm A}, {\rm B}$,
which are summarized as follows:
\begin{itemize}
\item[(i)] \emph{Ferromagnetic (FM) phase:} All spins are fully polarized, pointing to the same direction,
\begin{eqnarray}
{\bm T}^{\rm MF}_{\rm A} =  {\bm T}^{\rm MF}_{\rm B},\ \ \
|{\bm T}^{\rm MF}_\Lambda|=1,\ \ \  | {\bm N}^{\rm MF}_\Lambda | = 0
\label{eq:Opara_FM}
\end{eqnarray}
for $\Lambda={\rm A}, {\rm B}$.
The ground-state energy per bond of this phase is given by
\begin{eqnarray}
E_{\rm FM}=\frac{1}{2}\left( \frac{J_{\rm d}}{z}+J_\parallel+J_\times\right)+\frac{1}{16}J_4.
\label{eq:gsene_FM}
\end{eqnarray}

\item[(ii)] \emph{A-type antiferromagnetic (A-type AFM) phase:}
Two spins in each dimer are antiparallel to each other, and all of the staggered moments $\langle {\bm N}_j \rangle$ are in the same direction,
\begin{eqnarray}
 {\bm N}^{\rm MF}_{\rm A} =  {\bm N}^{\rm MF}_{\rm B} ,\ \ \
 |{\bm T}^{\rm MF}_\Lambda |=0,\ \ \  | {\bm N}^{\rm MF}_\Lambda | > 0
\label{eq:Opara_A-AFM}
\end{eqnarray}
for $\Lambda={\rm A}, {\rm B}$.
This state can be also regarded as two ferromagnetic layers whose moments are antiparallel to each other.
The spin moments $\langle {\bm S}_{l,j}\rangle$ may shrink, i.e.,
$| {\bm N}^{\rm MF}_\Lambda | \le 1$,
due to quantum fluctuation.

\item[(iii)] \emph{C-type antiferromagnetic (C-type AFM) phase:}
Two spins in each dimer point to the same direction, and the total spin moments $ {\bm T}^{\rm MF}_{\rm A} $ and
${\bm T}^{\rm MF}_{\rm B}$ are antiparallel, forming the N\'{e}el-type magnetic order in $\langle {\bm T}_j \rangle$,
\begin{eqnarray}
{\bm T}^{\rm MF}_{\rm A} =- {\bm T}^{\rm MF}_{\rm B},\ \ \
|{\bm T}^{\rm MF}_\Lambda|=1,\ \ \  |{\bm N}^{\rm MF}_\Lambda | = 0
\label{eq:Opara_C-AFM}
\end{eqnarray}
for $\Lambda={\rm A}, {\rm B}$.
Each local spin is fully polarized.
The ground-state energy per bond of this phase (in the mean-field approximation) is
\begin{eqnarray}
E^{\rm MF}_\text{C-AFM}=\frac{1}{2}\left( \frac{J_{\rm d}}{z}-J_\parallel-J_\times\right)+\frac{1}{16}J_4.
\label{eq:gsene_CAFM}
\end{eqnarray}

\item[(iv)] \emph{Spin nematic phase with ferro-quadrupolar order:}
All the spin-dipole moments vanish, while
the spin-quadrupolar moments take the same finite value on all dimers,
\begin{align}
&|{\bm T}^{\rm MF}_\Lambda|=0,\ \ \  |{\bm N}^{\rm MF}_\Lambda | = 0,
\nonumber\\
&{\bm Q}^{\rm MF}_{\rm A}={\bm Q}^{\rm MF}_{\rm B},\ \ \ | {\bm Q}^{\rm MF}_\Lambda | = \sqrt{\frac{4}{3}}
\label{eq:Opara_FQ}
\end{align}
for $\Lambda={\rm A}, {\rm B}$. The quadrupolar moments are saturated.
The ground-state energy per bond of this phase (in the mean-field approximation) is
\begin{eqnarray}
E^{\rm MF}_{\rm SNf}=\frac{1}{2} \frac{J_{\rm d}}{z}+\frac{3}{16}J_4.
\label{eq:gsene_FQ}
\end{eqnarray}
In the limit $J_{\rm d} \to -\infty$, this phase corresponds to the ferro-quadrupolar
phase of the spin-1 bilinear-biquadratic model.
\end{itemize}
Schematic illustration of spin structures representing these phases are shown in Figs.\ \ref{fig:state}(a)--(d).
We note that the fully-saturated nature of the C-type AFM and spin nematic phases is an artifact of
the approximation with product-state ansatz.
Indeed, we will show in Sec.\ \ref{subsec:mVMC} that, in the mVMC calculations,
the quantum reduction in the magnetic and spin-quadrupolar moments is observed also in these phases.

\begin{figure}
\begin{center}
\includegraphics[width=40mm]{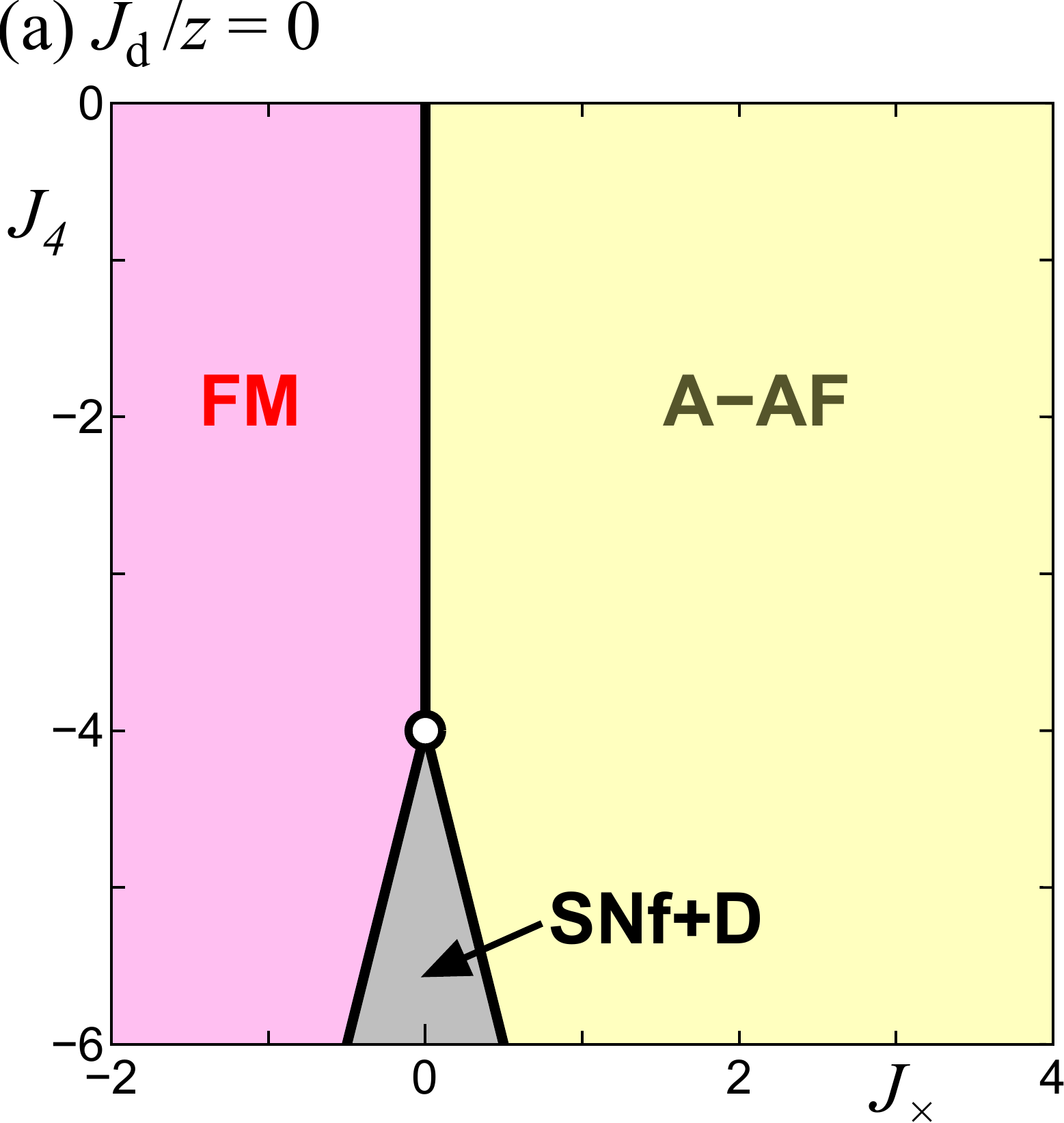}
\includegraphics[width=40mm]{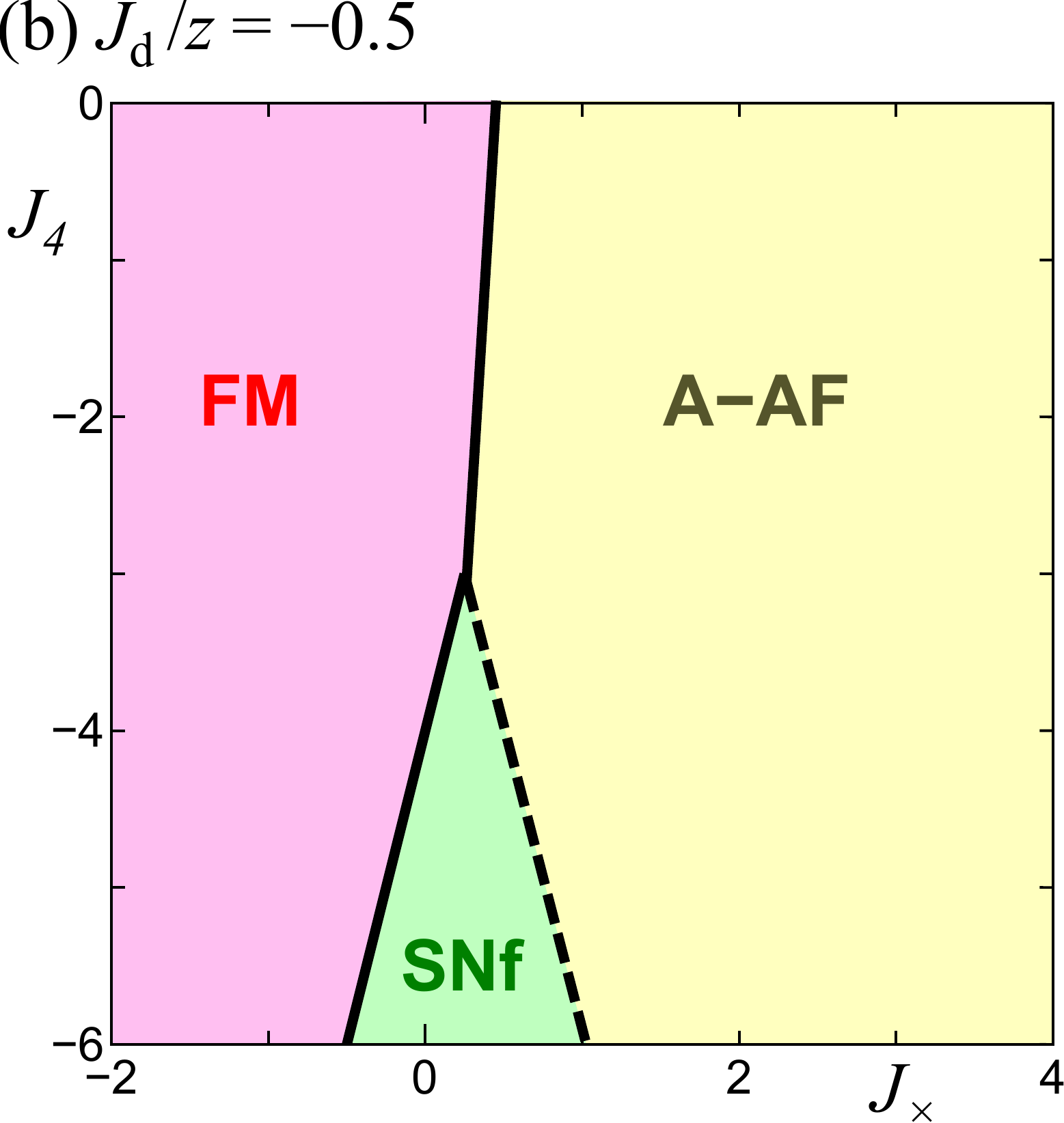}
\includegraphics[width=40mm]{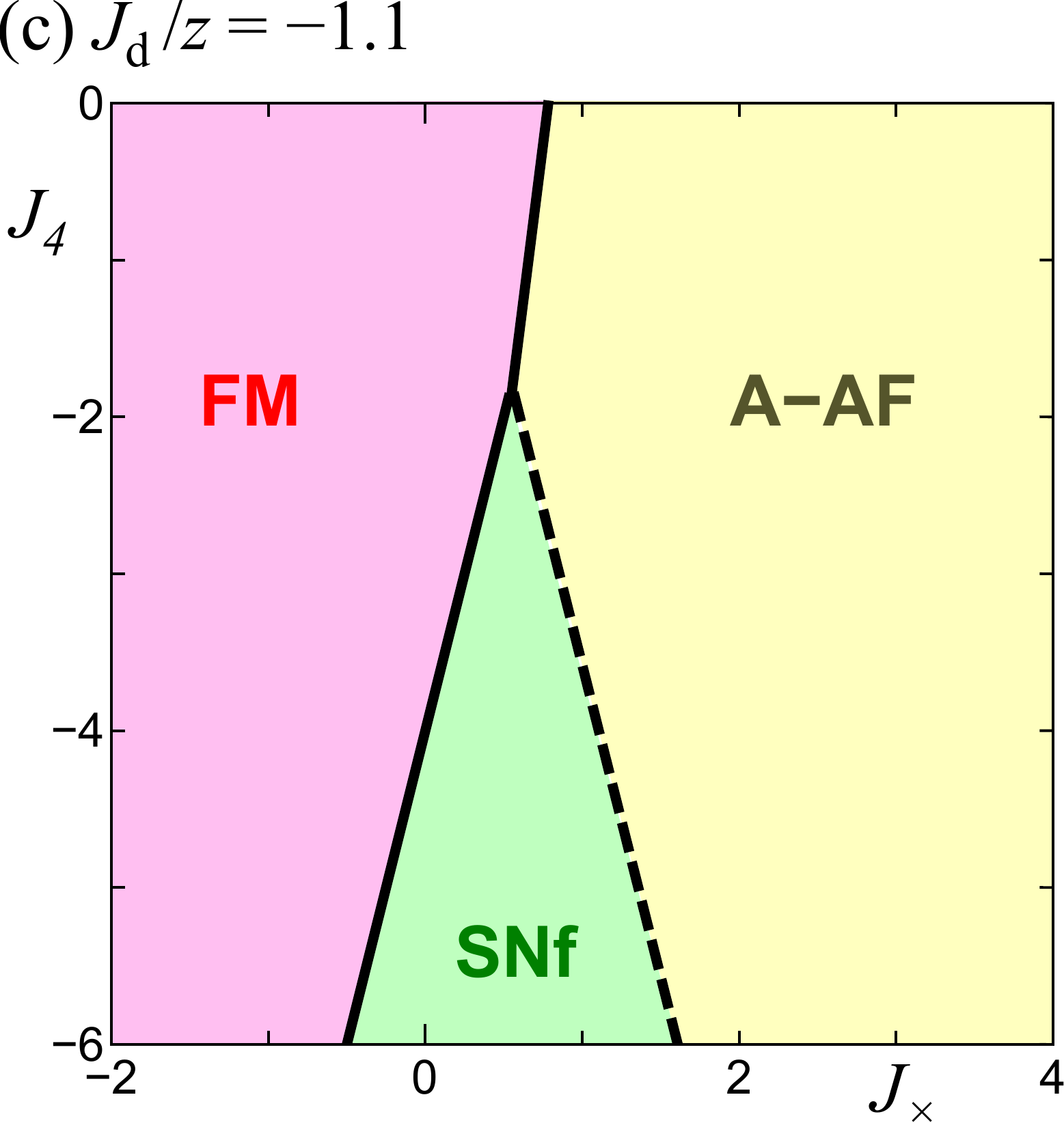}
\includegraphics[width=40mm]{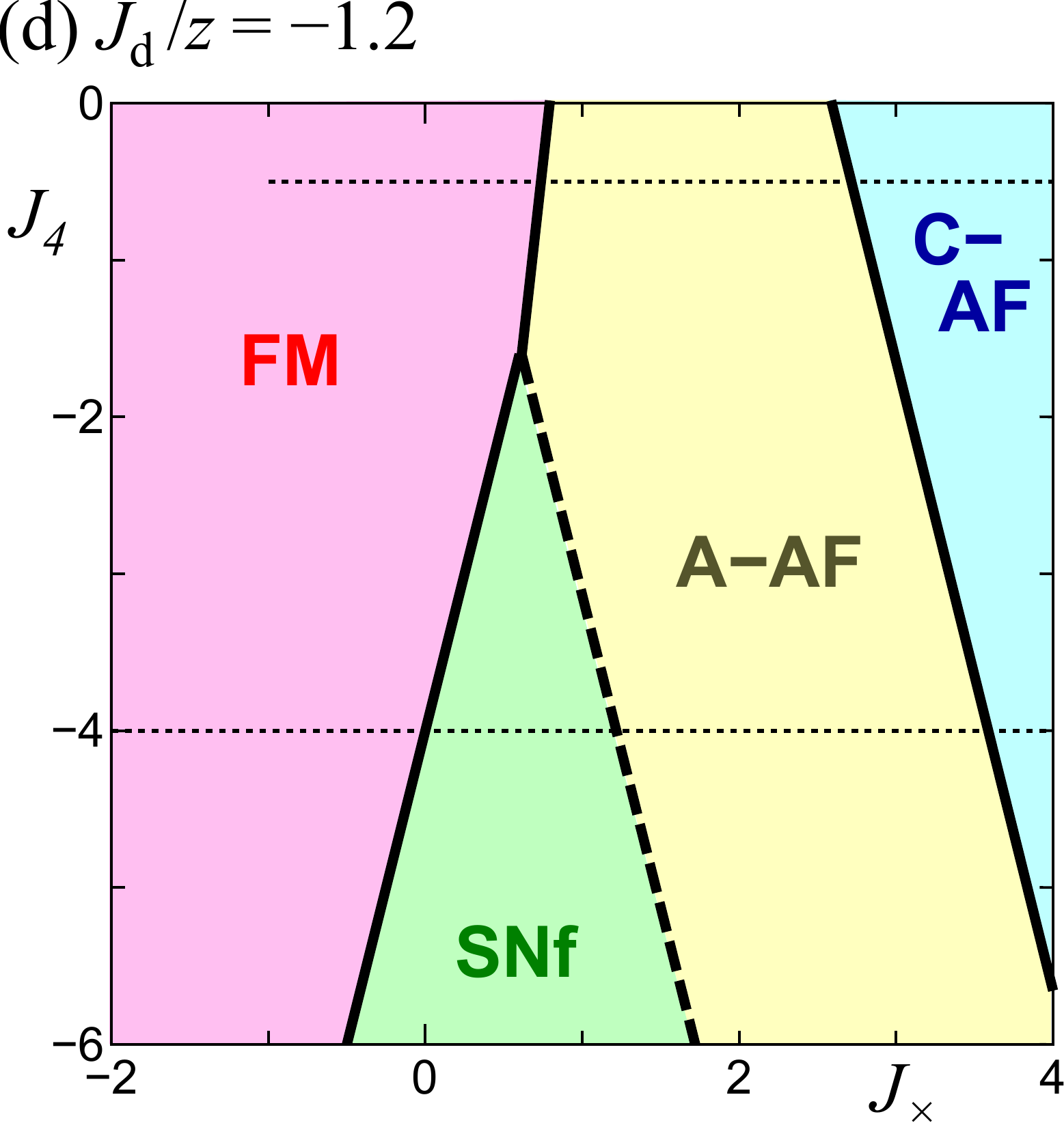}
\includegraphics[width=40mm]{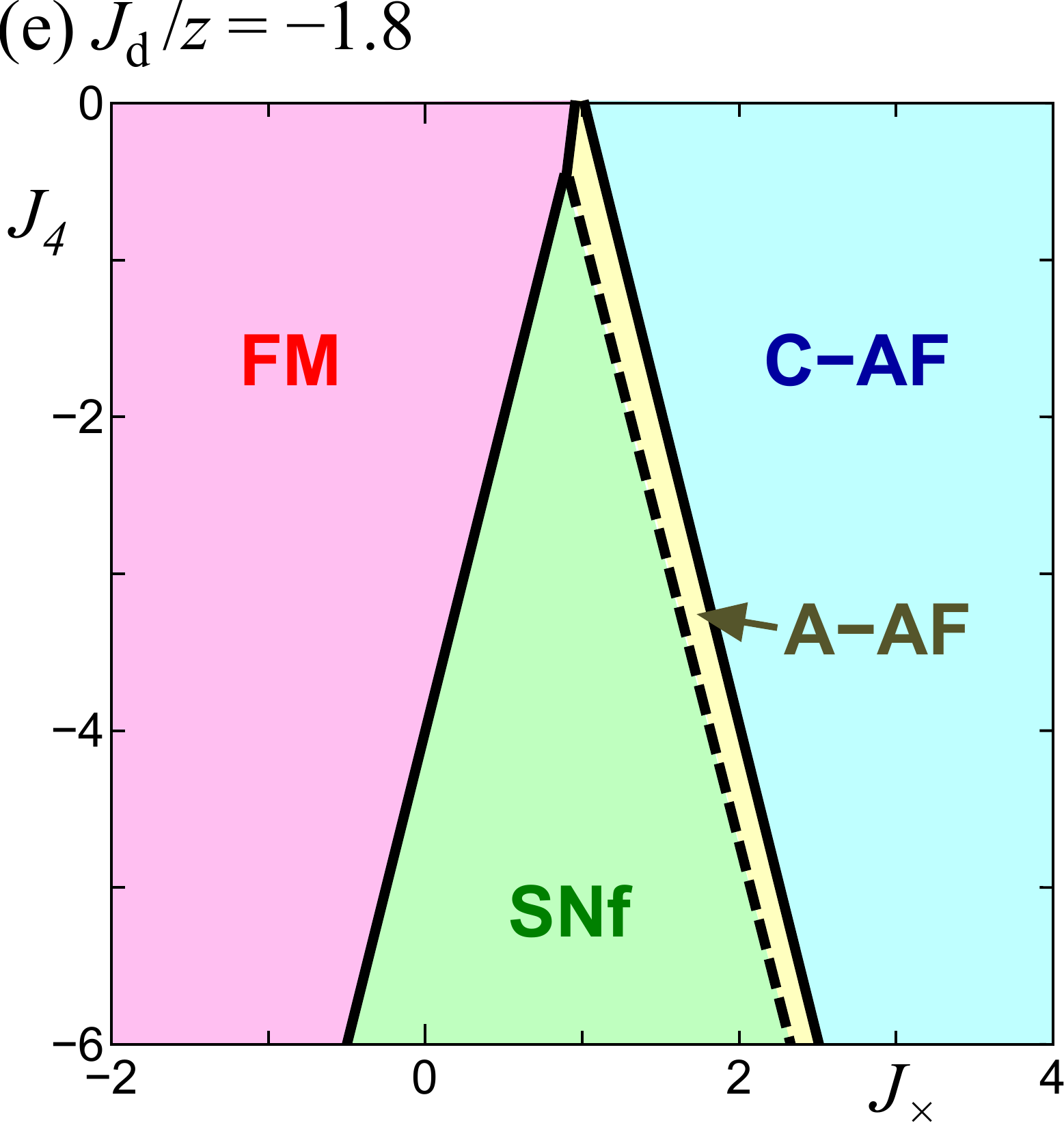}
\includegraphics[width=40mm]{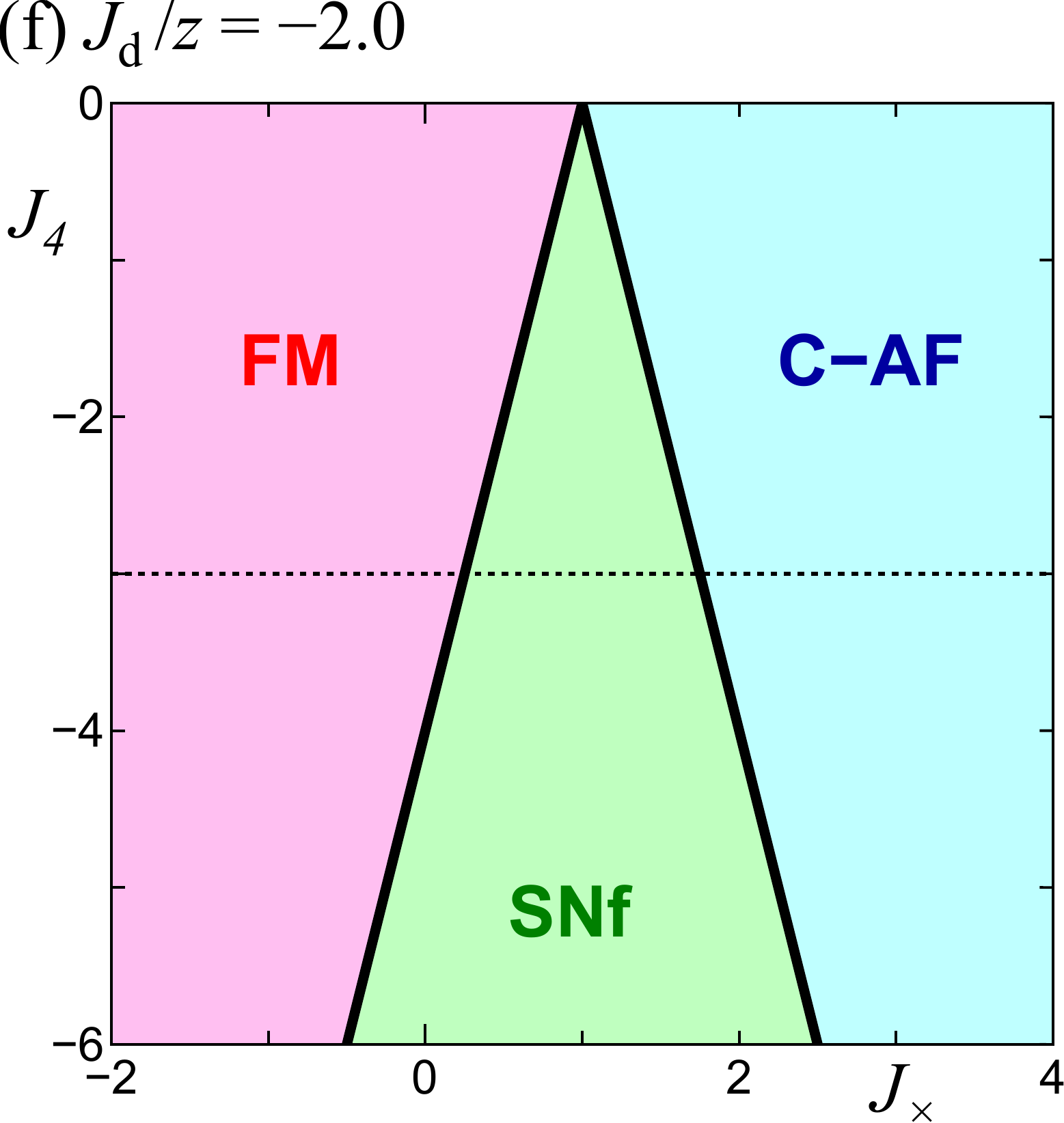}
\caption{
Phase diagrams for the two-sublattice structure. Parameters are set as
$J_\parallel=-1$ and
(a) $J_{\rm d}/z=0$,
(b) $J_{\rm d}/z=-0.5$,
(c) $J_{\rm d}/z=-1.1$,
(d) $J_{\rm d}/z=-1.2$,
(e) $J_{\rm d}/z=-1.8$, and
(f) $J_{\rm d}/z=-2.0$.
Solid and dashed lines, respectively, denote first-order and continuous transitions.
FM, A-AF, C-AF, and SNf represent the ferromagnetic phase, the A-type antiferromagnetic phase, the C-type antiferromagnetic phase, and the spin nematic phase with ferro-quadrupolar order, respectively.
Grey area in (a), labeled with ``SNf+D", is
%the region with a nontrivial degeneracy
the boundary between the spin nematic and dimer-singlet phases, which further has a nontrivial degeneracy.
%where the spin nematic phase extends to the region of $J_{\rm d}<0$.
Open circle in (a) represents the SU(4) symmetric point.
Horizontal dotted lines in (d) and (f) are the parameter lines
shown in Fig.\ \ref{fig:OPN2-Jtimes}.
}
\label{fig:PDN2}
\end{center}
\end{figure}

We have determined the phase diagrams for several values of $J_{\rm d}/z$.
Figure\ \ref{fig:PDN2} shows the results for some typical values of $J_{\rm d}/z$.

At $J_{\rm d}=0$, we find three regions; two are the FM and A-type AFM phases and the other corresponds to the boundary between the spin nematic phase, which appears for $J_{\rm d} <0$,
and dimer-singlet phase for $J_{\rm d} > 0$.
The quadruple point where these four phases coexist is the SU(4) symmetric point, given by
$J_\parallel=-1$, $J_4=-4$, and $J_{\rm d}=J_\times=0$.
We further show in Sec.\ \ref{sec:SU4}, using exact symmetry arguments on a generic model, that the spin nematic
state is naturally generated by SU(4) symmetry and the SU(4) model
is on the multiple point surrounded by, at least, five phases including the spin nematic phase.
The phase boundary between the FM and A-type AFM phases is on the line $J_\times=0$ for $J_4 > -4$,
while the region of the phase boundary between the spin-nematic and dimer-singlet phases
is surrounded
by the two lines $J_4=\pm 4 J_\times -4$.
On the boundary between the spin-nematic and dimer-singlet phases, there exists additional non-trivial degeneracy in the mean-field
solution, which
is a remnant of ${\rm SU}(2)\times {\rm SU}(2)$ symmetry in the case of $J_{\rm d}=J_\times=0$.
(See the Appendix\ \ref{subsec:nematic-singlet-boundary_at_zeroJ4} for more details.)

When $J_{\rm d}$ is negative, the spin nematic state with ferro-quadrupolar order is selected from the non-trivially degenerate ground states in the degenerate region, resulting in the spin nematic phase defined in Eq.\ (\ref{eq:Opara_FQ}).
Another signature of high SU(4) symmetry remains on the boundary between the FM and
spin-nematic phases for $J_{\rm d}<0$ in the mean-field approximation, where the boundary line $J_4 = 4J_\times -4$ is obtained from the
condition $E_{\rm FM}=E^{\rm MF}_{\rm SNf}$.
On this boundary, the mean-field solution of the ground state has non-trivial SU(3) degeneracy,
which we further explain in Appendix~\ref{sec:append_SU3}.

As $|J_{\rm d}|/z$ increases, the region of the spin nematic phase enlarges toward smaller $|J_4|$ regime, and the boundary between the FM and A-type AFM phases also moves toward large $J_\times$ [see Figs.\ \ref{fig:PDN2}(b) and (c)].
Then, at $J_{\rm d}/z \sim -1.15$, the C-type AFM phase enters the parameter space calculated [see Fig.\ \ref{fig:PDN2}(d)].
The appearance of the C-type AFM phase is due to the competition  between the exchange interactions
$J_{\rm d}$ and $J_\parallel$.
Indeed, at the limit of $J_\times \to \infty$ and $J_4=0$, the interactions $J_{\rm d}$ and $J_\parallel$ lead to
the A-type AFM phase for $|J_{\rm d}|/z < |J_\parallel|=1$ and the C-type AFM phase for $|J_{\rm d}|/z > |J_\parallel|$.

With further increasing $|J_{\rm d}|/z$, we see that the A-type AFM phase shrinks while the other phases enlarge.
[See Fig.\ \ref{fig:PDN2}(e).]
The A-type AFM phase eventually vanishes at $J_{\rm d}/z \simeq -2$ and the spin nematic phase touches with the C-type AFM
phase. [See Fig.\ \ref{fig:PDN2}(f).]  The spin nematic phase extends to the line of $J_4=0$, but touches to
the line $J_4=0$ only at the single point $J_\times=1$. Hence in the case of $J_{\rm d}/z \le -2$ and $J_4=0$, the phase diagram contains
only the FM and C-type AFM phases.
For $J_{\rm d}/z \le -2$,  the phase diagram, which is unaltered at least down to $J_{\rm d}/z = -10$ in our calculation,
is the same as the one obtained for the limit $J_{\rm d} \to -\infty$ in Sec.\ \ref{sec:perturbation}.
The boundary lines of the spin nematic phase are $J_4=\pm 4(J_\times-1)$,
 which are obtained from the conditions $E_{\rm FM}=E^{\rm MF}_{\rm SNf}$ and $E^{\rm MF}_\text{C-AFM}=E^{\rm MF}_{\rm SNf}$.

\begin{figure}[t]
\begin{center}
\includegraphics[width=70mm]{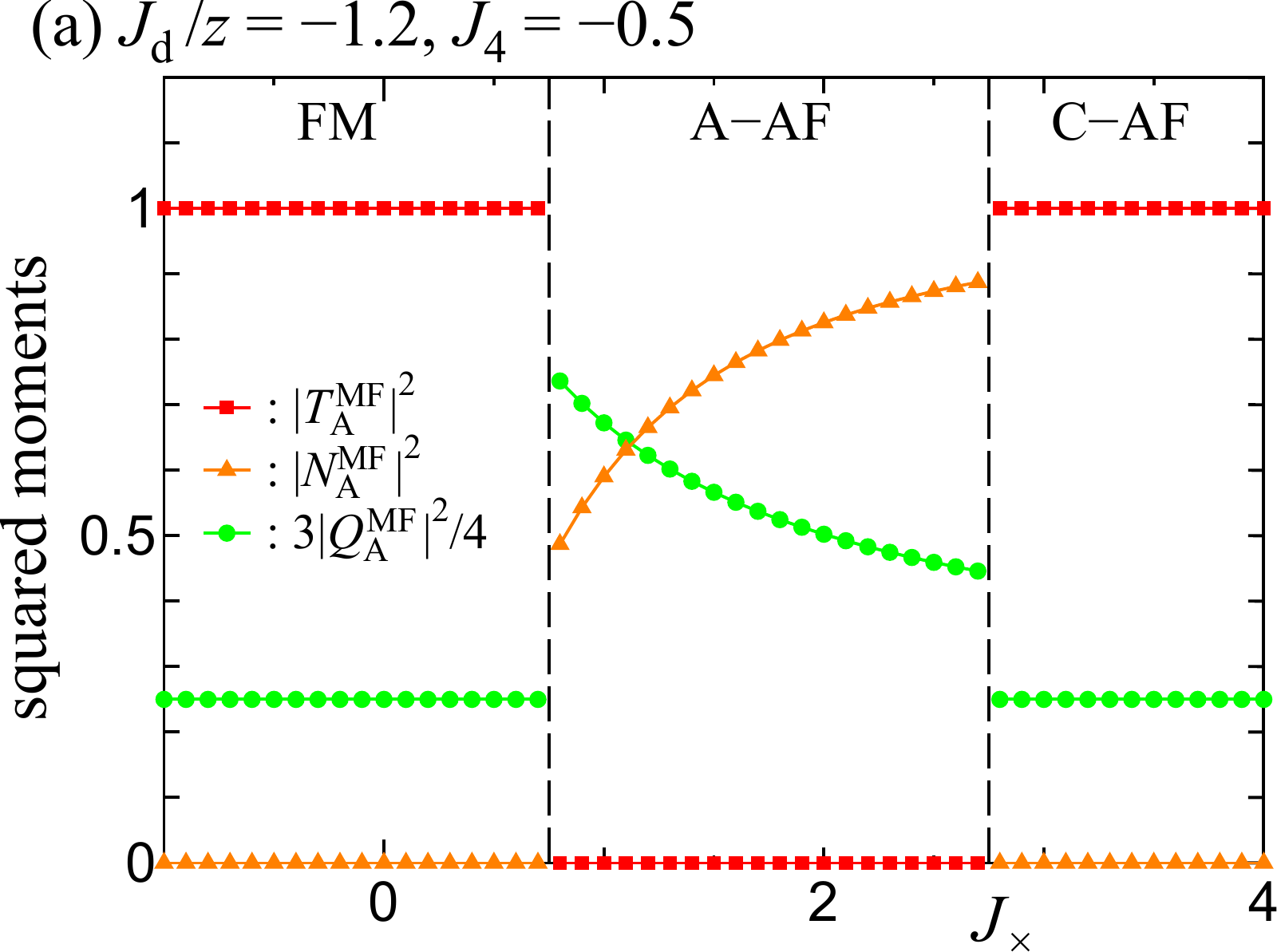}
\includegraphics[width=70mm]{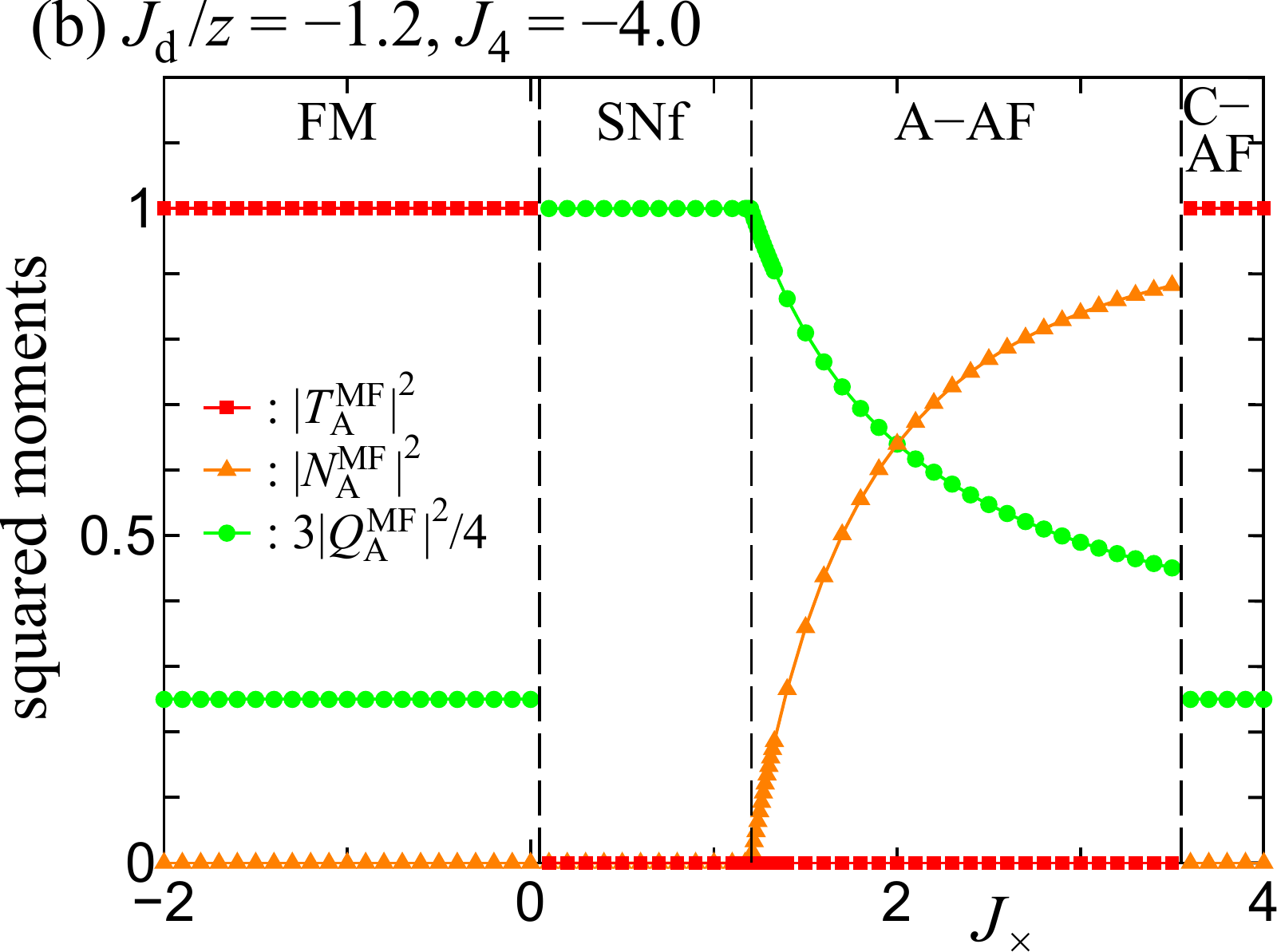}
\includegraphics[width=70mm]{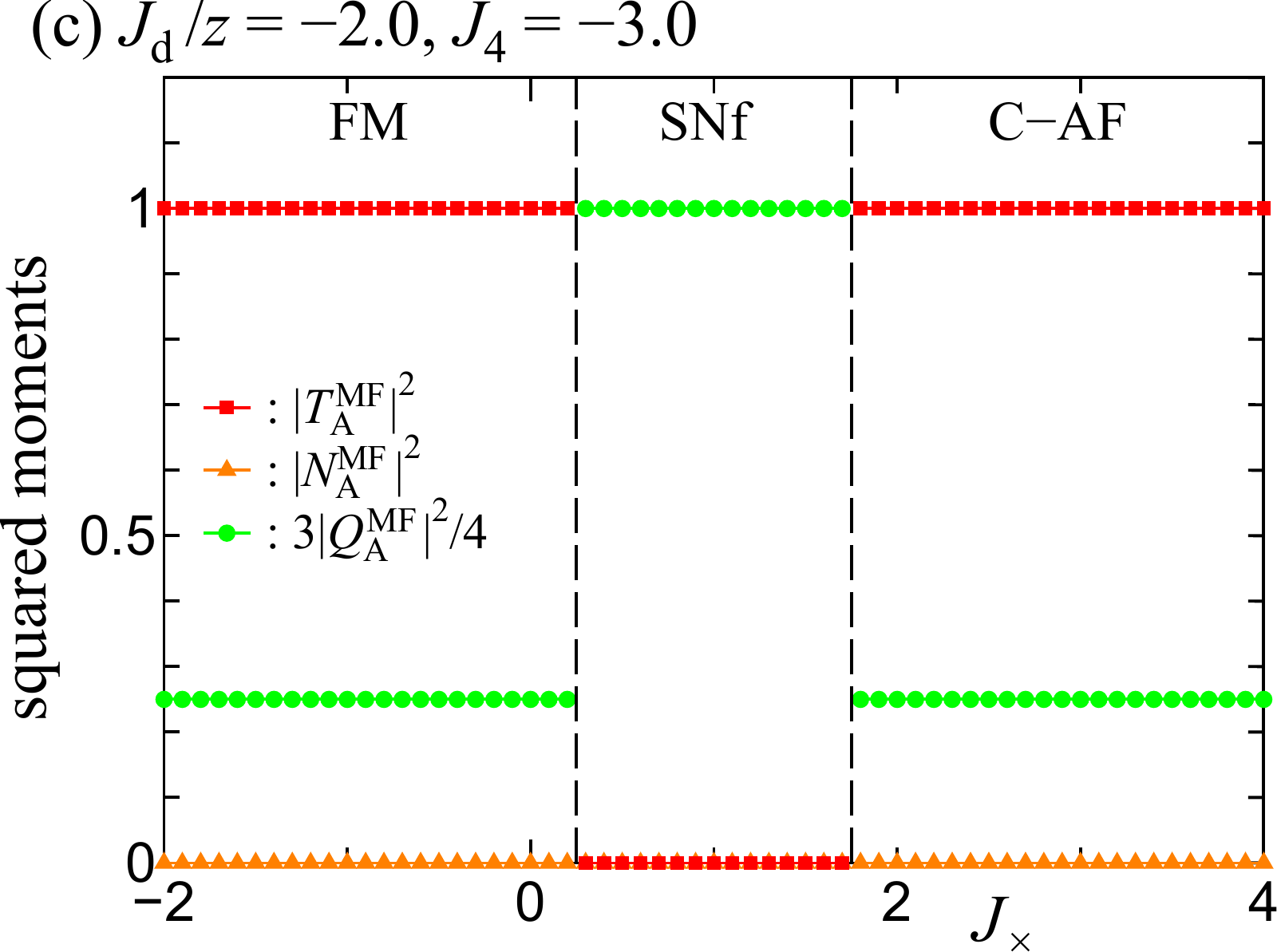}
\caption{
$J_\times$-dependence of squared magnetic moments $|{{\bm T}^{\rm MF}_{\rm A}}|^2$ and $|{\bm N}^{\rm MF}_{\rm A}|^2$
and normalized squared spin-quadrupolar moment $3 |{\bm Q}^{\rm MF}_{\rm A}|^2/4$ for the two-sublattice structure;
(a) $J_{\rm d}/z=-1.2$ and $J_4=-0.5$,
(b) $J_{\rm d}/z=-1.2$ and $J_4=-4.0$,
(c) $J_{\rm d}/z=-2.0$ and $J_4=-3.0$.
FM, A-AF, C-AF, and SNf represent the ferromagnetic phase, the A-type antiferromagnetic phase, the C-type antiferromagnetic phase, and the spin nematic phase with ferro-quadrupolar order, respectively.
Vertical dashed lines represent phase boundaries.
}
\label{fig:OPN2-Jtimes}
\end{center}
\end{figure}

Among the transitions occurring in the present parameter space,
the transition between the spin-nematic and A-type AFM phases is continuous.
In contrast,
the other transitions are always of first order,
accompanied with a jump in ${\bm T}^{\rm MF}_\Lambda$.
(See the Appendix\ \ref{subsec:append_MFA_SNf-AAFM} for detailed analysis of the order of the transitions.)
In Fig.\ \ref{fig:OPN2-Jtimes}, we plot $J_\times$-dependence of the order parameters on the parameter lines given by $J_4=-0.5$ and $-4.0$ in the plane with $J_{\rm d}/z=-1.2$ [dotted lines of Fig.\ \ref{fig:PDN2}(d)]
and the parameter line given by $J_4=-3.0$ in the plane with $J_{\rm d}/z=-2.0$ [dotted line in Fig.\ \ref{fig:PDN2}(f)].
In the case of $J_{\rm d}/z=-1.2$ and $J_4=-0.5$ [Fig.\ \ref{fig:OPN2-Jtimes}(a)], the system undergoes two successive transitions from the FM phase to the A-type AFM phase, and then, to the C-type AFM phase as $J_\times$ increases.
At both transitions, the order parameters exhibit finite jumps,
showing first-order phase transitions.
In the intermediate A-type AFM phase, the N\'{e}el-spin moment ${\bm N}^{\rm MF}_\Lambda$ shrinks from the saturated value due to the zero-point quantum reduction, while the total spin
${\bm T}^{\rm MF}_\Lambda$ on each dimer is fully polarized in the FM and C-type AFM phases.
When negative $J_4$ becomes stronger [see Fig.\ \ref{fig:OPN2-Jtimes}(b)], the system exhibits four phases, the FM, spin-nematic, A-type AFM, and C-type AFM phases.
It is found that the transition between the spin-nematic and A-type AFM phases is continuous while the other two transitions are of first order.
The N\'{e}el-spin moment exhibits the quantum reduction in the A-type AFM phase and vanishes continuously at the boundary to the spin nematic phase.
For $J_{\rm d}/z=-2.0$ and $J_4=-3.0$ [Fig.\ \ref{fig:OPN2-Jtimes}(c)], there occur two first-order transitions from the FM phase to the spin nematic phase, and then, to the C-type AFM phase with increasing $J_\times$.
Finally, we note that the transition between the FM and C-type AFM phases occurring on the line of $J_4=0$ is of first order.

\subsection{Three-sublattice case}\label{subsec:Vari_product_three_sublattice}

\begin{figure}
\begin{center}
\includegraphics[width=40mm]{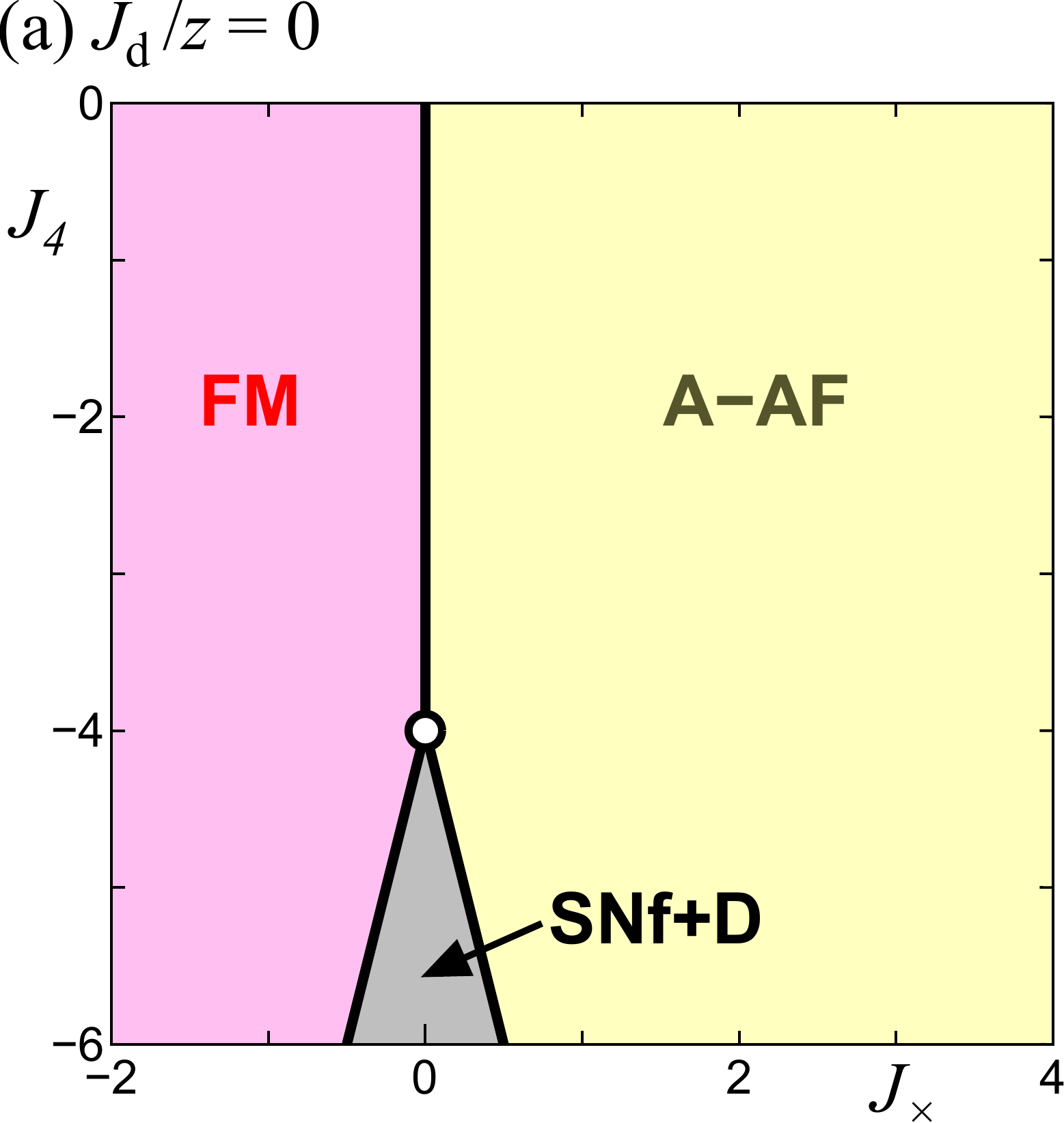}
\includegraphics[width=40mm]{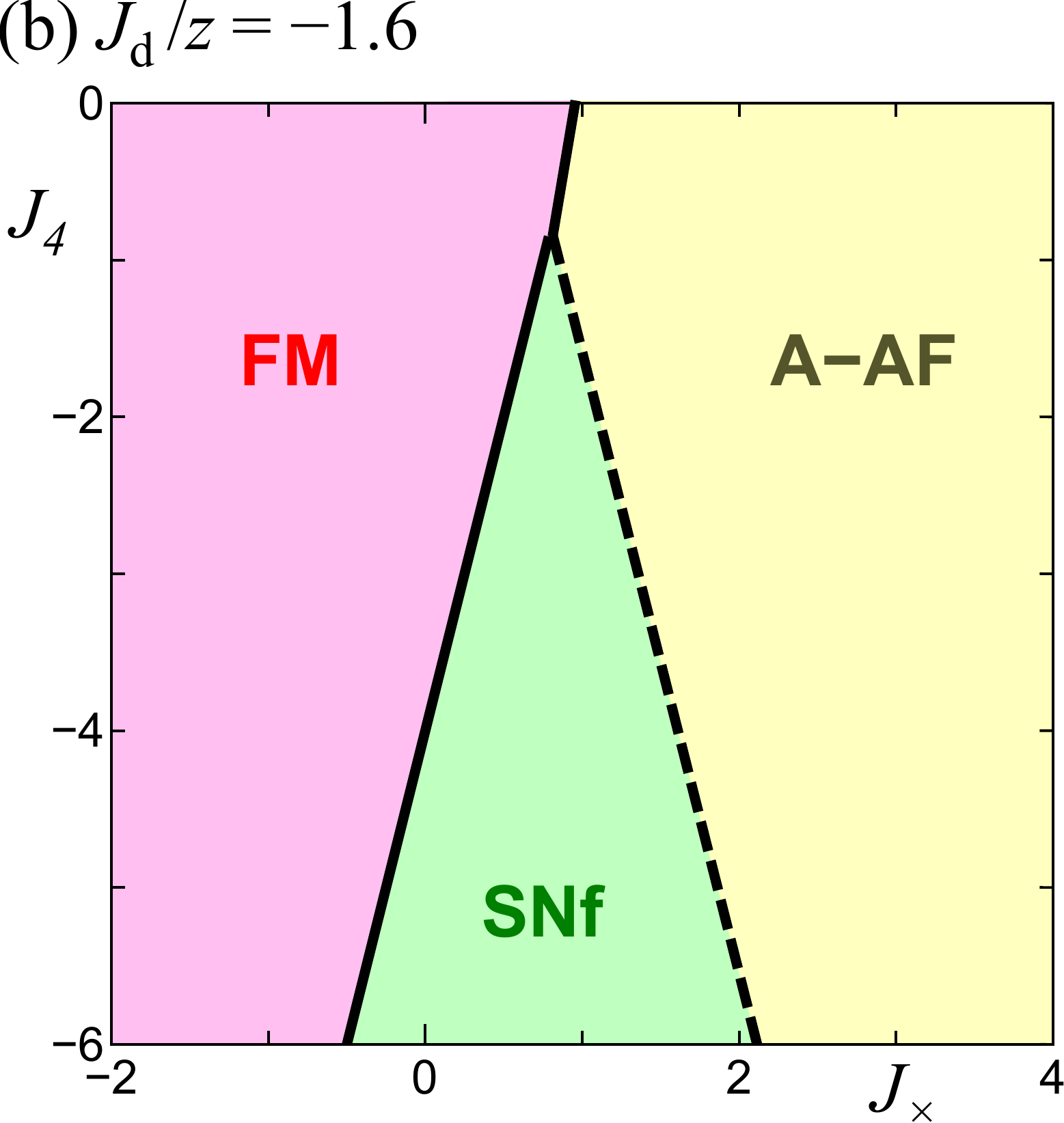}
\includegraphics[width=40mm]{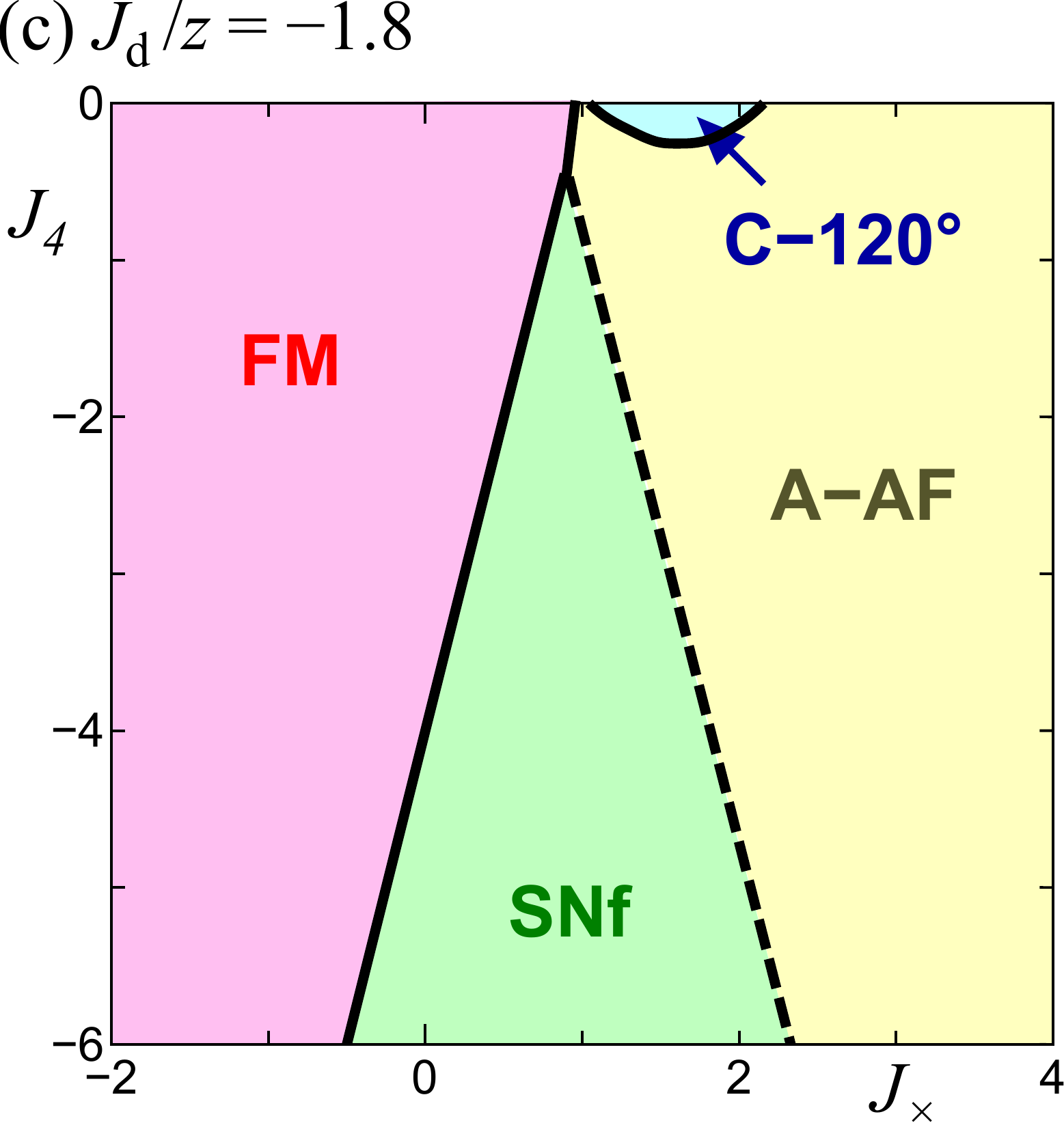}
\includegraphics[width=40mm]{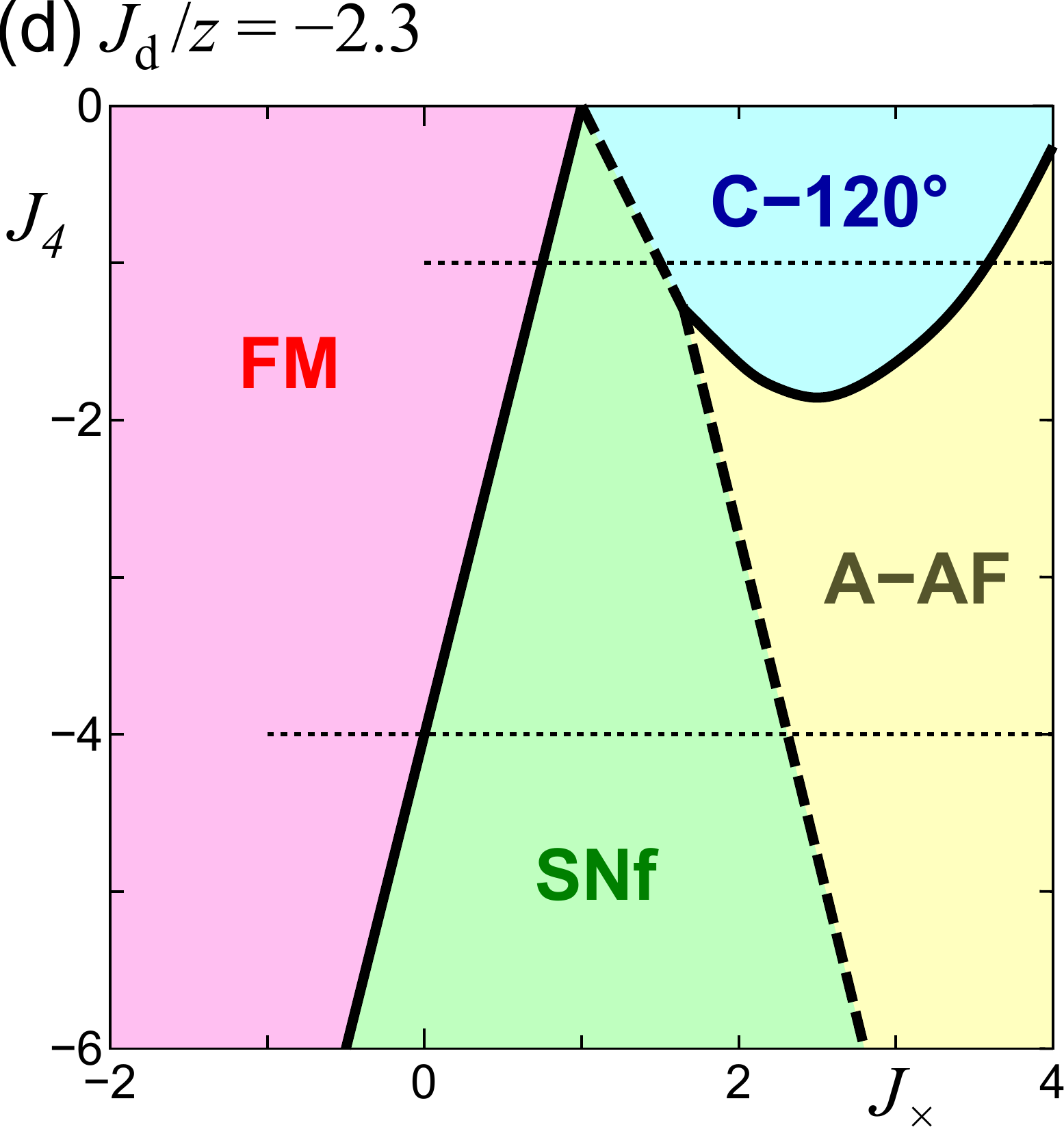}
\includegraphics[width=40mm]{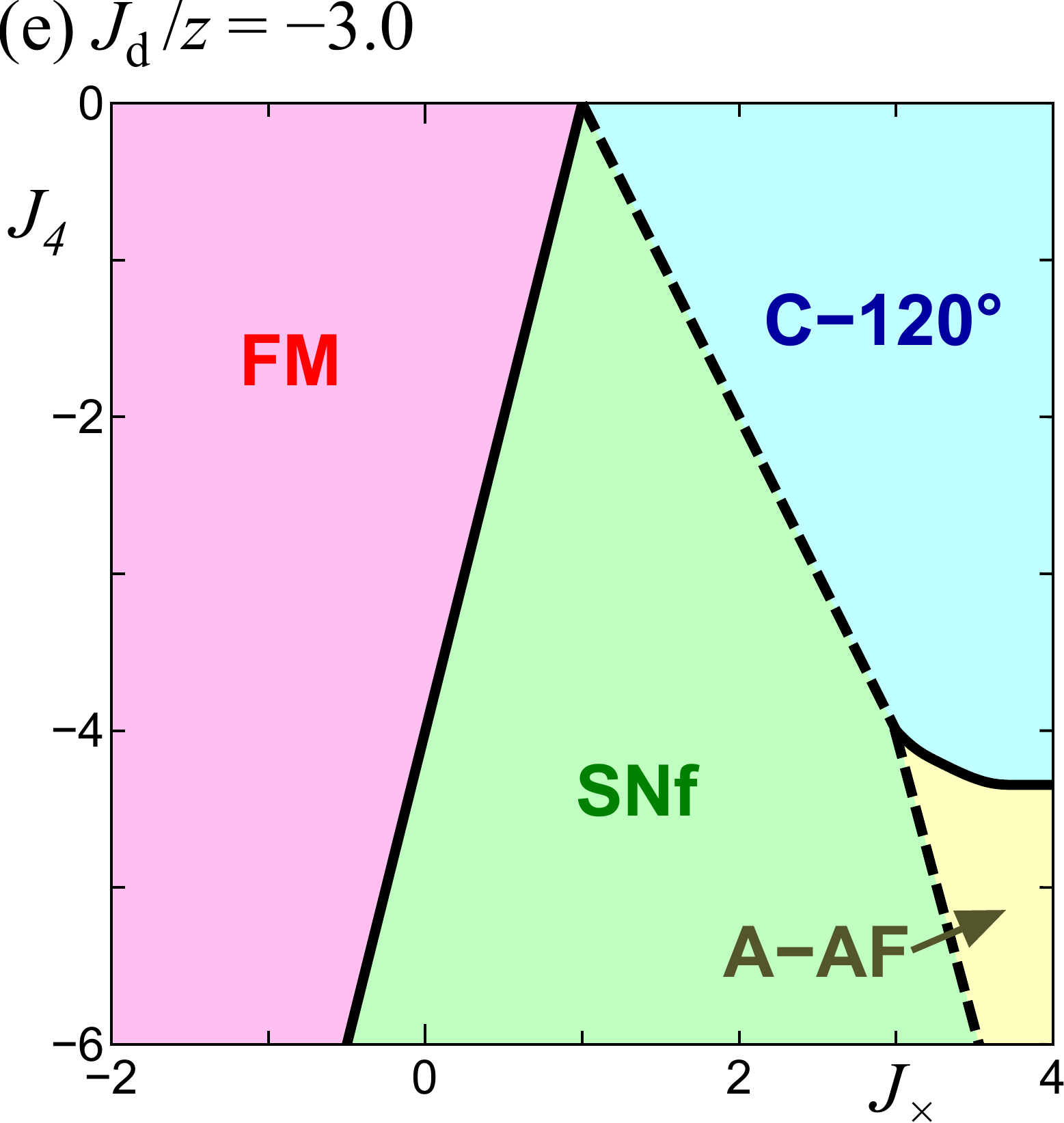}
\includegraphics[width=40mm]{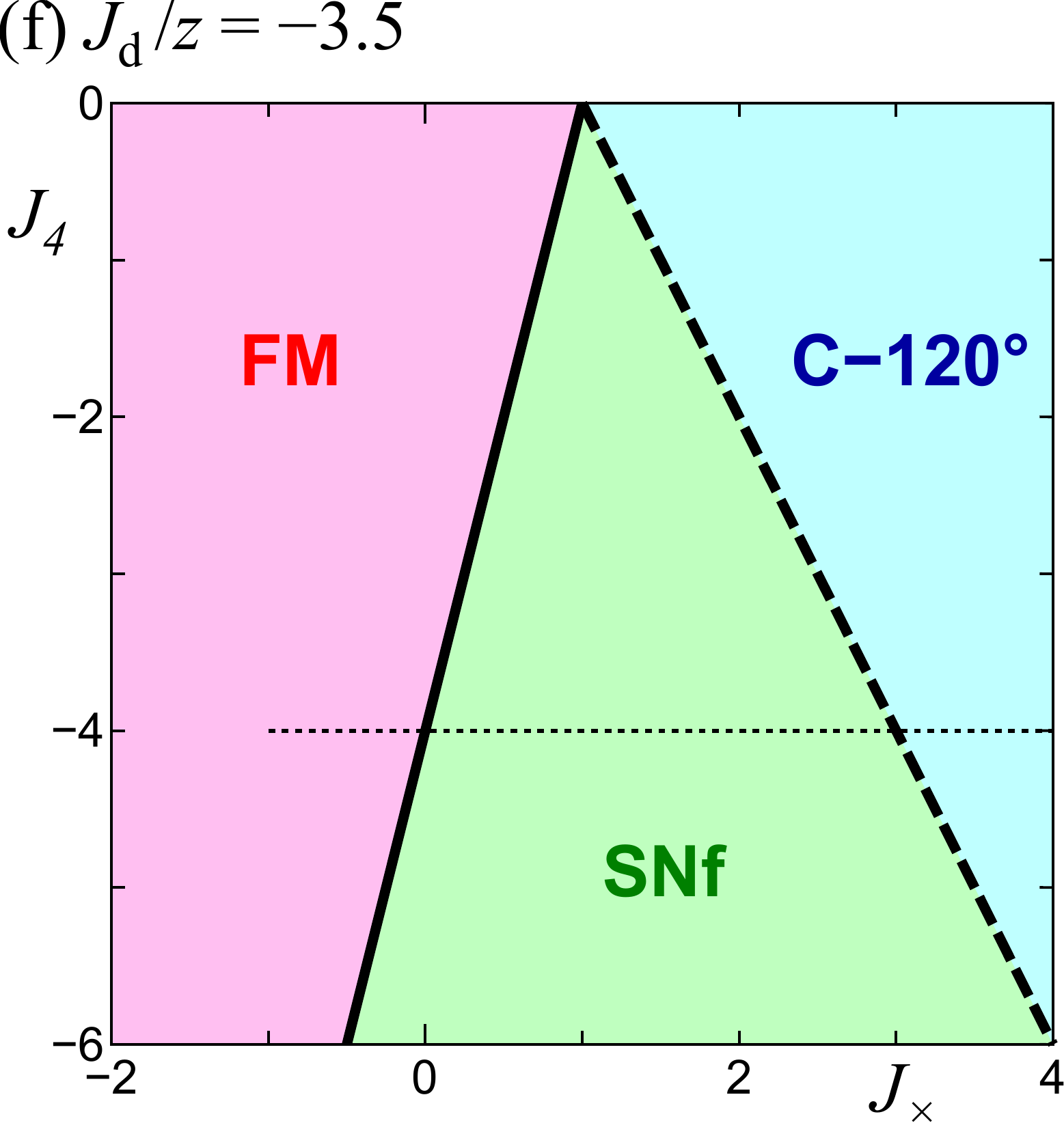}
\caption{
Phase diagrams for the three-sublattice structure. Parameters are set as
$J_\parallel=-1$ and
(a) $J_{\rm d}/z=0$,
(b) $J_{\rm d}/z=-1.6$,
(c) $J_{\rm d}/z=-1.8$,
(d) $J_{\rm d}/z=-2.3$,
(e) $J_{\rm d}/z=-3.0$, and
(f) $J_{\rm d}/z=-3.5$.
Solid and dashed lines denote  first-order and continuous transitions, respectively.
FM, A-AF, C-120$^\circ$, and SNf represent the ferromagnetic phase, the A-type antiferromagnetic phase, the C-type 120$^\circ$-structure antiferromagnetic phase, and the spin nematic phase with ferro-quadrupolar order, respectively.
Grey area in (a), labeled with ``SNf+D", is
%the region with a nontrivial degeneracy
the boundary between the spin nematic and dimer-singlet phases, which further has a nontrivial degeneracy.
%where the spin nematic phase extends to the region of $J_{\rm d}<0$.
Open circle in (a) represents the SU(4) symmetric point.
Horizontal dotted lines in (d) and (f) are the parameter lines
shown in Fig.\ \ref{fig:OPN3-Jtimes}.
}
\label{fig:PDN3}
\end{center}
\end{figure}

Next, we discuss the case of three-sublattice structure on the triangular lattice.
Figure \ref{fig:PDN3} presents the phase diagrams we obtained for various $J_{\rm d}/z$.
We found four distinct phases;
three of them are the FM, A-type AFM, and spin-nematic phases, which are translationally invariant and defined in the same manners as those
for the two-sublattice case, i.e., Eqs.\ (\ref{eq:Opara_FM}), (\ref{eq:Opara_A-AFM}), and (\ref{eq:Opara_FQ}), respectively.
The other phase is as follows:
\begin{itemize}
\item[(iii')] \emph{C-type 120$^\circ$-structure antiferromagnetic (C-type 120$^\circ$-AFM) phase:}
Two spins in each dimer are parallel to each other,
\begin{subequations}
\label{eq:C-type120}
\begin{eqnarray}
|{\bm T}^{\rm MF}_\Lambda |>0,\ \ \  | {\bm N}^{\rm MF}_\Lambda | = 0
\label{eq:C-type120_A}
\end{eqnarray}
for $\Lambda={\rm A}, {\rm B}, {\rm C}$,
and the total spins ${\bm T}^{\rm MF}_\Lambda$ on three sublattices form a 120$^\circ$ structure,
\begin{eqnarray}
{\bm T}^{\rm MF}_\Lambda \cdot  {\bm T}^{\rm MF}_{\Lambda'}
=-\frac{1}{2}| {\bm T}^{\rm MF}_\Lambda|~ | {\bm T}^{\rm MF}_{\Lambda'}|
\label{eq:C-type120_B}
\end{eqnarray}
\end{subequations}
for different sublattices $\Lambda$ and $\Lambda'$, which is the well-known spin structure in the triangular-lattice antiferromagnet.\cite{Miyashita1984,Kawamura1984}
The local spins shrink due to quantum fluctuation.
\end{itemize}
A schematic picture representing this phase is shown in Fig.\ \ref{fig:state}(e).

For several values of $J_{\rm d}/z$, we have determined the ground-state phase diagrams,
some of which are shown in Fig.\ \ref{fig:PDN3}.
For small $|J_{\rm d}|/z$, the resultant diagrams are the same as those for the two-sublattice case.
At $J_{\rm d}=0$, the diagram includes the FM phase, the A-type AFM phase, and the region with the degenerate ground states which is the phase boundary between the spin-nematic and dimer-singlet phases.
The phase boundaries are given by $J_4=\pm 4 J_\times-4$ ($J_4<-4$) and $J_\times=0$ ($J_4>-4$).
The SU(4) point ($J_\parallel=-1$, $J_4=-4$, $J_{\rm d}=J_\times=0$) is on the quadruple point,
where the spin nematic phase is generated
by SU(4) symmetry as shown in Sec.\ \ref{subsec:SU4GS}.
When a negative $J_{\rm d}$ is introduced, the spin nematic phase with ferro-quadrupolar order replaces the region of the degenerate ground states.
As $|J_{\rm d}|/z$ increases, the FM and spin-nematic phases enlarge, while the A-type AFM phase becomes smaller.
Then, at $J_{\rm d}/z \sim -1.7$, the C-type 120$^\circ$-AFM phase appears from the region around $1 \lesssim J_\times \lesssim 2$ and $J_4=0$.
The appearance of this phase can be understood from the strong coupling limit $J_{\rm d}\rightarrow -\infty$,
where 120$^\circ$-AFM phase appears in large $J_\times$ regime.
As $|J_{\rm d}|/z$ further increases, the spin nematic and C-type 120$^\circ$-AFM phases extend and eventually,  around $J_{\rm d}/z \sim -3.5$,
cover the A-type AFM phase region.
For $J_{\rm d}/z \le -3.5$, the phase diagram in our scope ($-2 \le J_\times \le 4$ and $-6 \le J_4 \le 0$) does not depend on $J_{\rm d}/z $, at least down to $J_{\rm d}/z = -10$ in our calculation.
This phase diagram coincides with the diagram obtained by the mean-field approximation\cite{LauchliMP2006}
to the spin-1
bilinear-biquadratic model (\ref{eq:1st_order_Ham}) derived with the first-order perturbation in Sec.\ \ref{sec:perturbation}
on the triangular lattice.
The spin nematic phase boundary touches to the $J_4=0$ line on the single point $J_\times=1$.
The two boundary lines are given by $J_4=4(J_\times-1)$ and $J_4=-2(J_\times-1)$.

\begin{figure}
\begin{center}
\includegraphics[width=70mm]{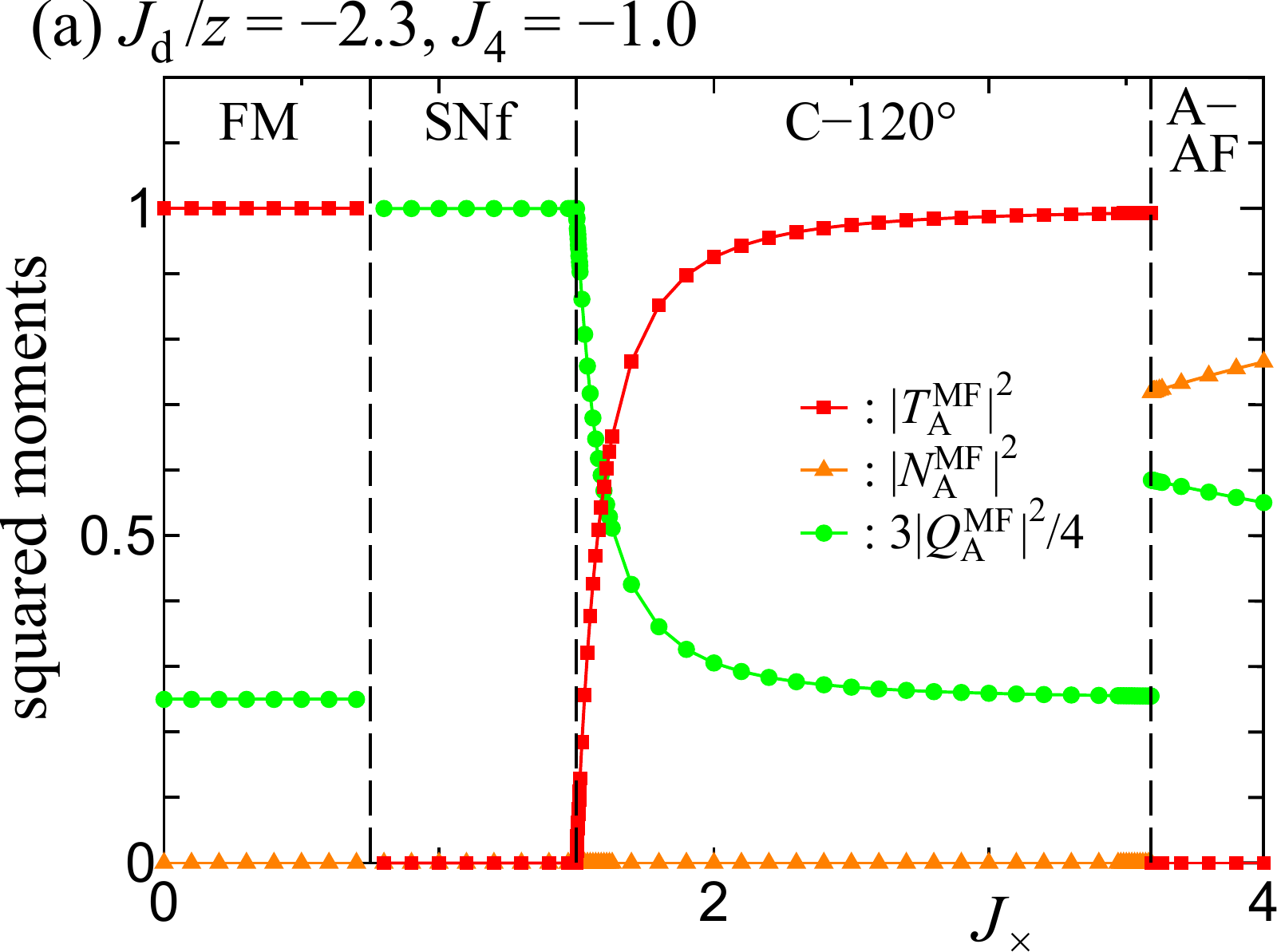}
\includegraphics[width=70mm]{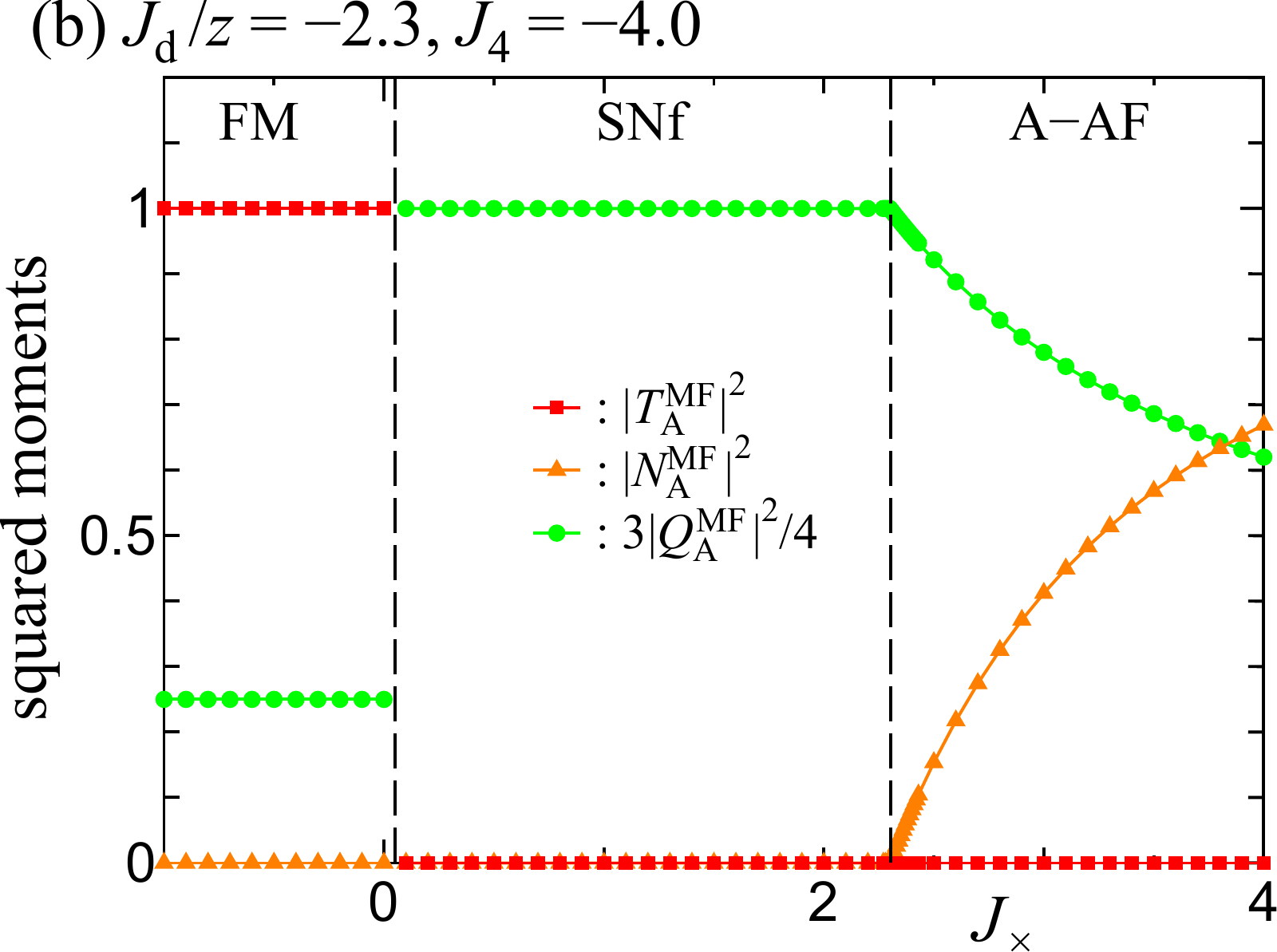}
\includegraphics[width=70mm]{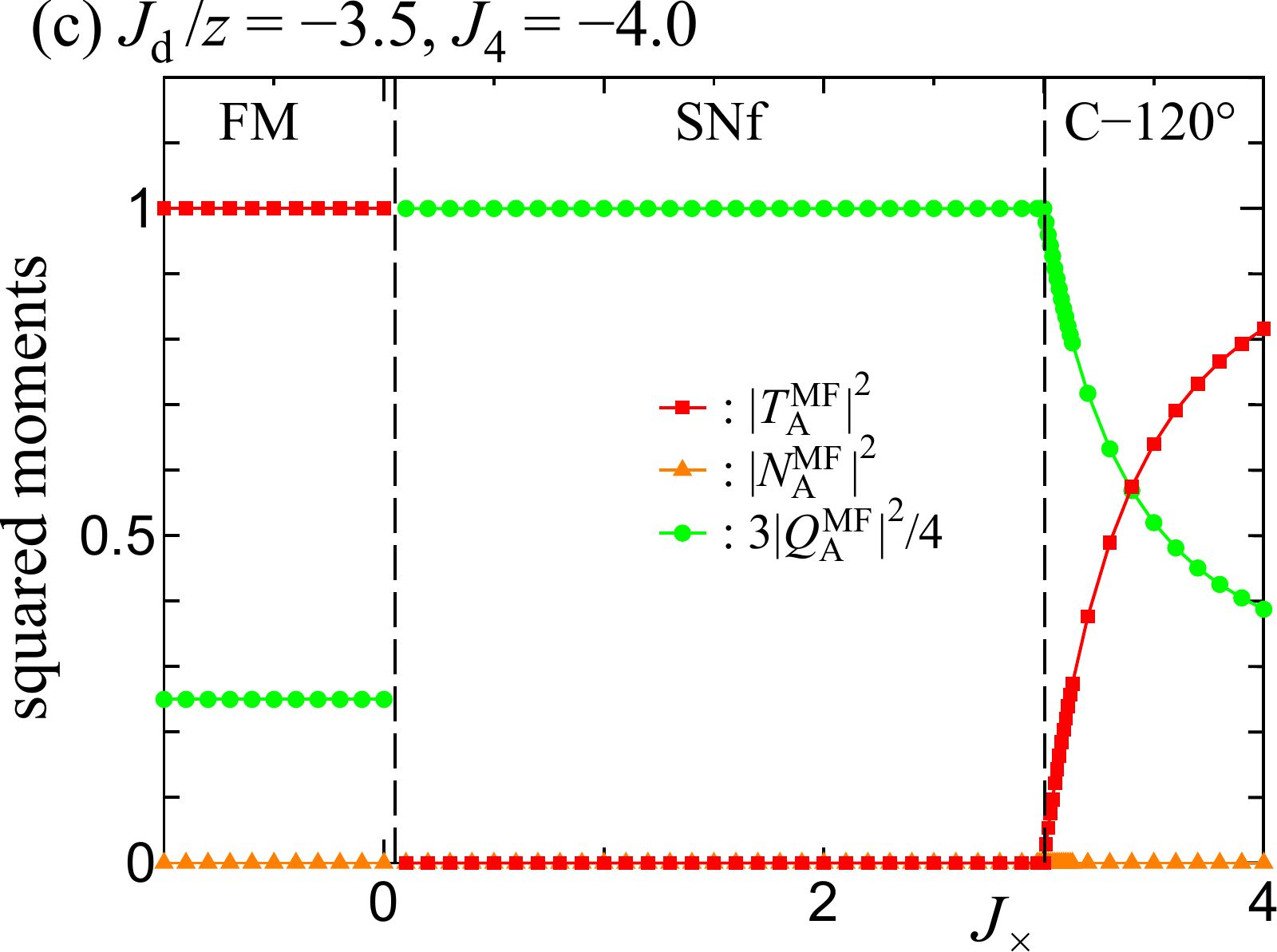}
\caption{
$J_\times$-dependence of squared magnetic moments $|{\bm T}^{\rm MF}_{\rm A}|^2$ and
$|{\bm N}^{\rm MF}_{\rm A}|^2$ and normalized squared spin-quadrupolar moment $3 |{\bm Q}^{\rm MF}_{\rm A} |^2/4$ for the three-sublattice structure;
(a) $J_{\rm d}/z=-2.3$ and $J_4=-1.0$,
(b) $J_{\rm d}/z=-2.3$ and $J_4=-4.0$,
(c) $J_{\rm d}/z=-3.5$ and $J_4=-4.0$.
FM, A-AF, C-120$^\circ$, and SNf represent the ferromagnetic phase, the A-type antiferromagnetic phase, the C-type 120$^\circ$-structure antiferromagnetic phase, and the spin nematic phase with ferro-quadrupolar order, respectively.
Vertical dashed lines represent phase boundaries.
}
\label{fig:OPN3-Jtimes}
\end{center}
\end{figure}

We also show the order of phase transitions in Fig.\ \ref{fig:PDN3}. The  transitions from the FM phase to the other phases are of first order, accompanied with jumps in  ${\bm T}^{\rm MF}_\Lambda$.
The transitions from the spin nematic phase to the two types of AFM phases, i.e., A-type AFM and C-type 120$^\circ$-AFM phases, are continuous while the one between these two AFM phases is of first order.
(See also Appendices\ \ref{subsec:append_MFA_SNf-AAFM}--\ref{subsec:append_MFA_AAFM-C120}
for detailed results.)
Figure~\ref{fig:OPN3-Jtimes} presents the $J_\times$-dependence of the order parameters on the parameter lines given by $J_4=-1.0$ and $-4.0$ in the plane with $J_{\rm d}/z=-2.3$ [dotted lines in Fig.\ \ref{fig:PDN3}(d)] and
the line given by $J_4=-4.0$ in the plane with $J_{\rm d}/z=-3.5$ [dotted line in Fig.\ \ref{fig:PDN3}(f)].
For $J_{\rm d}/z=-2.3$ and $J_4=-1.0$ [Fig.\ \ref{fig:OPN3-Jtimes}(a)], there appear four phases, the FM, spin-nematic, C-type 120$^\circ$-AFM, and A-type AFM phases.
The magnetic moments exhibit the quantum reduction in the C-type 120$^\circ$-AFM and A-type AFM phases and vanish in the spin nematic phase.
The order parameters exhibit jumps at the phase transitions between the FM and spin-nematic phases and between the C-type 120$^\circ$-AFM and A-type AFM phases, while they change continuously at the transition between the spin-nematic and C-type 120$^\circ$-AFM phases.
The continuous nature of the latter transition is also confirmed clearly in the results for $J_{\rm d}/z=-3.5$ and $J_4=-4.0$ in Fig.\ \ref{fig:OPN3-Jtimes}(c).
On the parameter line of $J_{\rm d}/z=-2.3$ and $J_4=-4.0$ [Fig.\ \ref{fig:OPN3-Jtimes}(b)], the system undergoes two transitions from the FM phase to the spin nematic phase, and then to the A-type AFM phase.
We find again that the transition between the FM and spin-nematic phases is of first order, while the transition between the spin-nematic and A-type AFM phases is continuous.

\subsection{Comments}\label{subsec:Vari_product_comment}
Three comments are in order here.
First, in our result by the mean-field approximation with the product-state ansatz, we find no finite region of the spin nematic phase in the case of $J_4=0$, contrary to the expectation from the perturbation theory in Sec.\ \ref{sec:perturbation}.
This result is attributed to the fact that the product-state ansatz, which completely ignores the entanglement between different dimers, is not able to include the effect of the second-order perturbation sufficiently.
Indeed, as shown in the subsequent section,  the mVMC method provides a result that the spin nematic phase emerges in a finite region in the $J_4=0$ case.

Second, it has been shown that, when applied to the spin-1 bilinear-biquadratic model (\ref{eq:Ham_BLBQ}) on the triangular lattice, the mean-field approximation yields an inaccurate result on the phase boundary between the spin-nematic and 120$^\circ$-AFM phases;
the mean-field approximation gives the critical point $\theta_{\rm c} = \arctan(-2) \sim -0.35\pi$, while the exact diagonalization method, which can take the entanglement between different sites into account,
gives $\theta_{\rm c} \sim -0.11\pi$.\cite{LauchliMP2006}
Hence, in our model (\ref{eq:Ham}) for large $|J_{\rm d}|/z$, the phase boundary between the spin-nematic and C-type 120$^\circ$-AFM phases
is also expected to move from the mean-field line $J_4=-2(J_\times-1)$ (corresponding to $\theta_{\rm c} \sim -0.35\pi$ at $J_{\rm d}/z \to -\infty$) toward the line $J_4=-0.61(J_\times-1)$ (corresponding to $\theta_{\rm c} \sim -0.11\pi$), in the direction to enlarge the spin nematic phase.

Third, we must be careful when applying the results for the three-sublattice case to the kagome lattice system.
For the spin-1 bilinear-biquadratic model in the kagome lattice, it was found that the 120$^\circ$-AFM phase is not present in the ground-state phase diagram and, instead, the trimerized valence-bond-crystal phase appears.\cite{Liuetal2015}
Therefore, in our result, the C-type 120$^\circ$-AFM phase should be replaced by a phase corresponding to the trimerized valence-bond-crystal phase, and the boundary lines of the phase as well as the nature of the phase transitions to the phase may also be different from the mean-field results.
We expect that our results for the FM, A-type AFM, and spin nematic phases remain valid semi-quantitatively for the kagome-lattice case, since the ground states of these ordered phases are discribed rather well by the mean-field approximation.

\section{Many-variable Variational Monte Carlo calculation}\label{subsec:mVMC}

In this section, we focus on the case of $J_4=0$ and large negative $J_{\rm d}$.
We numerically explore the ground-state phase diagram using
mVMC method, and establish the emergence of the spin nematic
phase with ferro-quadrupolar order suggested from
the second-order perturbation calculation.

\subsection{Method}\label{subsec:mVMC_method}
To analyze the Hamiltonian defined in Eq.~(\ref{eq:Ham}),
we use the mVMC method,\cite{Misawa2018,mVMC} which
can include the spatial correlations and
quantum fluctuations beyond the direct product of dimer states.
The variational wave function in mVMC is defined as
\begin{align}
|\psi\rangle =\mathcal{P}_{\rm G}|\phi_{\rm pair}\rangle.
\end{align}
Here, $\mathcal{P}_{\rm G}$
is the Gutzwiller factors
defined as
\begin{align}
\mathcal{P}_{\rm G}=e^{-g\sum_{l,j}n_{l,j,\uparrow}n_{l,j,\downarrow}},
\end{align}
where $n_{l,j,\sigma}$ is the number operator of electron at
the $l$th site
in the $j$th dimer with spin $\sigma$.
By taking $g \to \infty$, we completely exclude the doubly occupied states
and express the localized spin-1/2 systems at half filling.
The pair-product part $|\phi_{\rm pair}\rangle$ is
the generalized pairing wave function defined as
\begin{align}
|\phi_{\rm pair}\rangle= \Big[\sum_{l,j,l',j'} \sum_{\sigma,\sigma'}
F_{l,j,\sigma,l',j',\sigma'}c_{l,j,\sigma}^{\dag}c_{l',j',\sigma'}^{\dag}\Big]^{N_{\rm s}/2} |0 \rangle,
\end{align}
where $F_{l,j,\sigma,l',j',\sigma'}$ denotes the variational parameters,
$c_{l,j,\sigma}^{\dagger}$ represents the creation operator
of electron at the $l$th site in the $j$th dimer with spin $\sigma$,
$N_{\rm s}=2N$ is the number of spins (electrons) in the system,
and $|0\rangle$ is the vacuum of electrons.
In this form, we can express spin nematic states by using spin-triplet pairing wave functions.\cite{Shindou2009,Shindou2011}
In our calculations,
we have imposed $2\times2$ ($3\times3$) sublattice structure
in the pair-product part for the
square lattice (triangular lattice) to express the
C-type AFM (C-type 120$^\circ$-AFM) state.
All the variational parameters are simultaneously
optimized by using the stochastic
reconfiguration method.\cite{Sorella2001,Sorella2007}

We note that,
if one takes $F_{l,j,\sigma,l',j',\sigma'}$  as
\begin{align}
F_{l,j,\sigma,l',j',\sigma'} = \begin{cases}
&a_{\Lambda,\sigma\sigma'} ~~ (j=j') \\
&0~~ (j\ne j'),
\end{cases}
\end{align}
the wave function becomes
\begin{align}
|\phi_{\rm pair}\rangle
&= \Big[\sum_{j} \sum_{\sigma,\sigma'}a_{\Lambda}c_{1,j,\sigma}^{\dag}c_{2,j,\sigma'}^{\dag}\Big]^{N_{\rm s}/2} |0 \rangle
\nonumber \\
&\propto \prod_{j}\Big[\sum_{\sigma,\sigma'}a_{\Lambda}c_{1,j,\sigma}^{\dag}c_{2,j,\sigma'}^{\dag}\Big] |0 \rangle
= |\Phi_{\rm DP}\rangle.
\end{align}
This result shows that the pair-product state $|\phi_{\rm pair}\rangle$
includes the dimer-product state defined in Eq.\ (\ref{eq:variational_wf_DP}) as a special case.
Although the entanglement between
dimers is completely ignored in the dimer-product state,
the mVMC
can include such entanglement.

\subsection{Square Lattice}\label{subsec:mVMC_square}
Here, by using mVMC, we examine stability of the
spin nematic phase around $J_\parallel+J_{\times}=0$ in the square lattice.
As a typical case, we take $J_{\rm d}/z=-2~(J_{\rm d}=-8)$, $J_{\parallel}=-1$, and $J_{4}=0$.

To make the initial states of the
mVMC calculations,
we first impose the external fields that induce the
candidate states of the ground state.
In this calculation,
we consider the FM, C-type AFM, and spin nematic states
and take the external fields defined as
\begin{align}
&H_{\rm ex,FM}  = -h_{\rm FM}\sum_{j}T_{j}^{z}, \\
&H_{\rm ex,CAFM} = -h_{\rm CAFM}\sum_{j}T_{j}^{z}e^{i{\bm \pi}\cdot{\bm r}_{j}}, \\
&H_{\rm ex,FQ} = -h_{\rm FQ}\sum_{j}Q_{j}^{(2)},
\end{align}
where
${\bm \pi}=(\pi,\pi)$
is the wave vector of the C-type AFM ordering.
We typically take the amplitude of the external field as unity, i.e.,
$h_{\rm FM}=1$ for example.
We first optimize the variational parameters
using the stochastic reconfiguration
method under the external fields.
Then, by turning off the external fields,
we again optimize the variational parameters
and obtain the FM, C-type AFM, and spin nematic states.
We have also checked that the A-type AFM state, whose initial state can be prepared by imposing the external field $H_{\rm ex,AAFM}  = -h_{\rm AAFM}\sum_{j}N_{j}^{z}$, is unstable and becomes one of the other states
after the optimization without the external field.
We hence omit the result of the A-type AFM state in the following.

Using the optimized wave functions of the FM, C-type AFM,
and spin nematic states, we compute the energies,
the local total-spin moment $T$ defined by
\begin{align}
{T}=\frac{1}{N}\sum_{j}\sqrt{\langle {\bm T}_{j}\rangle^{2}},
\end{align}
and the local spin-quadrupolar moment $Q$ defined by
\begin{align}
Q=\frac{1}{N}\sum_{j}\sqrt{\langle {\bm Q}_{j}\rangle^{2}},
\end{align}
for each state.
The calculation was performed for finite systems with $N=L \times L$ sites under the periodic boundary condition.
We found that the finite-size effects are small for the FM and spin-nematic states, so that we could achieve convergence to thermodynamic-limit values with the data for the systems with up to $L=10$.
For the C-type AFM state, however, the system-size dependences of $T$ and $Q$ are large.
We therefore performed the calculation for the systems with up to $L=14$ and extrapolated the data of $T(L)$ and $Q(L)$ using
the least-square fitting with linear functions of $1/L$, such as $T(L)=T(L=\infty)+a/L$.
We note that, in our calculation, only $Q^{(2)}$ becomes finite
in the collinear magnetic ordered phases (FM phase and C-type AFM phase)
and the spin nematic phase.

\begin{figure}[tb]
\begin{center}
\includegraphics[width=60mm]{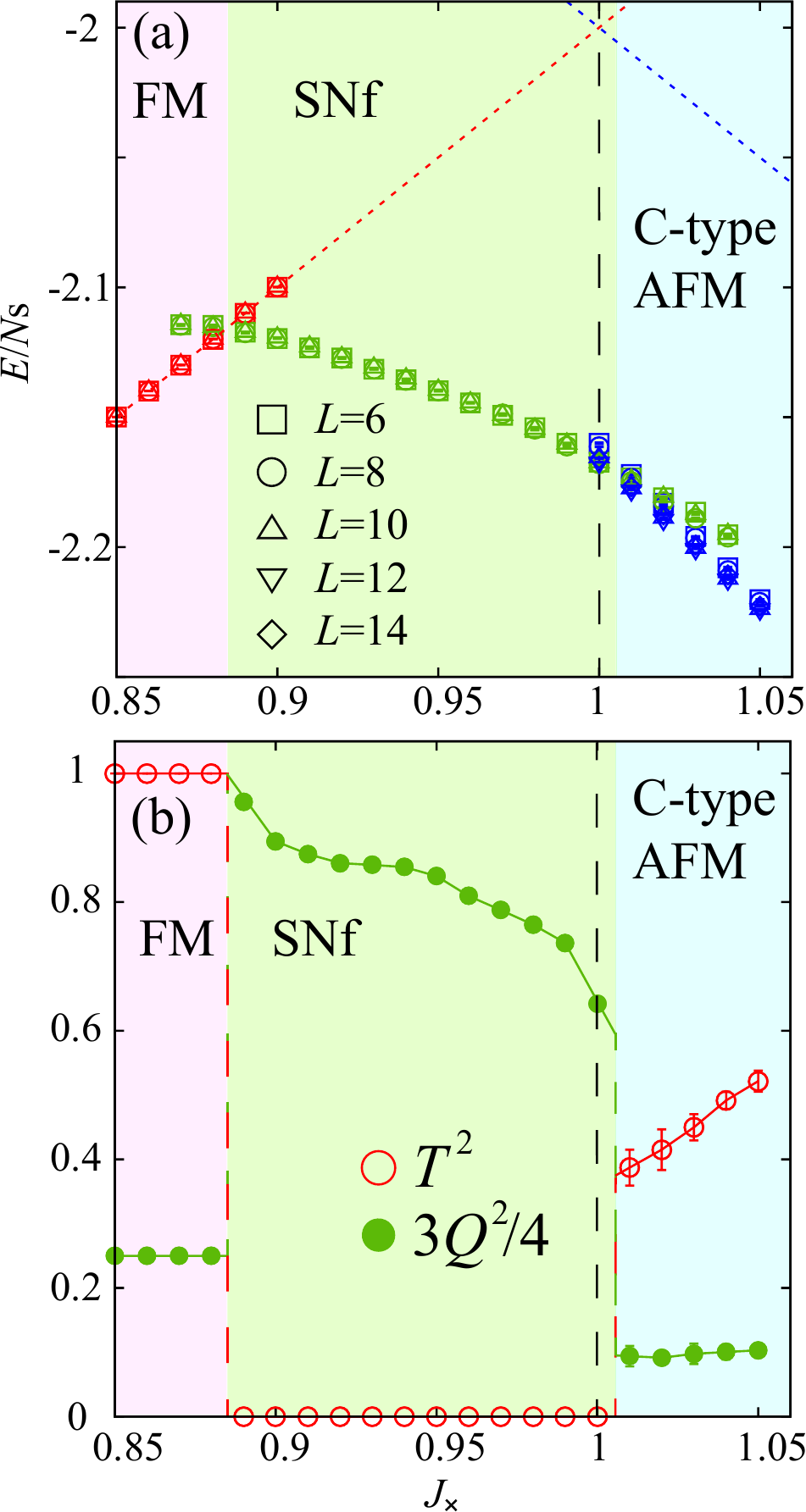}
\caption{
(a)~$J_{\times}$ dependence of the energies
for the FM, spin nematic, and C-type AFM states
for the square-lattice model
on the parameter line with $J_{\rm d}/z=-2~(J_{\rm d}=-8)$, $J_{\parallel}=-1$, and $J_{4}=0$.
The dotted line with positive (negative) slope
shows the energy of the direct product of
dimer states for the FM (C-type AFM) state, which is given by
$J_{\parallel}+J_{\rm d}/4+J_{\times}$ ($-J_{\parallel}+J_{\rm d}/4-J_{\times}$).
(b)~$J_{\times}$ dependence of
the squared local total-spin moment $T^2$ and the normalized squared spin-quadrupolar moment $3Q^2/4$
in the ground state.
For the FM and spin-nematic phases,
the data for $L=10$ are plotted as the thermodynamic-limit values,
while the extrapolated values are plotted for the C-type AFM phase.
Solid and broken lines connecting the data points are guide for the eye.
In both (a) and (b), the vertical broken line at $J_{\times}=1$
represents the degenerate point
($J_{\parallel}+J_{\times}=J_{4}=0$)
in the mean-field solutions.
}
\label{Fig_Sq}
\end{center}
\end{figure}

Figure~\ref{Fig_Sq} (a)
shows $J_{\times}$ dependence of the energies
for the FM, C-type AFM, and
spin nematic states,
while Fig.~\ref{Fig_Sq} (b) shows
the local total-spin moment $T$ and the ferro-quadrupolar moment $Q$
in the ground state.
The non-zero $T$ in Fig.~\ref{Fig_Sq} (b) indicates the appearance
of the magnetic ordered phase such as the FM or C-type AFM phase.
The spin nematic phase is
characterized by finite spin-quadrupolar moment ($Q>0$)
and absence of the magnetic order ($T=0$).

As shown in Fig.~\ref{Fig_Sq} (a),
for $J_{\times} \le 0.88$, we find that the
FM state is the ground state.
Its local moment is still fully polarized ($T=1$)
even when we seriously take
into account the interdimer correlations.

For $0.89 \le J_{\times} < 1.00$, we
find that the spin nematic state becomes
the ground state even when $J_{4}=0$.
In the spin nematic state, we confirm that
no spontaneous polarization occurs
in the spin degrees of freedom ($T=0$) and
the spin-quadrupolar moment $Q$ becomes finite as shown in Fig.~\ref{Fig_Sq} (b).
This result shows that
effects of the interdimer correlations included in mVMC
actually stabilize the spin nematic phase.
The spin nematic phase widely extends for $J_{\times}<1$ while
it does not for $J_{\times}>1$.
This is consistent with the result of the second-order
perturbation theory,
which indicates the stability of the spin nematic phase for
$J_\parallel+J_{\times}<0$.

At $J_\times=1.00$, the energy of the spin nematic state is slightly smaller
than that of the C-type AFM state within the system size treated, but they are almost equal, suggesting that the transition point between the phases is in the range $1.00 \le J_\times < 1.01$.
We note that the C-type AFM state is not stable for $J_{\times}<1$, i.e.,
even if we choose the C-type AFM state as an initial state,
the final state after optimization
becomes a spin nematic state, which has no spin order.
We therefore conclude that the
spin nematic phase exists at least for $J_{\times}<1$.

For $J_{\times} \ge 1.01$,
the C-type AFM state is the ground state.
In contrast to the FM state, the C-type AFM state
is largely affected by the interdimer correlations.
Due to the quantum fluctuations,
the energy of the C-type AFM state obtained by the mVMC method is significantly lower than that of the
direct product of the dimer states.
The local spin moment $T$ in the mVMC result
is also reduced from the saturated value.

From the results above, we conclude that the
system with $J_4=0$ and sufficiently
large negative $J_{\rm d}$ exhibits the
spin nematic phase in addition to the FM and C-type AFM phases.
Both transitions between the FM and spin-nematic phases and
between the spin-nematic and C-type AFM phases are of
first order accompanied by a jump of the magnetic moment.
These transition properties are the same as  the results of
the mean-field approximation
for those transitions occuring at $J_4 < 0$.

\subsection{Triangular Lattice}\label{subsec:mVMC_tri}
For the triangular lattice,
we perform basically the same calculations
as the case of the square lattice.
As a typical case, we take $J_{\rm d}/z=-2~(J_{\rm d}=-12)$,
$J_{\parallel}=-1$, and $J_{4}=0$.

In the triangular lattice,
it is expected that the C-type 120$^{\circ}$-AFM state becomes
the ground state
in addition to the FM and spin nematic states.
To prepare the initial state of the C-type 120$^{\circ}$-AFM state,
we impose the external field defined as
\begin{align}
H_{\rm ex,C120} = -h_{\rm C120}\sum_{j}
[T_{j}^{z}\cos\phi({\bm r}_j)
+T_{j}^{x}\sin\phi({\bm r}_j)],
\end{align}
where
$\phi({\bm r}_j)=\phi(x_j,y_j)=2\pi x_j/3+2\pi y_j/3$.
We have checked by the mVMC that the A-type AFM state is unstable for $J_{\rm d}/z = -2$.

We performed the calculation for finite systems
with $N=6 \times 6, 12 \times 6$, and $12 \times 12$ sites
under the periodic boundary condition.
As in the case of the square lattice, the system-size dependences of the data
are small for the FM and spin-nematic states, so that we could obtain
a good convergence in the data for the systems with up to $N=12\times12$ sites.
For the C-type 120$^\circ$-AFM state,
sizable system-size dependences still remain in the results of $T$ and $Q$.
We hence extrapolated the data of $T(N)$ and $Q(N)$ by using least-square fitting
to linear functions of $1/\sqrt{N}$,
{\it e.g.}, $T(N) = T(N=\infty)+a'/\sqrt{N}$.
We note that in our calculation, only $Q^{(2)}$ is finite in the FM and spin-nematic phases while $Q^{(1)}$ and $Q^{(2)}$
become finite in the C-type 120$^{\circ}$-AFM phase with the coplanar magnetic order.

\begin{figure}[tb]
\begin{center}
\includegraphics[width=60mm]{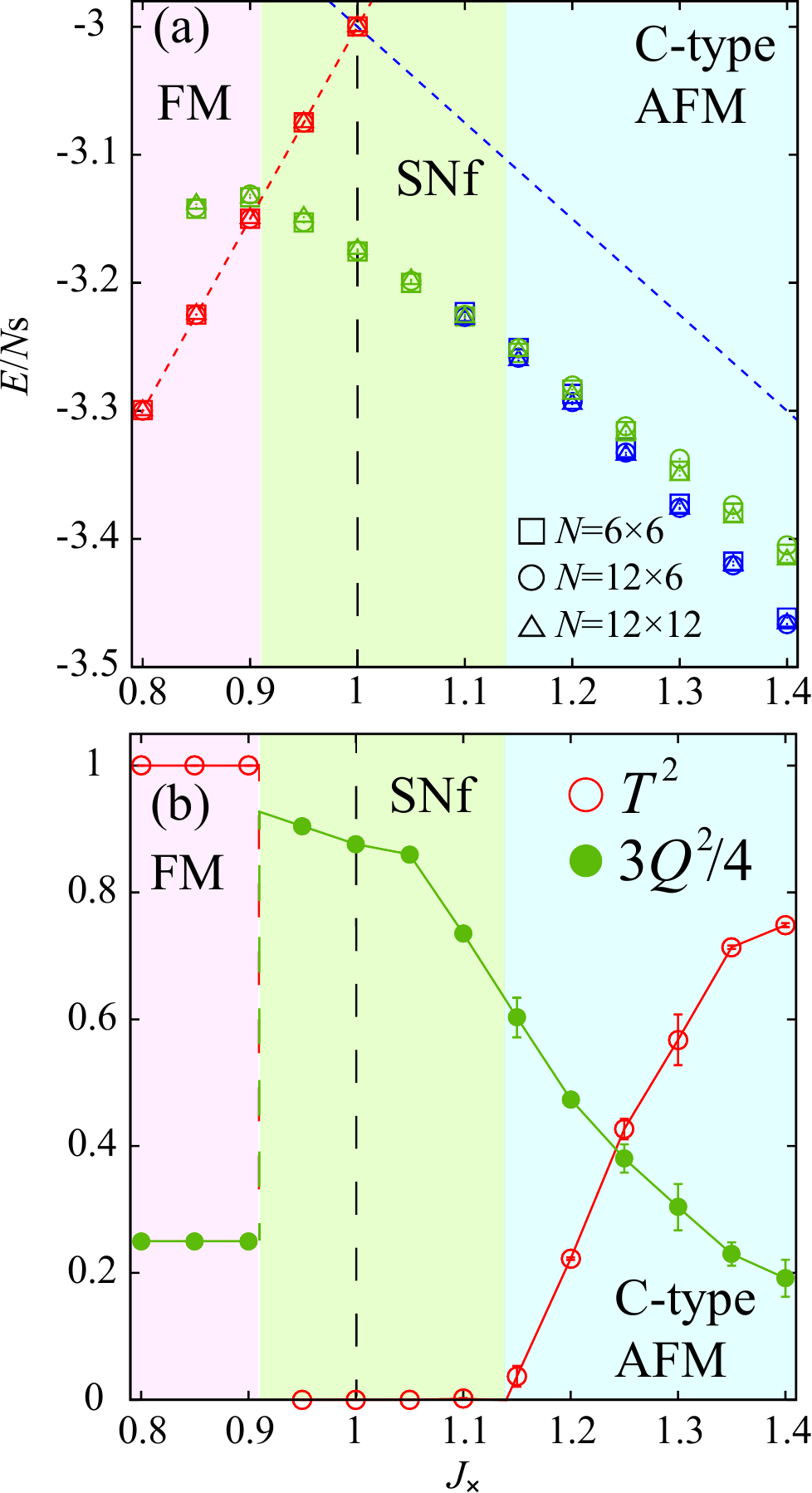}
\caption{
(a)~$J_{\times}$ dependence of the energies
for the FM, spin-nematic, and C-type 120$^{\circ}$-AFM states
for the triangular-lattice model
on the parameter line with $J_{\rm d}/z=-2~(J_{\rm d}=-12)$,
$J_{\parallel}=-1$, and $J_{4}=0$.
The dotted line with positive (negative) slope
shows the energy of the direct product of
dimer states for the FM (C-type 120$^{\circ}$-AFM) state, which is given by
$3J_{\parallel}/2+J_{\rm d}/4+3J_{\times}/2$ ($-3J_{\parallel}/4+J_{\rm d}/4-3J_{\times}/4$).
(b)~$J_{\times}$ dependences of the
squared local total-spin moment $T^2$ and the normalized squared spin-quadrupolar moment $3Q^2/4$
in the ground state.
The values for $N=12 \times 12$ are plotted
for the FM and spin-nematic phases as the thermodynamic-limit values,
while the extrapolated values are shown
for the C-type 120$^{\circ}$-AFM phase.
Solid and broken lines connecting the data points are guide for the eye.
In both (a) and (b), the vertical broken line at $J_{\times}=1$
represents the degenerate point
($J_{\parallel}+J_{\times}=J_{4}=0$)
in the mean-field solutions.
}
\label{Fig_Tri}
\end{center}
\end{figure}

In Fig.~\ref{Fig_Tri}(a),
we show $J_{\times}$ dependence of the energies
for the FM, C-type 120$^\circ$-AFM, and spin nematic states,
while we show in Fig.~\ref{Fig_Tri}(b)
the local total-spin moment $T$ and the spin-quadrupolar moment $Q$
in the ground state.

We find that the FM state is the ground state for small $J_\times$ ($J_{\times} \le 0.9$).
With increasing $J_\times$, the system undergoes a first-order
transition into the spin nematic phase, accompanied with a level cross
of the ground states, around $J_\times \sim 0.9$.
At the other side of the spin nematic phase,
it has turned out that the C-type 120$^\circ$-AFM state
is unstable for $J_\times \le 1.1$.
We can thus conclude safely the appearance of the spin nematic phase
for $0.9 \lesssim J_\times \lesssim 1.1$.
In contrast to the square lattice,
the spin nematic phase emerges
for both $J_{\times}>1$ and $J_{\times}<1$.
This is consistent with the result of the perturbation theory,
which indicates that the spin nematic phase extends to  both sides
of $J_\parallel+J_{\times}=0$.

At $J_{\times} = 1.1$, the energies of the spin nematic
and C-type 120$^\circ$-AFM states are
almost degenerate, and their slopes also seem to be equal.
Furthermore, the local spin moment $T$
in the C-type 120$^\circ$-AFM state for $J_{\times} > 1.1$
decreases with decreasing $J_\times$ and
seems to gradually vanish at $J_{\times} \simeq 1.1$.
These results indicate that the continuous phase transition occurs
between the C-type 120$^\circ$-AFM and spin nematic phases,
which is consistent with the conclusion
obtained by the mean-field approximation.

\section{Vicinity of the SU(4) symmetric point}\label{sec:SU4}
In this section, we present an exact analysis on the SU(4) symmetric point  of the model (\ref{eq:Ham})
with ferromagnetic coupling.
We show emergence of various phases
including the spin nematic phase and the vector chiral (p-type nematic) phase,
by adding perturbations to this symmetric point.
All of these phases are transformed to each other through SU(4) rotation.
We further argue that the spin nematic state can be stabilized, out of the ground-state manifold at the SU(4) point,
by adding a set of Ising interactions as a perturbation.

\subsection{Degenerate ground states of the SU(4) symmetric model}
\label{subsec:SU4GS}
The SU(4) symmetric model with ferromagnetic coupling is given by
$J_4=4J_\parallel<0$ and
$J_{\rm d}=J_\times=0$, as mentioned in Sec.\ \ref{sec:model}.
The Hamiltonian reads
\begin{eqnarray}
\mathcal{H}_{\rm su4} &=& -J_{\rm su4} \sum_{\langle j,j'\rangle}
\left[ {\bm S}_{1,j} \cdot {\bm S}_{1,j'} + {\bm S}_{2,j} \cdot {\bm S}_{2,j'}
\right.
\nonumber \\
&&\left.~~~~~~
+4 ({\bm S}_{1,j} \cdot {\bm S}_{1,j'})({\bm S}_{2,j} \cdot {\bm S}_{2,j'})
\right],
\label{eq:Ham_SU4}
\end{eqnarray}
where we set $J_{\rm su4}=-J_\parallel=-J_4/4>0$.
In this paper we call Eq.\ (\ref{eq:Ham_SU4}) \emph{ferromagnetic SU(4) model}.
We note that  the overall coupling constant is negative opposed to the SU(4) spin-orbital model for the two-orbital Hubbard model
at quarter filling,\cite{Kugel1973,Li1998} whose quantum criticality has been extensively investigated
in one dimension.\cite{Yamashita1998,Azaria1999,Azaria2000,Momoi2003,LecheminantT2005,Lecheminant2006}
We rigorously show in the following that the model (\ref{eq:Ham_SU4}) has various degenerate ground states which are transformed
to each other through the SU(4) rotation. One of  them  is a spin nematic state with ferro-quadrupolar order.

To analyze the SU(4) model, we use the following  fifteen local generators of SU(4) group on each dimer.
We first adopt the spin dipole operators
$T_j^\alpha$ and
$N_j^\alpha$ ($\alpha=x,y,z$), and the quadrupolar operators $Q_j^{(n)}$ ($n=1,\cdots, 5$).
We further use the vector chiral operators (or p-type nematic operators\cite{Andreev1984})
\begin{align}\label{eq:chirality}
  \chi_j^\alpha &= 2 \sum_{\beta,\gamma} \epsilon_{\alpha\beta\gamma} S_{1,j}^\beta S_{2,j}^\gamma
\end{align}
($\alpha=x,y,z$),  where $\epsilon_{\alpha\beta\gamma}$ is the Levi-Civita tensor,
 and the Heisenberg exchange operator
\begin{align}\label{eq:su4_gen0}
  {\cal O}_j &= \frac{2 \sqrt{2}}{\sqrt{3}} {\bm S}_{1,j} \cdot {\bm S}_{2,j}.
\end{align}
All of these fifteen operators,
 $T_j^\alpha $, $N_j^\alpha$, $\chi_j^\alpha$ ($\alpha=x,y,z$),  $Q_j^{(n)}$ ($n=1,\cdots,5$), and ${\cal O}_j$,
are the generators of SU(4) group on dimers.\cite{Lecheminant2006}
This is a useful representation of SU(4) generators for our analysis  of emerging phases.
We note that ${\cal O}_j$ is also the generator of the spin-chirality dual transformation
on dimers.\cite{Momoi2003}
See also Appendix~\ref{sec:append_su4} for a convenient definition of SU(4) generators.

Let us start our argument from the fact that the SU(4) Hamiltonian can be expressed in terms of permutation operators,
\begin{eqnarray}
\mathcal{H}_{\rm su4} &=& -J_{\rm su4} \sum_{\langle j,j'\rangle}
\left[
\sum_{\sigma_1\sigma_2\sigma'_1\sigma'_2}
|\sigma'_1 \sigma'_2\rangle_j |\sigma_1 \sigma_2\rangle_{j'}
\langle \sigma_1 \sigma_2|_j \langle \sigma'_1 \sigma'_2|_{j'} \right.
\nonumber \\
&&\left. ~~~~~~~~~~-\frac{1}{4}\right] .
\label{eq:Ham_SU4_perm}
\end{eqnarray}
From this, it immediately follows that eigenenergies of the ferromagnetic SU(4) model are
lower-bounded by $-3J_{\rm su4}N_{\rm b}/4$,
where $N_{\rm b}\equiv z N/2$ is the number of nearest-neighbor dimer pairs.

A trivial ground state
with the energy $-3J_{\rm su4}N_{\rm b}/4$ is the state in which all spins are down,
\begin{eqnarray}
|{\rm FM} \rangle = \prod_j |\downarrow\downarrow\rangle_j.
\label{eq:all-down}
\end{eqnarray}
Due to SU(4) symmetry, any state obtained through  global SU(4) rotation to Eq.~(\ref{eq:all-down}) also belongs to the ground states.
We summarize below five typical states of  degenerate ground states and their order parameters.
Among these states, the antiferromagnetic state and the vector chiral state are transformed
to each other
through the spin-chirality dual relation,\cite{Hikihara2003,Momoi2003}
whereas
the rest of states are invariant, i.e., self dual.
We note that, since all of these states have product wave functions,
the mean-field approximation presented in Sec.\ \ref{subsec:Vari_product} gives exact ground states  in the ferromagnetic
SU(4) model.

\subsubsection{Ferromagnetic state}
In the ferromagnetic state $|{\rm FM} \rangle$,
all spins are ferromagnetically ordered. The order parameter is given by $\sum_j {\bm T}_j$.
In this state, spin SU(2) and time-reversal symmetries are broken.

\subsubsection{Antiferromagnetic state}
Applying the SU(4) rotation $U=\prod_j (-i) \exp(i \pi  T_j^x/2)\exp (i \pi  N_j^x/2)$ to $|{\rm FM} \rangle$, we obtain
the A-type antiferromagnetic state
\begin{align}\label{eq:st_AF}
|\mbox{A-AF}\rangle=\prod_j  |\uparrow\downarrow\rangle_j.
\end{align}
The spins have an antiferromagnetic order detected with $\sum_j {\bm N}_j$.
In this state, spin SU(2) and time-reversal symmetries are broken.

\subsubsection{Spin nematic state}
Applying the rotation $U=\prod_j  \exp (i \pi Q_j^{(3)} /4)$ to $|{\rm FM}\rangle$, we obtain the spin nematic state with ferro-quadrupolar order,
\begin{align}\label{eq:quad}
  |{\rm SNf}\rangle &= \prod_j \frac{1}{\sqrt{2}}
  ( |\uparrow\uparrow\rangle_j+ |\downarrow\downarrow\rangle_j).
\end{align}
This state does not have any spin order
\begin{eqnarray}
\langle {\rm SNf} | {\bm S}_{l,j}  | {\rm SNf} \rangle =(0,0,0),
\end{eqnarray}
while  has a ferro-quadrupolar order
\begin{align}
&\langle {\rm SNf} | {\bm Q}_j  | {\rm SNf} \rangle = (1,1/\sqrt{3},0,0,0).
\end{align}
Thus only spin SU(2) symmetry is broken.

\subsubsection{Vector chiral state}
Applying the duality transformation\cite{Momoi2003}
$U=\prod_j \exp [i \pi (-\sqrt{6} {\cal O}_j + 1 )/8]$ to $|\mbox{A-AF}\rangle$,
we obtain the vector chiral (p-type nematic) state
\begin{align}\label{eq:st_VC}
|{\rm VC}\rangle=\prod_j \frac{1}{\sqrt{2}}(e^{i\pi/4}  |\uparrow\downarrow\rangle_j+ e^{-i\pi/4} |\downarrow\uparrow\rangle_j ).
\end{align}
Each dimer state is a linear combination of the spin singlet state and a spin triplet state with the complex coefficients.
The vector chiral state does not have any spin order
\begin{eqnarray}
\langle {\rm VC} | {\bm S}_{l,j}  | {\rm VC} \rangle = (0,0,0),
\end{eqnarray}
while it has a vector chiral order
\begin{align}
&\langle {\rm VC} | \bm{\chi}_j  | {\rm VC} \rangle = (0,0,1).
\end{align}
Thus SU(2) and reflection symmetries are broken.
This vector chiral order is also accompanied with a quadrupolar order, which corresponds to
a quadrupolar moment of the vector chirality.

\subsubsection{Dimer singlet state}
The dimer singlet state
\begin{equation}
\prod_j \frac{1}{\sqrt{2}}( |\uparrow\downarrow\rangle_j- |\downarrow\uparrow\rangle_j )
\label{eq:dimer_singlet}
\end{equation}
is also obtained, for example, through the SU(4) transformation $\prod_{j} e^{-i3 \pi/4}\exp (i \pi N_j^z/4)|{\rm VC}\rangle$.
No symmetry is broken in this state.

\subsection{Perturbations to lift the degeneracy}
A weak perturbation to the ferromagnetic SU(4) model can stabilize one of these degenerate ground states,
making its phase realize in a finite parameter region.
For example, the ferromagnetic phase is realized in the vicinity of the ferromagnetic SU(4) model by adding a weak
ferromagnetic Heisenberg interaction on interdimer  bonds,
\begin{eqnarray}
\mathcal{H}_{\rm su4}-\lambda \sum_{\langle j,j'\rangle} {\bm T}_{j}\cdot {\bm T}_{j'}
\end{eqnarray}
with $\lambda>0$. The fully polarized state (\ref{eq:all-down}) still belongs to the lowest-energy eigenstates of this
perturbed Hamiltonian. Another example of exact results is the stability of the dimer singlet phase.
One can show that the dimer singlet state
(\ref{eq:dimer_singlet}) is the lowest-energy eigenstate of the perturbed Hamiltonian
\begin{eqnarray}
\mathcal{H}_{\rm su4}-\lambda' \sum_{ j} {\bm S}_{1,j}\cdot {\bm S}_{2,j}
\end{eqnarray}
with $\lambda'>0$.
%For example, in Sec.\ \ref{subsec:Vari_product}, we have found within the mean-field approximation that the ferromagnetic and A-type antiferromagnetic phases are realised in the vicinity of the ferromagnetic SU(4) model by adding a weak
%ferromagnetic Heisenberg interaction on interdimer  bonds,
%$\mathcal{H}_{\rm su4}-\lambda \sum_{\langle j,j'\rangle} {\bm T}_{j}\cdot {\bm T}_{j'}$
%with $\lambda>0$ and $\lambda < 0$, respectively.
The other states are, however, not eigenstates usually, if we add SU(2) symmetric perturbations.
In the case of the spin nematic state (\ref{eq:quad}),
we infer that the spin nematic phase with the ferro-quadrupolar order is stabilized by adding
a perturbation of biquadratic interactions,
\begin{align}
&\mathcal{H}_{\rm su4}-\lambda'' \sum_{\langle j,j'\rangle} \sum_{n=1}^5 Q_j^{(n)} Q_{j'}^{(n)} \nonumber\\
&=\mathcal{H}_{\rm su4}-4 \lambda'' \sum_{\langle j,j'\rangle}
\biggl[ \left( {\bm S}_{1,j}\cdot {\bm S}_{1,j'}\right)\left( {\bm S}_{2,j}\cdot {\bm S}_{2,j'}\right) \nonumber\\
&  +\left( {\bm S}_{1,j}\cdot {\bm S}_{2,j'}\right)\left( {\bm S}_{2,j}\cdot {\bm S}_{1,j'}\right)
-\frac23 \left( {\bm S}_{1,j}\cdot {\bm S}_{2,j}\right)\left( {\bm S}_{1,j'}\cdot {\bm S}_{2,j'}\right) \biggr]
\end{align}
with $\lambda''>0$, though the state (\ref{eq:quad}) is not an exact eigenstate of this perturbed Hamiltonian any more.

In the rest of this section, we consider an Ising anisotropic perturbation and prove
%instead of exploring these complicated four-spin interactions,
that adding some Ising couplings to the ferromagnetic SU(4) model can stabilize the spin nematic phase.
We use the Hamiltonian of the ferromagnetic SU(4)  model with additional Ising couplings
\begin{eqnarray}
\label{eq:Ham_SU4_pert}
\mathcal{H}' &=& \mathcal{H}_{\rm su4}+ \mathcal{H}_{\rm d}^z + \mathcal{H}_{\rm int}^z,
\label{eq:Ham_SU4andPert}
\end{eqnarray}
where
\begin{subequations}
\begin{align}
&\mathcal{H}_{\rm d}^z =
- J^z_{\rm d} \sum_{ j} S^z_{1,j} S^z_{2,j},
\\
&\mathcal{H}_{\rm int}^z = \sum_{\langle j,j'\rangle}
\left[J^z_\parallel \sum_{l=1,2}  S^z_{l,j} S^z_{l,j'}
+ J^z_\times ( S^z_{1,j} S^z_{2,j'} + S^z_{2,j} S^z_{1,j'} ) \right]
\nonumber \\
\end{align}
\end{subequations}
with $J^z_{\rm d}>0$.
This Hamiltonian conserves  the number of dimers having the state $|\sigma_1 \sigma_2\rangle_j$, which we denote by $N_{\sigma_1\sigma_2}$, and hence is block-diagonalized into subspaces characterized by the quantum numbers $\{N_{\sigma_1\sigma_2}\} = \{ N_{\uparrow\uparrow}, N_{\downarrow\downarrow}, N_{\uparrow\downarrow}, N_{\downarrow\uparrow}\}$.
In the following, we consider the case that the number of dimers, $N=\sum_{\sigma_1 \sigma_2} N_{\sigma_1 \sigma_2}$, is even.

Our argument has some analogies to  the derivation of the so-called $\eta$-pairing superconductivity in an extended Hubbard model.\cite{Yang1989,EsslerKS1992,EsslerKS1993}
We introduce $\eta$ operators of spins
\begin{subequations}
\label{eq:eta}
\begin{align}
& \eta^x_j \equiv \frac{1}{2}Q_j^{(1)}=  Re (S^+_{1,j} S^+_{2,j}), \\
&  \eta^y_j \equiv \frac{1}{2}Q_j^{(3)} = Im (S^+_{1,j} S^+_{2,j}), \\
& \eta^z_j \equiv  \frac{1}{2}T_j^z = \frac{1}{2}\left( S^z_{1,j}+S^z_{2,j}\right).
\label{eq:eta_z}
\end{align}
\end{subequations}
The operators $\eta^\alpha_j$ satisfy the commutation relation
\begin{eqnarray}
\left[ \eta^\alpha_j, \eta^\beta_j \right] = i  \sum_\gamma \epsilon^{\alpha\beta\gamma} \eta^\gamma_j,
\end{eqnarray}
and form an SU(2) group.
The two-dimensional fundamental representation of this SU(2) is spanned with two states $ |\uparrow\uparrow\rangle_j$ and
$|\downarrow\downarrow\rangle_j$. These two states carry pseudo-spin-1/2 degrees of freedom\cite{HikiharaTOS2017}
\begin{eqnarray}
\eta^z_j |\uparrow\uparrow\rangle_j = \frac{1}{2} |\uparrow\uparrow\rangle_j,~~
\eta^z_j |\downarrow\downarrow\rangle_j = - \frac{1}{2} |\downarrow\downarrow\rangle_j,
\label{eq:action_eta_1}
\end{eqnarray}
and $\eta^+_j\equiv \eta^x_j + i \eta^y_j$ ($\eta^-_j \equiv \eta^x_j - i \eta^y_j$) is the raising (lowering) operator between them,
while the rest of states, $ |\uparrow\downarrow\rangle_j$ and $ |\downarrow\uparrow\rangle_j$,
are doubly degenerate singlet states,
\begin{eqnarray}
\eta^\alpha_j |\uparrow\downarrow\rangle_j = \eta^\alpha_j |\downarrow\uparrow\rangle_j=0.
\label{eq:action_eta_2}
\end{eqnarray}
Since $\eta_j^\alpha$ ($\alpha=x,y,z$)  are parts of SU(4) generators,
$\sum_j \eta^\alpha_j$
commutes with $\mathcal{H}_{\rm su4}$.
For later use, we rewrite $\mathcal{H}_{\rm su4}$ and $\mathcal{H}_{\rm int}^z$
with SU(4) generators and $\eta$ operators
introduced in Eqs.~(\ref{eq:eta}),
\begin{align}
\mathcal{H}_{\rm su4} &= -\frac{J_{\rm su4}}{2}\sum_{\langle j,j'\rangle}
\Bigg( 4 {\bm \eta}_j \cdot {\bm \eta}_{j'} +\sum_{\alpha=x,y} T^\alpha_j T^\alpha_{j'}
+  {\bm N}_j \cdot  {\bm N}_{j'}  \nonumber\\
&+\sum_{n=2,4,5} Q^{(n)}_j  Q^{(n)}_{j'} +{\bm \chi}_j \cdot {\bm \chi}_{j'}
+{\cal O}_j {\cal O}_{j'}  \Bigg) ,
\\
\mathcal{H}_{\rm int}^z
&= \frac{1}{2}\sum_{\langle j,j'\rangle}
[4(J^z_\parallel+J^z_\times) \eta^z_j \eta^z_{j'}
+ (J^z_\parallel-J^z_\times)  N^z_j  N^z_{j'} ].
\label{eq:Ham_SU4_pert2}
\end{align}

The degenerate ground states of the ferromagnetic SU(4) model come from all subspaces of distinct $\{ N_{\sigma_1 \sigma_2}\}$.
We first consider the case that only the intradimer Ising coupling term
$\mathcal{H}^z_{\rm d}$
is added to the  SU(4) model.
Since
$\mathcal{H}_{\rm su4}$ is block diagonalized for each  subspace of $\{ N_{\sigma_1\sigma_2}\}$ and
the eigenvalue of $\mathcal{H}^z_{\rm d}$ is $-(J^z_{\rm d}/4)(N_{\uparrow\uparrow}+N_{\downarrow\downarrow}-N_{\uparrow\downarrow}-N_{\downarrow\uparrow})$,
this Ising term partially lifts the ground-state degeneracy  of  the SU(4) model.
Hence, only the states in the subspace with
$N_{\uparrow\downarrow}=N_{\downarrow\uparrow}=0$ have the lowest energy  among the degenerate ground states of the SU(4) model.
All other states in the subspaces with $N_{\uparrow\downarrow}+N_{\downarrow\uparrow}>0$ acquire finite energy costs
of $J^z_{\rm d}(N_{\uparrow\downarrow}+N_{\downarrow\uparrow})/2$ compared to the lowest-energy states with  $N_{\uparrow\downarrow}=N_{\downarrow\uparrow}=0$ and are gapped out of the ground-state manifold.
We note that the role of $\mathcal{H}^z_{\rm d}$ in our argument is
only to wipe out the states with $N_{\uparrow\downarrow}+N_{\downarrow\uparrow}>0$ from the ground states.
Hence, $J^z_{\rm d}$ does not have to be small and can be comparable to or larger than $J_{\rm su4}$.

To see the nature of the ground states of the Hamiltonian $\mathcal{H}_{\rm su4}+\mathcal{H}^z_{\rm d}$,
we consider the projection  ${\cal P}$ to the subspace with $N_{\uparrow\downarrow}=N_{\downarrow\uparrow}=0$, to
which the ground states belong.
In this projected space, the terms containing ${\bm \eta}$ spins in the Hamiltonian give non-trivial operators and the rest of terms give constants.
As a result, the Hamiltonian $\mathcal{H}_{\rm su4}+\mathcal{H}^z_{\rm d}$ in this subspace
reduces to a ``ferromagnetic" Heisenberg model of ${\bm \eta}$ spins,
\begin{eqnarray}
{\cal P} (\mathcal{H}_{\rm su4}+\mathcal{H}^z_{\rm d}) {\cal P}=
-2J_{\rm su4}\sum_{\langle j,j'\rangle} {\cal P} {\bm \eta}_j \cdot {\bm \eta}_{j'} {\cal P},
\label{eq:Ham_su4_Ising_d}
\end{eqnarray}
except for a constant.
Thus the ground states of $\mathcal{H}_{\rm su4}+\mathcal{H}^z_{\rm d}$ are the perfectly ferromagnetic states of ${\bm \eta}$ spins.

We next add  an interdimer Ising coupling term $\mathcal{H}^z_{\rm int}$.
Note that $\mathcal{H}^z_{\rm int}$ also preserves the quantum numbers $\{ N_{\sigma_1\sigma_2}\}$.
In a subspace with $N_{\uparrow\downarrow}+N_{\downarrow\uparrow} > 0$, the change of the lowest energy  induced by $\mathcal{H}^z_{\rm int}$ (compared to that in the subspace with $N_{\uparrow\downarrow}=N_{\downarrow\uparrow}=0$) is, at most, of order $\mathcal{O}(N_{\uparrow\downarrow}+N_{\downarrow\uparrow})$.
Therefore, if the intradimer Ising term $\mathcal{H}^z_{\rm d}$ is sufficiently strong compared to the interdimer Ising term $\mathcal{H}^z_{\rm int}$, i.e., $|J^z_\parallel|, |J^z_\times| \ll J^z_{\rm d}$, the ground state of total Hamiltonian $\mathcal{H}'$ [in Eq.\ (\ref{eq:Ham_SU4andPert})]
still belongs to  the subspace with $N_{\uparrow\downarrow}=N_{\downarrow\uparrow}=0$.

To consider the ground state in the thermodynamic limit, a careful treatment of low-lying excited states is needed here.
With increasing the system size,
quasi-degenerate low-lying states
(also known as Anderson tower states) in finite-size systems whose excitation energies decay
in the form of ${\cal O}(1/N)$ also join to the lowest-energy state, forming a symmetry broken ground state
in the thermodynamic limit.\cite{Anderson1952,Bernu1992,Koma1994}
In our model, if the coupling $J_{\rm d}^z$ is sufficiently strong so that the energy gap from the ground states to the lowest excited states with $N_{\uparrow\downarrow}+N_{\downarrow\uparrow}>0$ is of order unity,
these excitations do not contribute to the formation of ground states in the thermodynamic limit and
hence  the ground states are properly reproduced by the Hamiltonian in the projected space.

In the projected space with $N_{\uparrow\downarrow}=N_{\downarrow\uparrow}=0$,
only $\eta$ operators remain non-trivial and the effective total Hamiltonian is
written  as
\begin{align}
{\cal P} \mathcal{H}' {\cal P}
=-2J_{\rm su4} \sum_{\langle j,j'\rangle}
{\cal P}\left( \eta^x_j \eta^x_{j'} + \eta^y_j \eta^y_{j'} + \Delta_\eta \eta^z_j \eta^z_{j'} \right){\cal P}
\label{eq:Ham_SU4_effective}
\end{align}
with
\begin{eqnarray}
\Delta_\eta = 1 - \frac{J^z_\parallel+J^z_\times}{J_{\rm su4}}.
\label{eq:Delta_eta}
\end{eqnarray}
We thus find that the ground state of $\mathcal{H}'$
is described with ${\bm \eta}$ spins
and the effective Hamiltonian is equivalent to the ferromagnetic XXZ model.
When $|\Delta_\eta| < 1$, i.e., $0 < J^z_\parallel+J^z_\times < 2J_{\rm su4}$, the anisotropy is of easy-plane type, and in the ground state all ${\bm \eta}$ spins point to the same direction in the $xy$ plane of the $\eta$ spin space.
This ground state is indeed a spin nematic state: ${\bm \eta}$ spins have an order
$\langle {\bm \eta}_j \rangle=(\rho \cos \vartheta , \rho \sin \vartheta, 0)$ with finite positive constant $\rho$,
which corresponds to the ferro-quadrupolar order,
\begin{align}
&\langle Q^{(1)}_j \rangle = 2 \rho \cos \vartheta,\nonumber\\
&\langle Q^{(3)}_j \rangle = 2 \rho \sin \vartheta,
\label{eq:Q_expect}
\end{align}
and $\langle S^z_{1,j} +  S^z_{2,j} \rangle = 0$.
This order is accompanied with the spontaneous breaking of the global spin-rotation symmetry around the spin $z$ axis.
Furthermore, since the ground state belongs to the subspace with $N_{\uparrow\downarrow}=N_{\downarrow\uparrow}=0$, the expectation values of $S^x_{l,j}$, $S^y_{l,j}$ ($l=1,2$), and  $N^z_{j}=S^z_{1,j}-S^z_{2,j}$  are always zeros and hence the ground state
does not have any spin (dipole) order,
\begin{eqnarray}
\langle {\bm S}_{1,j} \rangle &=& \langle {\bm S}_{2,j} \rangle = (0,0,0).
\label{eq:S_expect}
\end{eqnarray}
In this argument, the ferromagnetic Ising couplings $J^z_{\rm d}$ on dimers are important and the
choice of the other couplings $J^z_\parallel$ and $J^z_\times$ is relatively free inside the region
$0<J^z_\parallel+J^z_\times<2 J_{\rm su4}$
(and $|J^z_\parallel|, |J^z_\times| \ll J^z_{\rm d}$).

To summarize, we have shown that the ferromagnetic SU(4) model [Eq.\ (\ref{eq:Ham_SU4})] has the degenerate various ground states including the spin nematic state with ferro-quadrupolar order and the vector chiral state.
Each phase extends to a finite parameter region of a generalized model around the SU(4) symmetric point.
Furthermore, adding some appropriate Ising couplings to the SU(4) model can also stabilize the spin nematic phase out of the degenerate ground states of the SU(4) model.

\section{Concluding remarks}\label{sec:conc}
%%%% summary
In this paper, we have studied the frustrated spin-1/2 dimer model (\ref{eq:Ham})
composed of ferromagnetic dimers on two-dimensional lattices with a bilayer structure,  to explore the spin nematic phase.
We have used various approaches --- perturbative calculations, a mean-field approximation, mVMC method, and exact arguments.
We have succeeded to show the appearance of the spin nematic phase with ferroquadrupolar order in a wide parameter range,
which includes the model with only two-spin interactions and also the SU(4) symmetric model.
All the spin nematic states found in this paper are adiabatically connected to each other on each lattice, forming the single spin nematic phase.

%In particular, inside of the nematic phase,
%we analyzed the model in three parameter
%conditions,
%to confirm the appearance of the phase
%and to elucidate the mechanism of spin nematic ordering.
The appearance of the spin nematic phase in our model
can be understood from two mechanisms, which are active in two distinct regimes in the phase.
One is attributed to the effective biquadratic interactions between spin-triplet states
in dimers, which couple the spin-quadrupolar degrees of freedom on neighboring dimers.
This mechanism is elucidated by the perturbative analysis in the strong ferromagnetic dimer-coupling regime.
In this limit, our spin-1/2 model is mapped to the effective spin-1 bilinear-biquadratic model.
%, which is known to exhibit the spin nematic order
%when the biquadratic interaction dominates the bilinear one.
The effective biquadratic interaction derived perturbatively
comes from the four-spin exchange interaction in the original model in the first-order process
and also from the two-spin interactions in the second-order process.
When this effective biquadratic interaction dominates the effective bilinear
interaction, the spin nematic phase emerges.

This mechanism opens a possibility to realize the spin nematic phase
in the model with only two-spin interactions.
When there is no four-spin interaction in the original model,
the effective biquadratic interaction is usually not strong
as it comes from the second-order perturbation.
However, when the interdimer two-spin interactions in the original model
have a strong competition between ferromagnetic and
antiferromagnetic couplings, the effective bilinear interaction becomes very weak because of the
cancellation in perturbations and
hence the effective biquadratic interaction becomes the strongest one.
The spin nematic phase thus appears in between
the ferromagnetic and antiferromagnetic phases, even if the original Hamiltonian does not
contain any four-spin interaction.
We note that a similar mechanism to surpress the bilinear interaction
was argued for the multi-orbital Hubbard model.\cite{MilaZ2000,TanakaYH2018}
%Second, we numerically study the system without any four-spin interaction,
%i.e., $J_4=0$,
%to confirm the appearance of the spin nematic phase.
%We loosen the condition of the strong coupling limit and consider
%the moderately strong $J_{\rm d}$ model.
To confirm the emergence of the spin nematic phase in the model
with only two-spin interactions,
we have performed large-size numerical mVMC calculations
for our model (\ref{eq:Ham}) with only two-spin interactions ($J_4=0$)
and moderately strong ferromagnetic $J_{\rm d}$.
%This method naturally includes quantum fluctuations,
%which are for example described by the second-order perturbation.
The resultant phase diagrams for  the square and triangular lattices
indeed exhibit the ferro-quadrupolar spin-nematic phases in finite parameter ranges
in between the ferromagnetic  and antiferromagnetic phases.
%which is consistent with the perturbative analysis mentioned above.

The other mechanism to realize the spin nematic phase is
found in the ferromagnetic SU(4) symmetric model.
%we analyze the system in the vicinity of the ferromagnetic SU(4)
%symmetric point, applying exact arguments.
%In this regime, the coupling of the four-spin interaction is strong.
Applying exact arguments, we have shown that
%the mean-field approximation
%with product-state ansatz gives an exact ground state
%on the ferromagnetic SU(4) model.
a spin-nematic product state becomes one of the exact ground states on the ferromagnetic SU(4) model,
because the spin nematic order parameters are given by the generators of SU(4) symmetry.
This result suggests that the mean-field approximation
with product-state ansatz becomes exact on the ferromagnetic SU(4) model.
Due to the SU(4) symmetry, various phases are generated  in the vicinity
of this high symmetric point.
In addition to the spin nematic phase, the vector chiral
(p-type nematic) phase as well as the conventional ferromagnetic and antiferromagnetic phases and the dimer singlet
phase appear in the vicinity of  the SU(4) symmetric point if appropriate perturbative interactions are added.
Using an argument similar to the one on the $\eta$-pairing superconductivity, we have also proven that our model at the SU(4) point with appropriate Ising couplings can exhibit the spin nematic phase.

%*********
%We have determined the ground-state phase diagram of the model (\ref{eq:Ham})
%as functions of $J_{\rm d}/z$, $J_\times$, and $J_4$ (with fixed $J_\parallel =-1$) using the mean-field approximation.
%
%This mapping also enables us to determine the ground-state phase diagram
%of our model in this limit
%by using the results already known for the spin-1 bilinear-biquadratic model.
%**********

%%%% SU(2)-symmetric FQ phase
The spin nematic state found  in this paper is
stable in a wide parameter region of the SU(2) symmetric model
without any magnetic field.
This is because both the mechanisms for spin nematic ordering we showed
are valid at zero field.
%as it originates from the SU(4) symmetric model.
This is a clear contrast to the spin nematic state found in spin-1/2 frustrated ferromagnets,\cite{Shannon2006,Sindzingre2009,Hikihara2008,Sudan2009,Ueda2013}
which is caused by the two-magnon instability\cite{Shannon2006,Chubukov1991,Kecke2007}
at the saturation field and hence usually more stable in a strong external magnetic field.
Furthermore,
the spin nematic phase in our model  remains to exist  even without
any four-spin interaction in contrast to the spin nematic state  in the spin-1 bilinear-biquadratic model,
which requires a strong biquadratic coupling.

Our analysis also revealed that the phase transition
between the ferroquadrupolar spin nematic and antiferromagnetic phases
is continuous in many cases. For example, on the triangular lattice, the transition between the spin nematic phase and
the C-type antiferromagnetic phase with $120^\circ$ structure is always continuous, as far as we studied.
This is consistent with the former analysis
in the $J_{\rm d}\rightarrow -\infty$ limit.\cite{LauchliMP2006}

%This can be understood in the perturbation theory from the strong intradimer coupling limit;
%an effective biquadratic interaction appears from the second-order perturbation even without four-spin interaction,
%whereas the effective bilinear interaction is
%drastically weakened by  the competition between
%the ferromagnetic and antiferromagnetic interdimer interactions.

%%%% implication to experiments (real material)
Though our model is a toy model, there are a few candidate materials which might capture some
features of our model.
We can  find possible candidates in organic magnets, which realize spin systems with high flexibility in the control of exchange interactions.\cite{Yamaguchi2018a,Yamaguchi2018b}
In particular, several organic biradical molecules  are known to contain dimer structure of two $S=1/2$ spins coupled ferromagnetically.\cite{Shiomi1993,Hosokoshi1999,Iwase2013}
Arranging these dimers in a two-dimensional lattice with a bilayer structure may provide a playground
for searching for the spin nematic state.
Another candidate is an SU(4)-symmetric system.
It was recently proposed  that a Coulomb-impurity lattice on a graphene substrate can realize an SU(4)-symmetric spin-orbital model with a tunable coupling constant.\cite{DouKU2016}

Even after a material exhibiting the spin nematic phase  is prepared, the direct experimental detection of the phase is
not easy, but we can manage to find it by combining various experiments.
One of the most striking features of the spin nematic phase is  the absence of the spin order down to zero temperature,
which we can observe by neutron scattering and nuclear magnetic resonance measurements.
In addition, spin-wave analyses revealed that
 the ferroquadrupolar state in the spin-1 bilinear-biquadratic models has gapless excitations with a linear dispersion.\cite{Papanicolaou1988,LauchliMP2006}
These low-energy excitations result in  algebraic temperature dependence in thermodynamic  properties,
{\it e.g.,} $T^2$ dependence in specific heat data in two dimensions.
Those properties are also expected to appear in  the spin nematic phase in our spin-1/2 dimer models.
As for the detection  of the spin-quadrupolar order, although it is a challenging task, a  few  theories
were  recently proposed for
inelastic light scattering,\cite{Michaud2011} resonant inelastic x-ray
scattering,\cite{Savary2015} and electron spin resonance.\cite{FuruyaM2018}
We hope that our study stimulates further the search for the spin nematic phase  in real materials.

\acknowledgments
We thank to Yuko Hosokoshi, Chisa Hotta, Hosho Katsura, Kouichi Okunishi, Nic Shannon, and Hironori Yamaguchi
for fruitful discussions.
This work was partially supported by JSPS KAKENHI Grant Numbers JP15K05198, JP16H06345, JP16K17746, JP17H02931, and JP16K05425.
This work was also supported by Building of Consortia for the Development
of Human Resources in Science and Technology from the MEXT of Japan.
Our calculation was partly carried out at the Supercomputer Center,
Institute for Solid State Physics, University of Tokyo.

\appendix

\section{Numerical results of the mean-field approximation}
\label{sec:append_Num_MFA}

In this Appendix, we present details of the numerical procedure in the analysis of the mean-field approximation discussed in Sec.\ \ref{subsec:Vari_product} as well as the results of phase transitions in the approximation.

\subsection{Details of the calculation}
\label{subsec:append_detail_MFA}

The mean-field solution was obtained by optimizing numerically the complex coefficients $\{ a_{\Lambda,\sigma_1\sigma_2}\}$ in the product state $|\Phi_{\rm DP}\rangle$ defined in Eqs.\ (\ref{eq:variational_wf_DP}) and (\ref{eq:dimerstate_DP}) so that the state has the minimum expectation value of the bond Hamiltonian $\mathcal{H}_{jj'}$ [Eq.\ (\ref{eq:Ham_bond})];
The minimized function is $\langle \Phi_{\rm DP} | \mathcal{H}_{jj'} | \Phi_{\rm DP}\rangle$ ($j \in {\rm A}, j' \in {\rm B}$) for the two-sublattice case and $\langle \Phi_{\rm DP} | (\mathcal{H}_{jj'} + \mathcal{H}_{j'j''} + \mathcal{H}_{j''j} ) | \Phi_{\rm DP} \rangle$ ($j \in {\rm A}, j' \in {\rm B}, j'' \in {\rm C}$) for the three-sublattice case.
In the calculation, the normalization condition $\sum_{\sigma_1\sigma_2} |a_{\Lambda,\sigma_1\sigma_2}|^2=1$ for each sublattice $\Lambda$ was imposed.
From the arbitrariness of the global phase factor of $|\Phi_{\rm DP}\rangle$, we further imposed the constraint that $a_{\Lambda,\downarrow\downarrow}$ is real for each $\Lambda$, without loss of generality.

The minimization was achieved by using the steepest descent method.
Since the optimization process often becomes slow and is trapped in a local minimum in the steepest descent method, we performed 1000 calculations starting from randomly-prepared initial states for each parameter point.
The calculations were continued until the optimization converged or the method reached 10000 iterations.
We note that we achieved the convergence for all the 1000 runs for most of the parameter points treated and for 165 runs even at the worst case.
We then adopted the state giving the lowest energy as the ground state.

For determining the phase diagrams, we performed the calculation on $J_\times$ versus $J_4$ planes with several values of
$J_{\rm d}/z$, varying $J_\times$ and $J_4$ with intervals of $0.1$.
(Note that $J_\parallel$ is fixed to be $J_\parallel=-1$.)
Figures\ \ref{fig:PDN2} and \ref{fig:PDN3} show typical examples of the obtained phase diagrams.
In order to explore the nature of the phase transitions, we performed the calculation on several parameter lines with fixed $J_{\rm d}/z$ and $J_4$ ($J_\times$),   varying $J_\times$ ($J_4$) with intervals of $0.01$ or $0.001$.

\subsection{Transition between spin-nematic and A-type AFM phases}
\label{subsec:append_MFA_SNf-AAFM}

\begin{figure}
\begin{center}
\includegraphics[width=70mm]{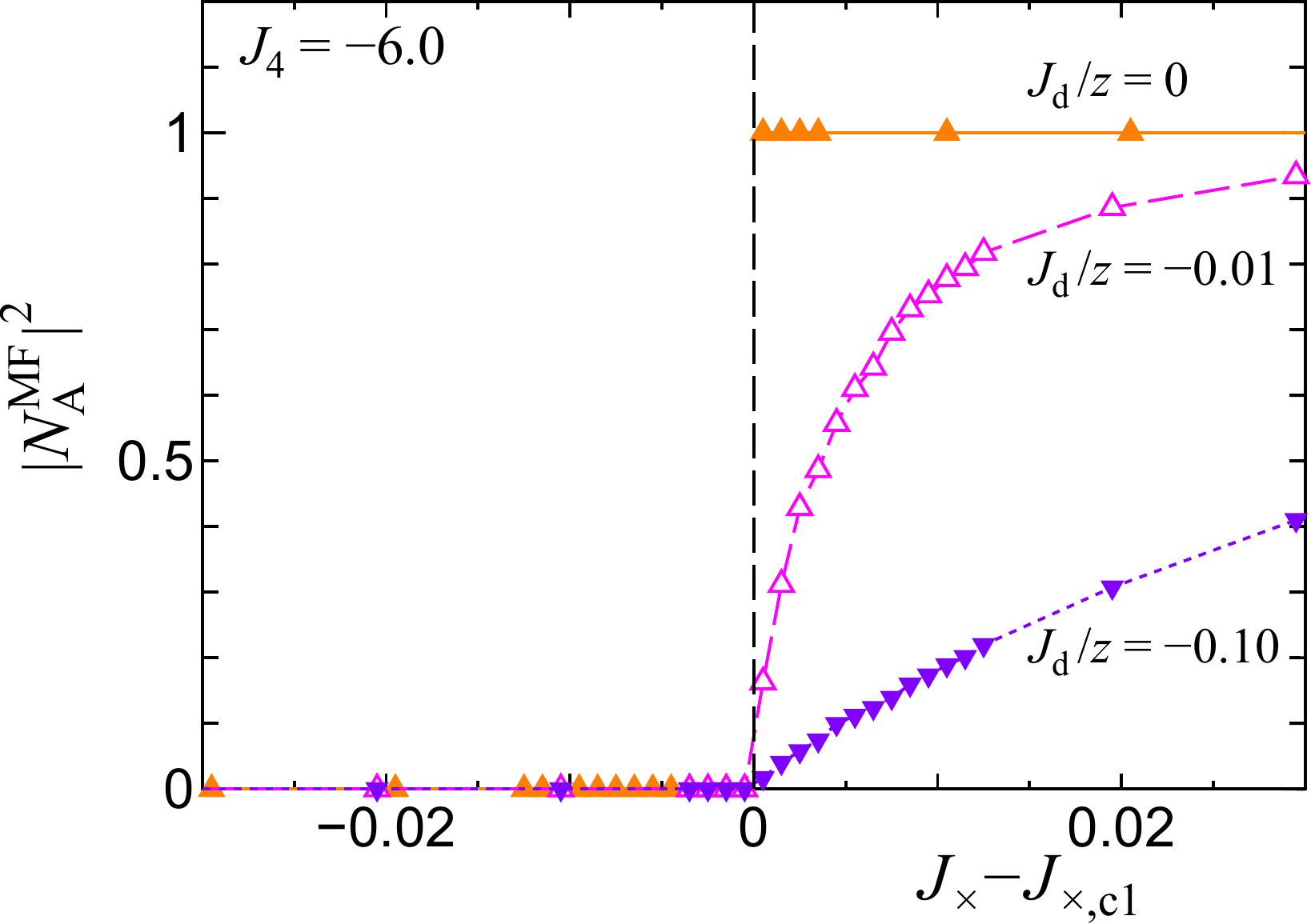}
\caption{
Squared N\'{e}el-spin moment $|{\bm N}^{\rm MF}_{\rm A}|^2$ for the two-sublattice structure for the cases
$J_{\rm d}/z=0$, $-0.01$ and $-0.10$ as a function of $J_\times - J_{\times, {\rm c1}}$,
where $J_{\times, {\rm c1}}$ is the critical value of $J_\times$ at the boundary between the spin nematic phase and
the A-type AFM phase.
The other parameters are set as $J_\parallel=-1$ and $J_4=-6.0$.
}
\label{fig:TO-SNf-AAFM}
\end{center}
\end{figure}

The phase transition between the spin nematic phase and the A-type AFM phase appears in both two- and three-sublattice cases.
We find that this transition is continuous for all parameter lines studied.
Figure\ \ref{fig:TO-SNf-AAFM} shows the $J_\times$-dependence of the N\'{e}el-spin moment around the transition in the parameter lines $J_4=-6.0$ and $J_{\rm d}/z = 0, -0.01, -0.10$ in the two-sublattice case.
For $J_{\rm d}/z \lesssim -0.10$, the N\'{e}el-spin moment rises from zero continuously with a moderate slope.
The slope is steeper as $|J_{\rm d}|/z$ is smaller, however, the transition is still continuous even for $J_{\rm d}/z = -0.01$.
At $J_{\rm d}=0$, the spin nematic phase reaches to  the boundary region between the spin-nematic and dimer-singlet phases with degenerate ground states, and the order parameters exhibit  finite jumps when the system moves from this region into the A-type AFM phase.
We note that the steep rise of the N\'{e}el-spin moment at $J_{\rm d}/z \to -0$ was observed in both two- and three-sublattice cases.
For not too small $|J_{\rm d}|/z$, $|{\bm N}^{\rm MF}_\Lambda|^2$ rises linearly with a moderate slope, suggesting
$|{\bm N}^{\rm MF}_\Lambda| \propto \sqrt{J_\times - J_{\times,{\rm c1}}}$, where $J_{\times,{\rm c1}}$ is the critical value.

\subsection{Transition between spin-nematic and C-type 120$^\circ$-AFM phases}
\label{subsec:append_MFA_SNf-C120}

\begin{figure}
\begin{center}
\includegraphics[width=70mm]{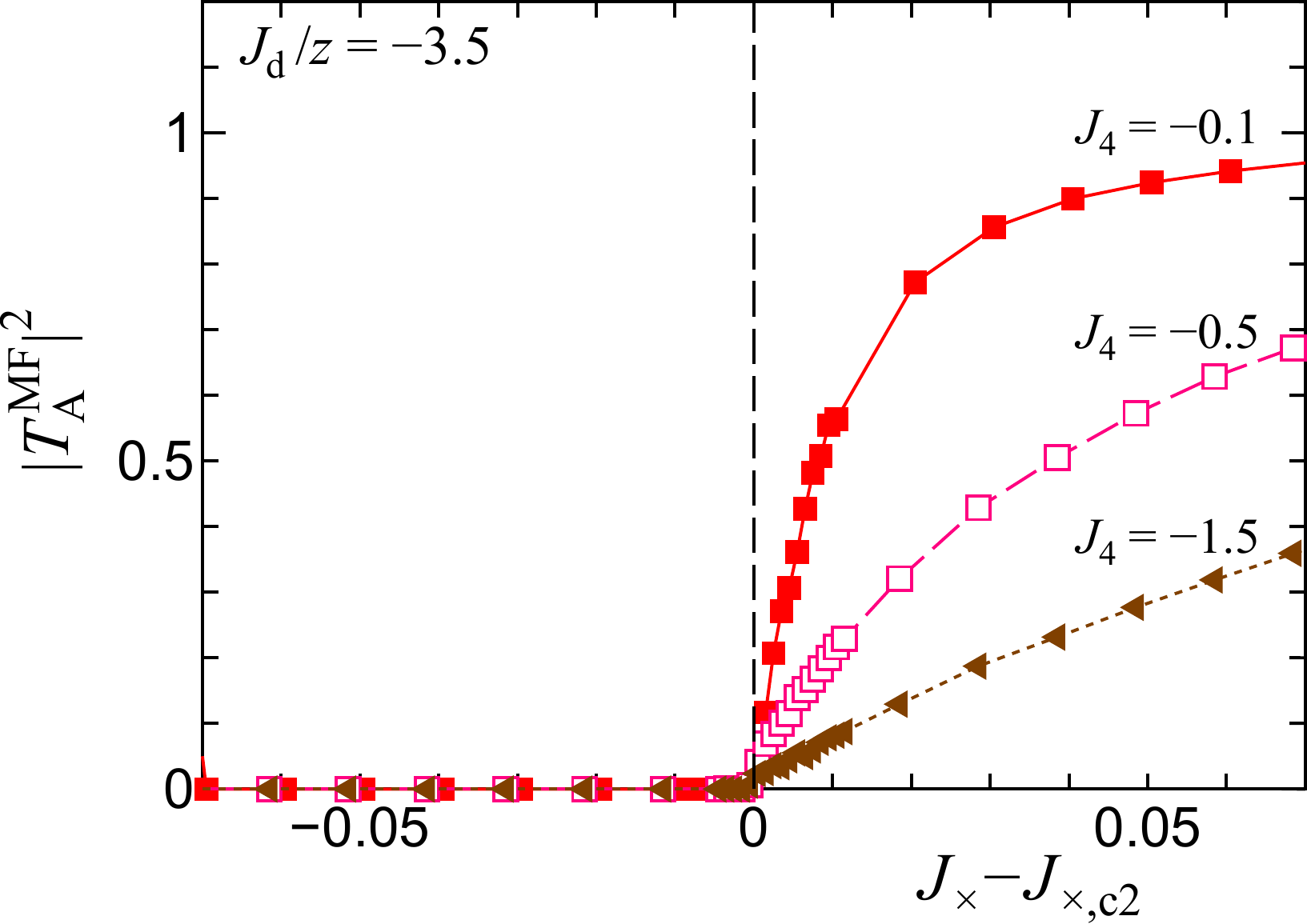}
\caption{
Squared local total-spin moment $|{\bm T}^{\rm MF}_{\rm A}|^2$ for the three-sublattice structure for the cases
$J_4=-0.1$, $-0.5$ and $-1.5$ as a function of $J_\times - J_{\times, {\rm c2}}$, where $J_{\times, {\rm c2}}$ is the critical value of $J_\times$ at the transition point between the spin nematic phase and the C-type 120$^\circ$-AFM phase.
The other parameters are set as $J_\parallel=-1$ and $J_{\rm d}/z=-3.5$.
}
\label{fig:TO-SNf-C120}
\end{center}
\end{figure}

The phase transition between the spin nematic phase and the C-type 120$^\circ$-AFM phase occurs in the three-sublattice case for large $|J_{\rm d}|/z$.
This transition is found to be continuous.
Figure\ \ref{fig:TO-SNf-C120} presents the data of the squared total-spin moment in a dimer, $|{\bm T}^{\rm MF}_{\rm A}|^2$, for $J_{\rm d}/z=-3.5$ and $J_4=-0.1, -0.5, -1.5$.
When $|J_4|$ is not too small, the total-spin moment rises from zero continuously with a finite slope, indicating
${\bm T}^{\rm MF}_\Lambda \propto \sqrt{J_\times - J_{\times,{\rm c2}}}$, where $J_{\times,{\rm c2}}$ is the critical value.
The slope becomes steeper as $J_4$ approaches zero.
At $J_4=0$, the spin nematic phase vanishes (within the mean-field approximation) and there occurs a direct transition between the FM and C-type 120$^\circ$-AFM phases via a special point with the degenerate ground states at $J_\times=-J_\parallel=1$. (See Appendix\ \ref{subsec:append_MFA_zeroJ4}.)

\subsection{Transition between A-type AFM and C-type 120$^\circ$-AFM phases}
\label{subsec:append_MFA_AAFM-C120}

\begin{figure}
\begin{center}
\includegraphics[width=70mm]{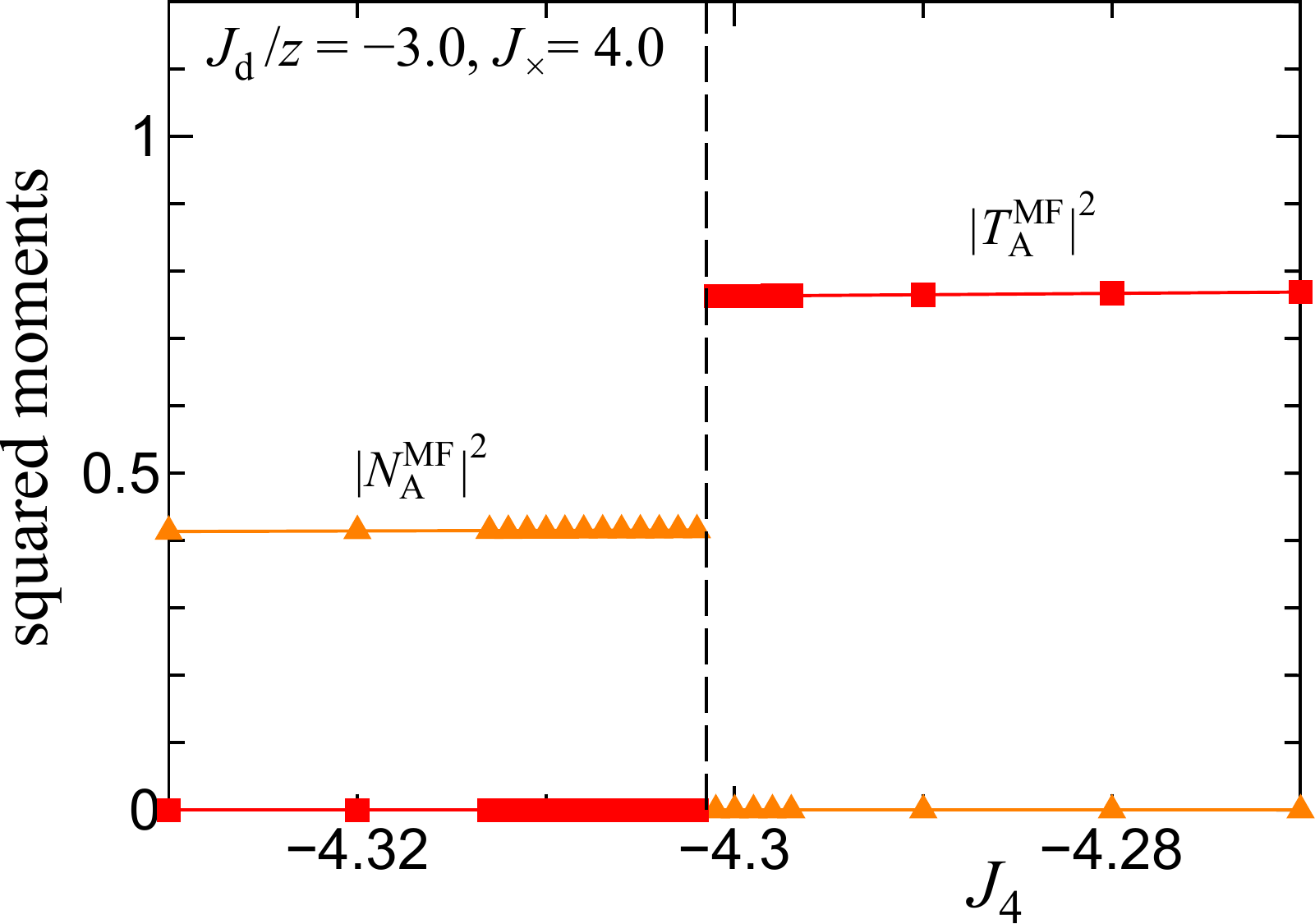}
\caption{
$J_4$-dependence of squared magnetic moments $|{\bm T}^{\rm MF}_{\rm A}|^2$ and $|{\bm N}^{\rm MF}_{\rm A}|^2$ for the three-sublattice structure. The other parameters are set as $J_\parallel=-1$, $J_\times=4.0$, and $J_{\rm d}/z=-3.0$.
Vertical dashed line represents the transition point between the A-type AFM phase and the C-type 120$^\circ$-AFM phase.
}
\label{fig:TO-AAFM-C120}
\end{center}
\end{figure}

The phase transition between the A-type AFM phase and the C-type 120$^\circ$-AFM phase occurs in the three-sublattice case for large $|J_{\rm d}|/z$.
Figure\ \ref{fig:TO-AAFM-C120} shows the $J_4$-dependence of the total-spin and N\'{e}el-spin moments on the parameter line with $J_{\rm d}/z=-3.0$ and $J_\times=4.0$.
The order parameters exhibit a clear jump at the transition.
The transition thus turns out to be the first-order one, occuring between two magnetically-ordered phases with distinct symmetries.

\section{SU(4) transformation on dimers}
\label{sec:append_su4}

In this Appendix, we briefly describe SU(4) transformation on a dimer.
The fifteen generators of SU(4) group\cite{Lecheminant2006} on a dimer are given by
the spin operators $T_j^\alpha $ and $N_j^\alpha$, the vector chiral operators $\chi_j^\alpha$ ($\alpha=x,y,z$), the quadrupolar operators $Q_j^{(n)}$ ($n=1,\cdots,5$), and the spin exchange
operator ${\cal O}_j$. Hereafter we omit the dimer index $j$.
Using these operators, we conveniently define the fifteen generators $\lambda_n$ ($n=1,\cdots,15$) of SU(4)
group as follows:
\begin{align}\label{eq:su4gen}
  &\lambda_1  = -Q^{(3)}, \ \ \ \    \lambda_2 = T^z, \ \ \ \
  \lambda_3  = -Q^{(1)}, \ \ \ \   \lambda_4  = -Q^{(5)},   \nonumber\\
  &\lambda_5 = -T^y, \ \ \ \   \lambda_6  = -Q^{(4)}, \ \ \ \
  \lambda_7 = T^x, \ \ \ \   \lambda_8  = Q^{(2)}, \nonumber\\
  &\lambda_9 =-  N^x, \ \ \ \   \lambda_{10} =-  \chi^x, \ \ \ \
  \lambda_{11} =-  N^y, \ \ \ \   \lambda_{12} =-  \chi^y, \nonumber\\
  &\lambda_{13} =-  N^z, \ \ \ \   \lambda_{14} =-  \chi^z, \ \ \ \
  \lambda_{15} = {\cal O} .
\end{align}
As for the orthonormal bases on a dimer, we use the following states
\begin{align}\label{eq:bases}
&|x\rangle  = \frac{1}{\sqrt2} (|\uparrow \uparrow\rangle
  -|\downarrow \downarrow\rangle),\nonumber\\
&|y\rangle  = \frac{1}{\sqrt2 i} (|\uparrow \uparrow\rangle
  +|\downarrow \downarrow\rangle),\nonumber\\
&|z\rangle  = -\frac{1}{\sqrt2} (|\uparrow \downarrow\rangle
  + |\downarrow \uparrow\rangle),\nonumber\\
&|0\rangle = \frac{1}{\sqrt2} (|\uparrow \downarrow\rangle
  - |\downarrow \uparrow\rangle).
\end{align}
Using these definitions, one can explicitly show  that the $4\times 4$ matrix representation of the generators $\lambda_n$
on the basis vector
$(|x\rangle, |y\rangle, |z\rangle, |0\rangle)$ coincides with
the $4 \times 4$ generalized Gell-Man matrices\cite{Greiner1994} which generate SU(4) algebra,
\begin{align}
&\lambda_1=\left[
  \begin{array}{cccc}
    0 & 1 & 0 & 0 \\
    1 & 0 & 0 & 0 \\
    0 & 0 & 0 & 0 \\
    0 & 0 & 0 & 0 \\
  \end{array}
\right],\ \ \
\lambda_2=\left[
  \begin{array}{cccc}
    0 & -i & 0 & 0 \\
    i & 0 & 0 & 0 \\
    0 & 0 & 0 & 0 \\
    0 & 0 & 0 & 0 \\
  \end{array}
\right],\nonumber\\
&\lambda_3=\left[
  \begin{array}{cccc}
    1 & 0 & 0 & 0 \\
    0 & -1 & 0 & 0 \\
    0 & 0 & 0 & 0 \\
    0 & 0 & 0 & 0 \\
  \end{array}
\right],\ \ \
\lambda_4=\left[
  \begin{array}{cccc}
    0 & 0 & 1 & 0 \\
    0 & 0 & 0 & 0 \\
    1 & 0 & 0 & 0 \\
    0 & 0 & 0 & 0 \\
  \end{array}
\right],\nonumber\\
&\lambda_5=\left[
  \begin{array}{cccc}
    0 & 0 & -i & 0 \\
    0 & 0 & 0 & 0 \\
    i & 0 & 0 & 0 \\
    0 & 0 & 0 & 0 \\
  \end{array}
\right],\ \ \
\lambda_6=\left[
  \begin{array}{cccc}
    0 & 0 & 0 & 0 \\
    0 & 0 & 1 & 0 \\
    0 & 1 & 0 & 0 \\
    0 & 0 & 0 & 0 \\
  \end{array}
\right],\nonumber\\
%\end{align}
%\begin{align}
&\lambda_7=\left[
  \begin{array}{cccc}
    0 & 0 & 0 & 0 \\
    0 & 0 & -i & 0 \\
    0 & i & 0 & 0 \\
    0 & 0 & 0 & 0 \\
  \end{array}
\right],\ \ \
\lambda_8=\frac{1}{\sqrt{3}}\left[
  \begin{array}{cccc}
    1 & 0 & 0 & 0 \\
    0 & 1 & 0 & 0 \\
    0 & 0 & -2 & 0 \\
    0 & 0 & 0 & 0 \\
  \end{array}
\right],\nonumber\\
&\lambda_9=\left[
  \begin{array}{cccc}
    0 & 0 & 0 & 1 \\
    0 & 0 & 0 & 0 \\
    0 & 0 & 0 & 0 \\
    1 & 0 & 0 & 0 \\
  \end{array}
\right],\ \ \
\lambda_{10}=\left[
  \begin{array}{cccc}
    0 & 0 & 0 & -i \\
    0 & 0 & 0 & 0 \\
    0 & 0 & 0 & 0 \\
    i & 0 & 0 & 0 \\
  \end{array}
\right],\nonumber
\end{align}
\begin{align}
&\lambda_{11}=\left[
  \begin{array}{cccc}
    0 & 0 & 0 & 0 \\
    0 & 0 & 0 & 1 \\
    0 & 0 & 0 & 0 \\
    0 & 1 & 0 & 0 \\
  \end{array}
\right],\ \ \
\lambda_{12}=\left[
  \begin{array}{cccc}
    0 & 0 & 0 & 0 \\
    0 & 0 & 0 & -i \\
    0 & 0 & 0 & 0 \\
    0 & i & 0 & 0 \\
  \end{array}
\right],\nonumber\\
&\lambda_{13}=\left[
  \begin{array}{cccc}
    0 & 0 & 0 & 0 \\
    0 & 0 & 0 & 0 \\
    0 & 0 & 0 & 1 \\
    0 & 0 & 1 & 0 \\
  \end{array}
\right],\ \ \
\lambda_{14}=\left[
  \begin{array}{cccc}
    0 & 0 & 0 & 0 \\
    0 & 0 & 0 & 0 \\
    0 & 0 & 0 & -i \\
    0 & 0 & i & 0 \\
  \end{array}
\right],\nonumber\\
&\lambda_{15}=\frac{1}{\sqrt{6}}\left[
  \begin{array}{cccc}
    1 & 0 & 0 & 0 \\
    0 & 1 & 0 & 0 \\
    0 & 0 & 1 & 0 \\
    0 & 0 & 0 & -3 \\
  \end{array}
\right].
\end{align}
Arbitrary SU(4) transformation is given by these generators parameterized by generalized Euler angles.\cite{Tilma2002}

\section{Non-trivial degeneracies in the mean-field approximation}
\label{sec:append_degeneracy_MFA}
We describe non-trivial degeneracies in the mean-field solutions in the two types of phase boundaries and a triple point found in Sec.\ \ref{subsec:Vari_product}.
One boundary exists between the ferromagnetic and spin-nematic phases, which indicates emergent
SU(3) symmetry. The other exists
in the boundary between the spin-nematic and dimer-singlet phases.

\subsection{Emergent SU(3) symmetry in the boundary between the FM and spin-nematic
phases}\label{sec:append_SU3}

In the phase boundary between the FM phase and the spin nematic phase with ferro-quadrupolar order,
the ground state manifold has non-trivial degeneracy corresponding to SU(3) rotation.
We first briefly summarize SU(3) rotation on a dimer. The eight  generators $\lambda_n$ ($n=1,\cdots,8$)
of SU(3) rotation are given by
$T^\alpha$ ($\alpha=x,y,z$) and
$Q^{(n)}$ ($n=1,\cdots,5$)  as
in Eqs.~(\ref{eq:su4gen}).
Using these generators, we can write arbitrary SU(3) rotation in the form\cite{Byrd1997}
\begin{align}
 &U(\alpha,\beta,\gamma,\theta,a,b,c,\phi)= \nonumber\\
 &\exp(-i \alpha Q_j^{(1)})
    \exp(i \beta T_{j}^z)
    \exp(-i \gamma Q_j^{(1)}) \exp(-i \theta T_{j}^y)\nonumber\\
  & \times
    \exp(-i a Q_j^{(1)}) \exp(i b T_{j}^z) \exp(-i c Q_j^{(1)})
    \exp(i \phi Q_j^{(2)})
    \label{eq:su3trans}
\end{align}
with parameters ($\alpha,\beta,\gamma,\theta,a,b,c,\phi$).
This gives arbitrary unitary transformation among the spin triplet.
As an initial state, we use the state
 $ |z\rangle_j$.
Since this state is invariant under the last (right) four rotations,
the transformed state is simply written as
\begin{align}
  U (\alpha,\beta,&\gamma,\theta,a,b,c,\phi)  |z\rangle_j
  = e^{-\frac{2}{\sqrt3} i \phi}\{  \cos \theta |z\rangle_j \nonumber\\
  &+
  e^{i \gamma} \sin \theta (e^{i \alpha } \cos \beta  |x\rangle_j  -e^{-i \alpha} \sin \beta |y\rangle_j )
    \}.
   \label{eq:any_su3}
\end{align}

In the  mean-field approximation performed in Sec.~\ref{subsec:Vari_product},
the FM state and the spin nematic state are degenerate in energy in the boundary between these two phases.
We further find that, in the mean-field solutions,  any state obtained by arbitrary global
SU(3) rotation to the FM  state,
which is a translationally invariant product state of Eq.\ (\ref{eq:any_su3}), also takes the exactly same energy.
This is emergent non-trivial degeneracy associated with the global SU(3) rotation.
%On the SU(4) symmetric point, which exists inside of the parameter space
%of the phase boundary,
%this degeneracy comes from the inherent exact SU(3) symmetry of
%the Hamiltonian.
As mentioned in Sec.\ \ref{subsec:symmetries}, the Hamiltonian has
the exact global SU(3) symmetry and the corresponding degenerate ground states
only in the parameter space (\ref{eq:SU3_line}),
which exists inside of  the mean-field phase boundary.
This degeneracy thus remains in the whole phase boundary in the mean-field approximation even though the model
Hamiltonian does not possess the  SU(3) symmetry.

\subsection{Boundary between the spin-nematic and  dimer-singlet phases}
\label{subsec:nematic-singlet-boundary_at_zeroJ4}
In the mean-field approximation, the spin nematic phase touches with the dimer singlet phase in a finite parameter plane at $J_{\rm d}=0$,
as shown in Figs.\ \ref{fig:PDN2}(a) and \ref{fig:PDN3}(a).
In this phase boundary, the ground state manifold has non-trivial degeneracy
corresponding to ${\rm SU(2)}\times {\rm SU(2)}$ rotation, even though the Hamiltonian does not have this symmetry except for the
special case of $J_{\rm d}=J_\times=0$. Among the spin-nematic ground states, we consider the product state of
$|z\rangle$ without
loss of generality. All other states are related with the global SU(2) rotation. We next apply SU(2) rotation only to one spin of each dimer,
\begin{align}
 & \exp (i {\bm \omega}\cdot  {\bm S}_{1,j}) |z\rangle_j
  =  \cos \frac{|{\bm \omega}|}{2} |z \rangle_j \nonumber\\
 &+ \sin \frac{|{\bm \omega}|}{2} (-\hat{{\bm \omega}}\cdot \hat{{\bm y}} | x \rangle_j
 +\hat{{\bm \omega}} \cdot \hat{{\bm x}} | y \rangle_j - i \hat{{\bm \omega}}\cdot \hat{{\bm z}} |0\rangle_j )  ,
\end{align}
where $\hat{{\bm \alpha}}$ ($\alpha=x,y,z$) denotes the unit vector parallel to $\alpha$-axis and
$\hat{{\bm \omega}}$ the unit vector parallel to ${\bm \omega}$.
The quadrupolar state ($|{\bm \omega}|=0$) is
continuously transformed to the vector chiral states
($0<|{\bm \omega}|<\pi$) and the dimer singlet state ($|{\bm \omega}|=\pi$)
when $\hat{\bm \omega}\parallel \hat{\bm z}$.
By a straightforward calculation, one can show that the translationally invariant product state
of these dimer bases also has the same energy as the
ferro-quadrupolar state.
Thus the ground state manifold has the same degrees of freedom as global ${\rm SU(2)}\times {\rm SU(2)}$ rotation.
This degeneracy exists in the whole phase boundary between the spin-nematic and  dimer-singlet
phases in the mean-field approximation.

\subsection{Triple point for the FM, spin nematic, and C-type AFM phases
}
\label{subsec:append_MFA_zeroJ4}

For  large $|J_{\rm d}|/z$,
the triple point for the FM, spin nematic, and C-type (120$^\circ$-)AFM phases exists at
$J_\times=-J_\parallel$ and $J_4=0$ in both two-sublattice and three-sublattice cases
in the mean-field approximation.
On this triple point,
any product state in which each dimer independently take
an arbitrary superposition of spin-triplet states
has the same energy to form the massively-degenerate ground state manifold
in the mean-field level.
%the ground states are  massively degenerate,  in which each dimer can
%independently take an arbitrary superposition of spin-triplet states.

\bibliographystyle{apsrev4-1}
\bibliography{FMdim_resub}

\end{document}